# ABSTRACT

Title of Dissertation: Efficient learning-based sound propagation
for virtual and real-world audio processing applications

Anton Jeran Ratnarajah
Doctor of Philosophy, 2024

Dissertation Directed by: Professor Dinesh Manocha
Department of Electrical and Computer Engineering

Sound propagation is the process by which sound energy travels through a medium, such as air, to the surrounding environment as sound waves. The room impulse response (RIR) describes this process and is influenced by the positions of the source and listener, the room's geometry, and its materials. Physics-based acoustic simulators have been used for decades to compute accurate RIRs for specific acoustic environments. However, we have encountered limitations with existing acoustic simulators. For example, they require a 3D representation and detailed material knowledge of the environment.

To address these limitations, we propose three novel solutions. First, we introduce a learning-based RIR generator that is two orders of magnitude faster than an interactive ray-tracing simulator. Our approach can be trained to input both statistical and traditional parameters directly, and it can generate both monaural and binaural RIRs for both reconstructed and synthetic 3D scenes. Our generated RIRs outperform interactive ray-tracing simulators in speech-processing applications, including Automatic Speech Recognition (ASR), Speech Enhancement, and Speech Separation, by 2.5%, 12%, and 48%, respectively.

Secondly, we propose estimating RIRs from reverberant speech signals and visual cues in the absence of a 3D representation of the environment. By estimating RIRs from reverberant speech, we can augment training data to match test data, improving the word error rate of the ASR system. Our estimated RIRs achieve a 6.9% improvement over previous learning-based RIR estimators in real-world far-field ASR tasks. We demonstrate that our audio-visual RIR estimator aids tasks like visual acoustic matching, novel-view acoustic synthesis, and voice dubbing, validated through perceptual evaluation.

Finally, we introduce IR-GAN to augment accurate RIRs using real RIRs. IR-GAN parametrically controls acoustic parameters learned from real RIRs to generate new RIRs that imitate different acoustic environments, outperforming Ray-tracing simulators on the Kaldi far-field ASR benchmark by 8.95%.

# Efficient learning-based sound propagation for virtual and real-world audio processing applications

by

Anton Jeran Ratnarajah

Dissertation submitted to the Faculty of the Graduate School of the University of Maryland, College Park in partial fulfillment of the requirements for the degree of Doctor of Philosophy
2024

Advisory Committee:
Professor Dinesh Manocha, Chair/Advisor
Professor Nikhil Chopra, Dean's Representative
Professor Carol Espy-Wilson
Professor Ramani Duraiswami
Professor Sanghamitra Dutta

# Acknowledgments

My PhD journey has been filled with challenges, scientific discoveries, and personal growth. I am deeply grateful to many individuals who have encouraged and supported me throughout these five years. First and foremost, I extend my heartfelt thanks to Professor Dinesh Manocha, my PhD supervisor, for his invaluable experience and knowledge that guided me in conducting high-quality research. His motivation, intellectual freedom, and constant encouragement were instrumental in making my PhD journey a great success.

I am also thankful to Tencent AI Lab, META, and Dolby for providing the funding and resources needed to tackle practical research problems related to my PhD. Additionally, I appreciate Professor Dinesh Manocha and Professor Shuvra Bhattacharyya for offering me a Research Assistantship during my PhD. I am grateful to the Electrical and Computer Engineering Department and the Computer Science Department for their support through a Teaching Assistantship.

My gratitude extends to my industry collaborators Dr. Shi-Xiong Zhang, Dr. Paul Calamia, Dr. Dong Yu, Dr. Meng Yu, Dr. Ishwarya Ananthabhotla, Dr. Vamsi Krishna Ithapu, Dr. Scott Norcorss, and Dr. Pablo Hoffmann for their dedication and time in brainstorming preliminary ideas and transforming them into meaningful research findings. I am equally grateful to my academic collaborators Dr. Zhenyu Tang, Rohtih Aralikatti, Sreyan Ghosh, Sonal Kumar, and Purva Chiniya for helping to validate my research findings. A special thanks to Dr. Zhenyu Tang for introducing me to acoustic research and for his support in sharing his knowledge and experience during the early days of my PhD.

I extend my sincere appreciation to Professor Carol Espy-Wilson, Professor Behtash Babadi, Professor K J Ray Liu, Professor Furong Huang, and Professor Manoj Franklin for allowing me to serve as their Teaching Assistant and for aiding in the development of my teaching skills by supporting me in leading discussions and lab sessions.

Pursuing a PhD far from my home country, Sri Lanka, has been mentally and physically challenging. I am grateful to my Sri Lankan friends at UMD for their unwavering support during tough times and for standing by me throughout my PhD journey. Finally, I am deeply thankful for the continuous love and support of my parents and sisters, which has helped me stay strong during difficult moments.

# Table of Contents

# List of Tables

List of Figures

# List of Abbreviations

| | |
|---|---|
| 3D | Three dimensional |
| AC | Absorption Coefficient |
| AV | Audio Visual |
| AMT | Amazon Mechanical Turk |
| API | Application Programming Interface |
| AR | Augmented Reality |
| ASR | Automatic Speech Recognition |
| BIR | Binaural Impulse Response |
| CAD | Computer-Aided Design |
| CGAN | Conditional Generative Adversarial Network |
| CLIP | Contrastive Language Image Pre-training |
| CPU | Central Processing Unit |
| CRIP | Contrastive RIR Image Pre-training |
| CTE | Early-to-late index |
| DAS | Diffuse Acoustic Simulator |
| dB | Decibels |
| DRR | Direct-to-Reverberant Ratio |
| ED | Energy Decay |
| EDC | Energy Decay Curve |
| EDR | Energy Decay Relief |
| EDT | Early-Decay-Time |
| ERE | Early Reflection Energy |
| EQ | Equalization |
| FiNS | Filtered Noise Shaping |
| FIR | Finite Impulse Response |
| GAN | Generative Adversarial Network |
| GAS | Geometric Acoustic Simulator |
| GPU | Graphical Processing Unit |
| GT | Ground Truth |
| GWA | Geometric-Wave Acoustic |
| IHM | Individual Headset Microphones |
| ILD | Interaural Level Difference |
| IR | Impulse Response |
| ITD | Interaural Time Difference |
| MFCC | Mel-Frequency Cepstral Coefficients |
| MSE | Mean Square Error |
| NAC | Neural Audio Codecs |
| OB | Overlapped Binaural speech |
| RIR | Room Impulse Response |
| RGB-D | Red Green Blue – Depth |
| RVQ | Residual Vector Quantization |
| SD | Standard Deviation |
| SDM | Single Distant Microphones |
| SOTA | State Of The Art |
| STFT | Short-Time Fourier Transform |
| $T_{60}$ | Reverberation Time |
| VAM | Visual Acoustic Matching |
| VR | Virtual Reality |
| WER | Word Error Rate |
| WGAN | Wasserstein Generative Adversarial Networks |

# Chapter 1: Introduction

Rapid advancements in interactive applications (e.g., games, virtual environments, speech recognition) require realistic sound effects in complex dynamic environments with multiple sound sources. Creating these realistic sound effects remains challenging due to the geometric, material and auditory complexity of real-world settings. Geometric complexity refers to the number of vertices and faces required to represent the intricate environment as a 3D mesh accurately. Material complexity involves the material properties of each object in the environment. Auditory complexity depends on the number of sound sources, the presence of static and dynamic objects, and various acoustic effects (e.g., early reflection, late reverberation, diffraction, scattering). The way the sound propagates from the speaker to the listener is characterized by a transfer function known as room impulse response (RIR) [10]. RIR represents the intensity and time of arrival of direct sound, early reflections and late reverberation (Figure 1.1).

The RIR can be recorded from a physical environment using different techniques [4,11–13]. Recording real RIRs requires substantial human labor and specialized hardware. Consequently, many interactive applications use synthetic RIRs, which can be generated using physics-based acoustic simulators [14–16], to train speech processing systems such as automated speech recognition, speech enhancement and speech separation [1, 17–19].

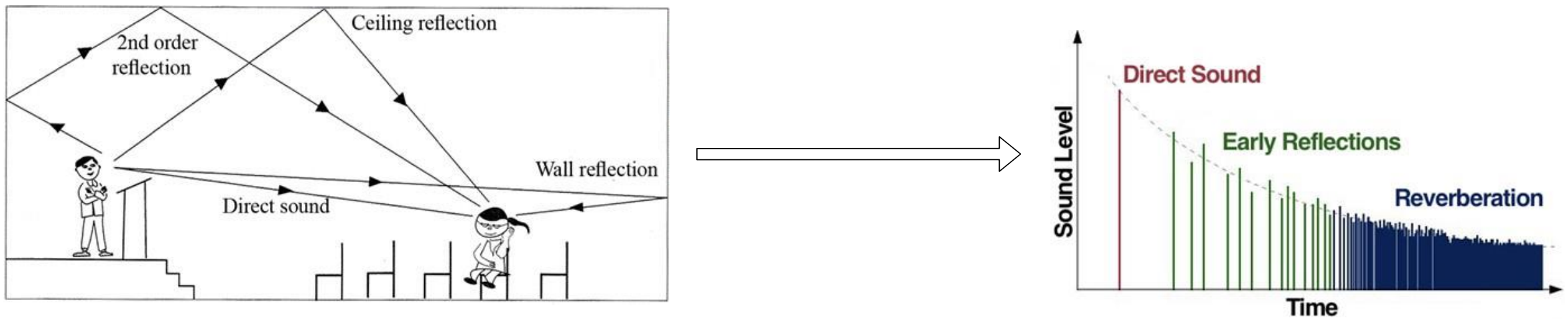

Figure 1.1: The time-domain plot of a typical Room Impulse Response (RIR) illustrates its decomposition into three components: direct response, early reflections, and late reverberation. The direct response depends on the line of sight between the sound source and the listener. Early reflections are sounds that arrive shortly after the direct response, bouncing off surfaces like the ceiling and walls. Late reverberation consists of higher-order reflections, which are more diffuse and arrive later.

Sound propagation can be accurately simulated using wave-based methods, which solve wave equations using various numerical solvers such as the boundary-element method [20], the finite-difference time-domain simulation [21], etc. The complexity of the wave-based approach increases with the fourth power of the frequency and these methods are mostly limited to static scenes. Geometric acoustic algorithms reduce the complexity of realistic wave-based sound propagation by approximating sound waves as rays (ray assumption) [16]. These algorithms can handle complex environments with dynamic objects [22] and multiple sources [23]. However, geometric acoustic methods do not accurately model the low-frequency components of impulse responses (IRs) due to the ray assumption. Sound waves can be treated as rays when their wavelength is smaller than the obstacles in the environment. At low frequencies under 500 Hz, this assumption is invalid for most scenes, resulting in significant simulation errors [24]. Hybrid sound propagation algorithms [25, 26] combine IRs from both wave-based and geometric techniques to generate accurate IRs across the human auditory range in complex dynamic scenes. However, generating IRs for thousands of sound sources at an interactive rate is not feasible with the hybrid method due to its complexity.

Most commonly used acoustic simulators cannot model all the acoustic effects in an environment that real RIRs can capture. For example, geometric acoustic simulators [1] can only accurately simulate high-frequency acoustic effects but are less accurate for low-frequency effects like diffraction or interference. Additionally, simulating accurate RIRs requires accurate 3D mesh representations of the underlying scenes [3] and comprehensive knowledge of material properties [26, 27]. Consequently, it is impractical to precisely simulate RIRs for a real-world environment without a detailed 3D mesh representation.

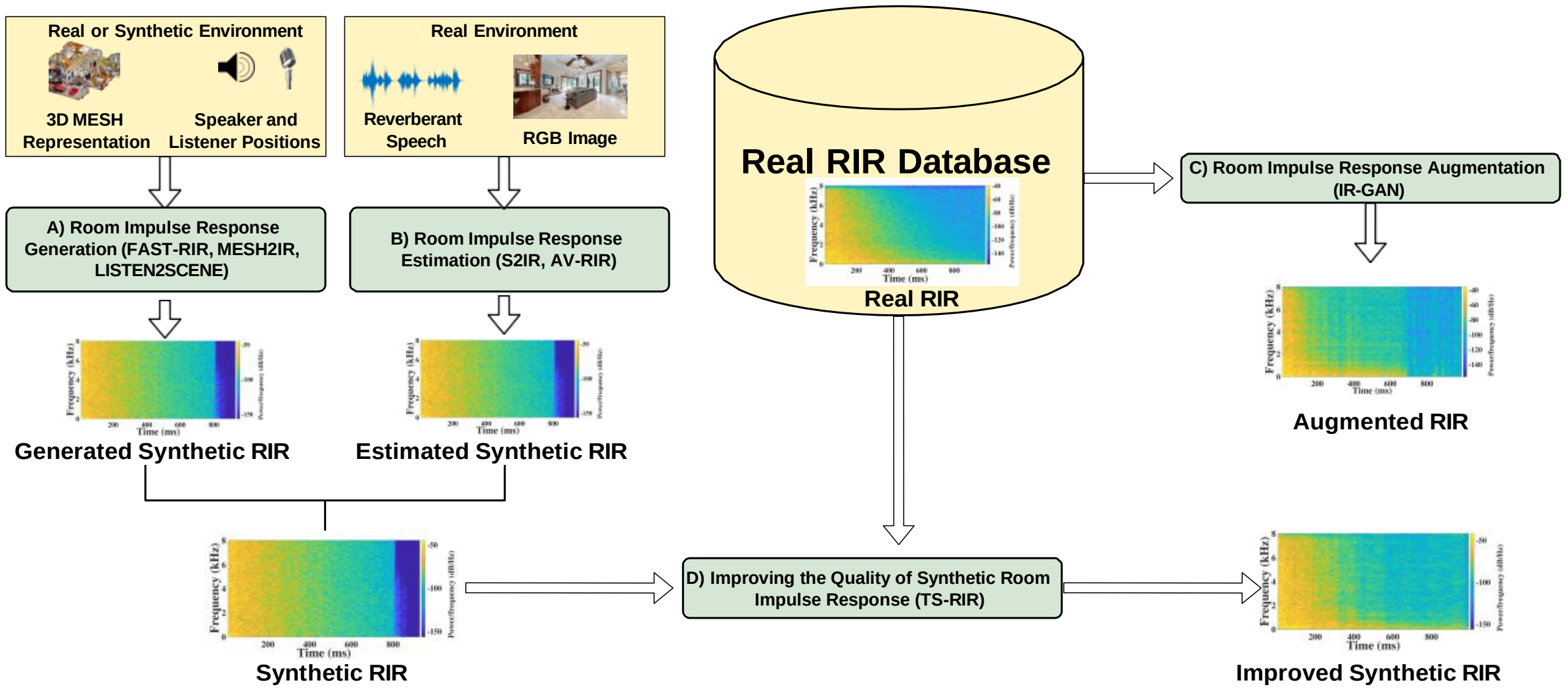

Figure 1.2: The overall research goal is to generate high-quality Room Impulse Responses (RIRs). We present learning-based RIR generators (FAST-RIR, MESH2IR, and LISTEN2SCENE) to generate thousands of synthetic RIRs for specific speaker and listener positions for a given 3D mesh representation of the environment [A]. We also propose RIR estimators (S2IR and AV-RIR) to estimate synthetic RIRs from reverberant speech signals and RGB images of the environment [B]. Additionally, we introduce a GAN-based model, IR-GAN, to augment RIRs using a real RIR database [C]. Finally, we propose TS-RIR to enhance the quality of synthetic RIRs using real RIRs [D].

## 1.1 Our Research Goals

The primary aim of the research proposed for our dissertation is to create a series of learning-based algorithms and tools to generate or estimate high-quality room impulse responses (RIRs) at interactive rates for both virtual and real-world 3D scenes. Additionally, we utilize the acoustic effects captured from real RIRs to enhance the accuracy of the generated synthetic RIRs. Figure 1.2 illustrates the overall goal of our research. Our main contributions can be summarized in the following four areas:

1. **Room Impulse Response Generation**

   We propose two neural-network-based RIR generators for automatic speech recognition applications (FAST-RIR) and AR/VR applications (MESH2IR and Listen2Scene).

   - **Fast Room Impulse Response Generator for Automatic Speech Recognition**

     We present a neural-network-based fast diffuse room impulse response generator (FAST-RIR) for generating room impulse responses (RIRs) for a given acoustic environment. Our FAST-RIR takes rectangular room dimensions, listener and speaker positions, and reverberation time ($T_{60}$) as inputs and generates specular and diffuse reflections for a given acoustic environment. Our FAST-RIR is capable of generating RIRs for a given input $T_{60}$ with an average error of 0.02s. We evaluate our generated RIRs in automatic speech recognition (ASR) applications using Google Speech API, Microsoft Speech API, and Kaldi tools. We show that our proposed FAST-RIR with batch size 1 is 400 times faster than a state-of-the-art diffuse acoustic simulator (DAS)

on a CPU and gives similar performance to DAS in ASR experiments. Our FAST-RIR is 12 times faster than an existing GPU-based RIR generator (gpuRIR) [28]. We show that our FAST-RIR outperforms gpuRIR by 2.5% in an AMI far-field ASR benchmark.

- **Fast Room Impulse Response Generator for AR/VR Applications**

  We present a novel neural-network-based sound propagation method to render audio for virtual and real 3D scenes in real-time for virtual reality (VR) and augmented reality (AR) applications. Our approach is general and can generate RIRs for arbitrary topologies and material properties in the 3D scenes, based on the source and the listener locations. Any clean audio or dry audio can be convolved with the generated acoustic effects to render audio corresponding to the real environment. We propose a graph neural network that uses both the material and the topology information of the 3D scenes and generates a scene latent vector. Moreover, we use a conditional generative adversarial network (CGAN) to generate acoustic effects from the scene latent vector. Our network can handle holes or other artifacts in the reconstructed 3D mesh model. We present an efficient cost function for the generator network to incorporate spatial audio effects. Given the source and the listener position, our learning-based binaural sound propagation approach can generate an acoustic effect in 0.1 milliseconds on an NVIDIA GeForce RTX 2080 Ti GPU and can easily handle multiple sources. We have evaluated the accuracy of our approach with binaural acoustic effects generated using an interactive geometric sound propagation algorithm and captured real acoustic effects. We also performed a perceptual evaluation and observed that the audio rendered by our approach is more plausible than audio rendered using prior learning-based sound propagation algorithms.

2. **Room Impulse Response Augmentation**

We present a Generative Adversarial Network (GAN) based room impulse response generator (IR-GAN) for augmenting realistic synthetic RIRs using real-world RIRs. IR-GAN extracts acoustic parameters from captured real-world RIRs and uses these parameters to generate new synthetic RIRs. We use these generated synthetic RIRs to improve far-field automatic speech recognition in new environments that are different from the ones used in training datasets. In particular, we augment the far-field speech training set by convolving our synthesized RIRs with a clean LibriSpeech dataset [29]. We evaluate the quality of our synthetic RIRs on the far-field LibriSpeech test set created using real-world RIRs from the BUT ReverbDB [30] and AIR [31] datasets. Our IR-GAN reports up to an 8.95% lower error rate than the Geometric Acoustic Simulator (GAS) in far-field speech recognition benchmarks. We further improve the performance when we combine our synthetic RIRs with synthetic impulse responses generated using GAS. This combination can reduce the word error rate by up to 14.3% in far-field speech recognition benchmarks.

3. **Improving the Quality of Synthetic Room Impulse Responses**

Over the past decades, several physics-based acoustic simulators have been proposed to generate synthetic RIRs. However, there is still a gap between the performance of RIRs generated using acoustic simulators and the performance of real RIRs. Most commonly used acoustic simulators are unable to model all the acoustic effects in the environment, which can be captured by real RIRs. For example, ray-tracing-based acoustic simulators cannot model low-frequency effects accurately because of ray assumptions.

We present a novel approach to improve the accuracy of synthetic RIRs. We design a TS-RIR-gan architecture to translate the synthetic RIR to a real RIR. TS-RIR-gan takes synthetic RIRs as audio samples to translate them into real RIRs and uses multiple loss functions. We also perform real-world sub-band room equalization to the translated RIRs to further improve their quality. We demonstrate the benefits of our post-processed RIRs in far-field automatic speech recognition systems.

4. **Room Impulse Responses Estimation**

We developed two RIR estimators for automatic speech recognition application (S2IR-GAN) and AR/VR applications (AV-RIR).

- **Room Impulse Responses Estimator for Automatic Speech Recognition**

  We propose to characterize and improve the performance of blind room impulse response (RIR) estimation systems in the context of a downstream application scenario, far-field automatic speech recognition (ASR). We first draw the connection between improved RIR estimation and improved ASR performance, as a means of evaluating neural RIR estimators. We then propose a GAN-based architecture that encodes RIR features from reverberant speech and constructs an RIR from the encoded features, and uses a novel energy decay relief loss to optimize for capturing energy-based properties of the input reverberant speech. We show that our model outperforms the state-of-the-art baselines on acoustic benchmarks (by 72% on the energy decay relief and 22% on an early-reflection energy metric), as well as in an ASR evaluation task (by 6.9% in word error rate).

- **Audio-Visual Room Impulse Response Estimator for AR/VR Applications**

  Accurate estimation of Room Impulse Response (RIR), which captures an environment's acoustic properties, is important for speech processing and AR/VR applications. We propose AV-RIR, a novel multi-modal multi-task learning approach to accurately estimate the RIR from a given reverberant speech signal and the visual cues of its corresponding environment. AV-RIR builds on a novel neural codec-based architecture that effectively captures environment geometry and materials properties and solves speech dereverberation as an auxiliary task by using multi-task learning. We also propose Geo-Mat features that augment material information into visual cues and CRIP that improves late reverberation components in the estimated RIR via image-to-RIR retrieval by 86%. Empirical results show that AV-RIR quantitatively outperforms previous audio-only and visual-only approaches by achieving 36% - 63% improvement across various acoustic metrics in RIR estimation. Additionally, it also achieves higher preference scores in human evaluation. As an auxiliary benefit, dereverbed speech from AV-RIR shows competitive performance with the state-of-the-art in various spoken language processing tasks and outperforms reverberation time error score in the real-world AVSpeech dataset. Qualitative examples of both synthesized reverberant speech and enhanced speech are available online[1].

[1] https://anton-jeran.github.io/AVRIR/

**THESIS STATEMENT**

*By utilizing learning-based techniques, realistic audio can be rendered for both real and virtual environments at interactive rates, enhancing the performance of gaming, augmented reality, virtual reality, and speech applications.*

## 1.2 Organization

Our dissertation is organized as follows: Chapter 2 provides a comprehensive background and a detailed review of related work relevant to our research objectives. Chapter 3 introduces our fast room impulse response generator for automated speech recognition. Chapters 4 and 5 focus on our fast room impulse response generator for AR/VR applications. Chapter 6 presents our method for room impulse response augmentation. In Chapter 7, we explore ways to improve the quality of synthetic room impulse responses. Chapters 8 and 9 detail our room impulse response estimator, with Chapter 8 focusing on automated speech recognition and Chapter 9 on AR/VR applications. Finally, Chapter 10 summarizes our findings and proposes directions for future research.

# Chapter 2: Related Works

## 2.1 Physics-Based RIR Computation

Many wave-based [32–34], geometric-based [35–38], and hybrid [26, 39] interactive RIR simulation algorithms have been proposed to simulate RIRs for complex scenes [40]. The wave-based algorithms are computationally expensive and their runtime is proportional to the third or fourth power of the highest simulation frequency [15]. Wave-based and hybrid algorithms precompute the RIRs for a static scene and, at runtime, the RIR for an arbitrary listener position is calculated by efficient interpolation techniques [25,41,42]. These precomputation-based interactive RIR simulation algorithms can be used only for static scenes [40]. Geometric sound propagation algorithms are based on ray tracing or its variants are proposed for interactive RIR simulation in dynamic scenes [22, 43, 44]. The limitations of geometric-based techniques, such as simulation error at lower frequencies, are inherent in these interactive geometric-based algorithms. There are no general physics-based algorithms known for computing accurate RIRs, including low-frequency components, for general dynamic scenes.

## 2.2 Learning-Based RIR Computation

Learning-based RIR computation have been proposed to generate RIRs based on a single image of the environment [5, 45, 46], reverberant speech signal [4, 6], or shoe-box shaped room geometry [2]. Neural networks are also used to translate synthetic RIRs to real RIRs and to augment RIRs [47,48] and estimate room acoustic parameters [49–51]. Learning-based approaches are proposed to learn the implicit representation of RIRs for a given 3D scene and predict RIRs for new locations on the same training scene [52, 53]. MESH2IR [3, 54] is a sound propagation network that takes the complete 3D mesh of a 3D scene and the source and the listener positions as input and generates monaural RIRs in real-time on a high-end GPU. However, the audio rendered using these learning-based sound propagation methods may not be smooth and can have artifacts. Prior learning-based binaural sound propagation methods require a few RIRs captured in a new 3D scene to generate new RIRs for different source and listener locations in the same 3D scene [46].

## 2.3 Indoor 3D Scenes Datasets for RIR Computation

The indoor 3D scenes can be either captured using RGB-D videos and reconstructed as meshes (real-world scenes) [55–57] or created by humans using professionally designed software (synthetic models) [58–63]. The mesh quality in the real-world scene datasets is not as good as the quality of the synthetic models because 3D scene reconstruction with accurate geometric details is challenging with existing computer vision methods and capturing hardware. Among the synthetic model datasets, 3D-FRONT [58] contains large-scale synthetic furnished indoor 3D scenes with fine geometric and texture details. The 3D-FRONT dataset has 6813 CAD houses, where 18,968 rooms are furnished with 13,151 3D furniture objects. The furniture is placed in varying numbers in meaningful locations in each room (e.g., living room, dining room, kitchen, bedroom, etc.)

## 2.4 Material Estimation for RIR Computation

The materials in the real scene influence the acoustic effects corresponding to the scene. The material information can be estimated from images and videos of real scenes and given as input to sound propagation algorithms using material acoustic coefficients [1, 64, 65]. Other methods are based on capturing reference audio samples or IRs in real scenes and the simulated IRs are adjusted to match the materials using reference audios or IRs [66–68]. In recent works, real scenes are annotated using crowd-sourcing [56] and material acoustic coefficients can be estimated by mapping the real scenes' annotated material labels to materials in the existing acoustic coefficient database [26, 27].

## 2.5 Measured RIR

To overcome the limitations of synthetic RIRs, real RIRs are recorded in a controlled environment using different techniques [11–13]. The maximum length sequence method [69], the time-stretched pulses method [11], and the exponential sine sweep method [12] are common methods to measure real RIRs. Among these approaches, the exponential sine sweep method is robust to changing loudspeaker output volume and performs well in automatic speech recognition tasks. The real RIRs in BUT ReverbDB [30] are collected using the exponential sine sweep method. Since collecting real RIRs is time-consuming and technically difficult, only a limited number of real RIR datasets are available.

## 2.6 Techniques of Improving Synthetic RIR

The geometric acoustic simulators are unable to model low-frequency wave effects such as diffraction [70] and room resonance [68] because of the ray assumption. On the other hand, we observe a boost or diminishing effect in the frequency response at different frequency bands in real RIRs due to wave modes created by room resonance. Some methods tend to compensate for the missing room response in synthetic RIRs using a sub-band room equalization approach [24].

## 2.7 RIR Datasets

The publicly available RIR datasets are either recorded in a real-world environments (recorded IR) [30, 71–80] or generated using RIR simulation tools (synthetic IR) [26, 54, 81–83]. The recorded RIR datasets are limited in size (fewer than 5000 RIRs) and number of environments (fewer than 10 rooms), and no sufficient information on the recording conditions is provided to train a deep learning model. Synthetic RIR datasets like SoundSpaces [83] and GWA [26] contain millions of RIRs. The RIRs in SoundSpaces is simulated using a geometric acoustic method [35] and GWA dataset contains high-quality RIRs computed using a hybrid method.

## 2.8 RIR Estimation from Reverberant Speech

Several algorithms have been proposed to blindly estimate an RIR from a reverberant source signal using traditional signal-processing approaches [84–88]. For some methods that take an $\ell_1$-norm-based approach, performance depends significantly on the choice of a regularization parameter corresponding to a real-world scenario [88]; some require multi-channel speech signals [88]; and most assume that either the source signal is a modulated Gaussian pulse [86,87], or that the speaker and microphone characteristics are known [84, 85]. For far-field automatic speech recognition tasks, however, we are required to estimate RIRs from reverberant speech source signals independent of speaker and microphone characteristics.

Recently, a neural network model was proposed to estimate the RIR from single-channel reverberant speech (FiNS) [4]. The FiNS model directly estimates early RIR components, and estimates late components as a combination of decaying filtered noise signals.

Estimating energy-based acoustic parameters such as reverberation time ($T_{60}$) and direct-to-reverberant ratio ($DRR$) and incorporating them in speech dereverberation and speech recognition systems have shown improved performance [89–91]. Therefore, we expect that accurately estimating the energy distribution in the estimated RIRs helps to improve the performance in far-field automatic speech recognition tasks.

## 2.9 Audio Processing using RIR

RIRs are used in a wide range of practical applications such as audio-visual navigation [52, 83,92–94], acoustic matching [95,96], sound rendering [23], novel-view audio generation [97],sound source localization [98], floorplan reconstruction [99,100], speech enhancement [101,102], speech recognition [17–19, 103, 104], and sound separation [105–108]. In most applications, learning-based networks are trained on large, diverse synthetic datasets and tested in real-world environments [1]. The performance of the deep neural network trained for a particular application is depended on the similarity of the synthetic training dataset and the real-world test environment [48]. Synthetic training datasets are created using RIR generators. Therefore, the accuracy of the RIR generator plays an important role in practical applications that depends on RIRs.

## 2.10 Audio Visual Learning

The advancement of multi-modal architectures [109,110] enables improving the performance of several applications by using audio-visual inputs. Recently, several algorithms are proposed for audio-visual sound separation [111–114], audio-visual action recognition [115–118], audio-visual navigation [119–121] and audio-visual localization [122–125]. However, no prior works investigate audio-visual RIR estimation.

## 2.11 Speech Dereverberation

The human auditory cortex's ability to adaptively filter out reverberation in various acoustic environments is well-documented [126]. Inspired by this, researchers have developed speech enhancement systems capable of transforming reverberant speech to anechoic speech [127–129]. Initially focused on multi-microphone inputs [130–132], recent deep learning techniques have shown promise with single-channel inputs [133–136]. While visual cues have been explored for speech enhancement, most studies have concentrated on near-field ASR using visible lip movements of the speaker [137–139]. Recent works use panoramic room images for speech dereverberation [140–142].

# Chapter 3: Fast Room Impulse Response Generator for Automatic Speech Recognition

## 3.1 Motivation

In recent years, an increasing number of RIR generators have been introduced to generate a realistic RIR for a given acoustic environment [1, 16, 44, 143] for speech applications. Accurate RIR generators can generate RIRs with various acoustic effects (e.g., diffraction, scattering, early reflections, late reverberations) [40]. A limitation of accurate RIR generators is that they are computationally expensive, and the time taken to generate RIRs depends on the geometric complexity of the acoustic environment. Also, traditional RIR generators rely on the empirical Sabine formula [144] to generate RIRs with expected reverberation time. Reverberation time ($T_{60}$) is the time required for the sound energy to decay by 60 decibels [10].

With advancements in deep neural-network-based far-field speech processing, the demand for on-the-fly simulation of far-field speech training datasets with hundreds of thousands of room configurations similar to the testing environment is increasing [28, 145–147]. The CPU-based offline simulation of far-field speech with balanced $T_{60}$ distribution requires a lot of computation time and disk space [1, 16], thus it is not scalable for production-level ASR training.

To overcome this problem, we propose a learning-based FAST-RIR architecture trained to generate both specular and diffuse reflections for a given acoustic environment. FAST-RIR can generate more than 20,000 RIR per second with balanced $T_{60}$ distribution and perform similarly to the RIRs generated using accurate RIR generators.

## 3.2 Main Contributions

We propose a neural-network-based fast diffuse room impulse response generator (FAST-RIR) that can be directly controlled using rectangular room dimension, listener and speaker positions, and $T_{60}$. $T_{60}$ implicitly reflects the characteristics of the room materials such as the floor, ceiling, walls, furniture etc. Our FAST-RIR takes a constant amount of time to generate an RIR for any given acoustic environment and yields accurate $T_{60}$.

## 3.3 Our Approach

To generate RIRs for a given acoustic environment, we propose a one-dimensional conditional generator network. Our generator network takes room geometry, listener and speaker positions, and $T_{60}$ as inputs, which are the common input used by all traditional RIR generators, and generates RIRs as raw-waveform audio (Figure 8.1). Our FAST-RIR generates RIRs of length 4096 at 16 kHz frequency.

### 3.3.1 Modified Conditional GAN

We propose a modified conditional GAN architecture to precisely generate an RIR for a given condition. GAN [148] consists of a generator ($G$) and a discriminator ($D$) networks that are alternatingly trained to compete. The network $G$ is trained to learn a mapping from noise vector samples ($z$) from distribution $p_z$ to the data distribution $p_{data}$. The network $G$ is optimized to produce samples that are difficult for the $D$ to distinguish from real samples ($x$) taken from true data distribution, while $D$ is optimized to differentiate samples generated from $G$ and real samples. The networks $G$ and $D$ are trained to optimize the following two-player min-max game with value function $V(G, D)$.

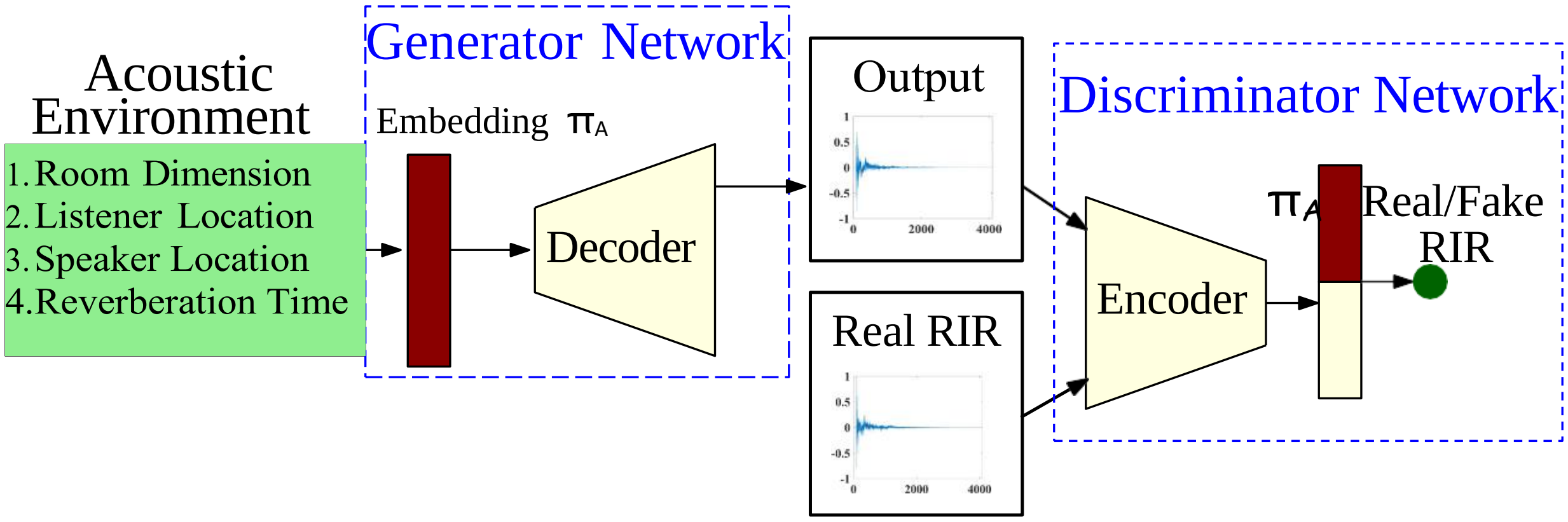

Figure 3.1: The architecture of our FAST-RIR. Our Generator network takes acoustic environment details as input and generates the corresponding RIR as output. Our Discriminator network discriminates between the generated RIR and the ground truth RIR for the given acoustic environment during training.

$$\min_G \max_D V(G, D) = \mathbb{E}_{x \sim p_{data}}[\log D(x)] + \mathbb{E}_{z \sim p_z}[\log(1 - D(G(z)))]. \tag{3.1}$$

Conditional GAN (CGAN) [149, 150] is an extended version of GAN where both the generator and discriminator networks are conditioned on additional information $y$. The generator network in CGAN is conditioned on the random noise $z$ and $y$. The vector $z$ is used to generate multiple different samples satisfying the given condition $y$. In our work, we train our FAST-RIR to generate a single sample precisely for a given condition. Our FAST-RIR is a modified CGAN architecture where the generator network is only conditioned on $y$.

### 3.3.2 FAST-RIR

We combine rectangular room dimension, listener location, and source location represented using 3D Cartesian coordinates $(x, y, z)$ and $T_{60}$ as a ten-dimensional vector embedding $\pi_A$. We normalize the vector embedding within the range -1.2 to 1.2 using the largest room dimension in the training dataset.

For each $\pi_A$, we generate RIR using Diffuse Acoustic Simulator (DAS) ($R_D$) and use it as ground truth to train our network. Our objective function for the generator network ($G_N$) consists of modified CGAN error, mean square error and $T_{60}$ error. The discriminator network is trained using the modified CGAN objective function.

**Generator Modified CGAN Error:** The $G_N$ is trained with the following modified CGAN error to generate RIRs that are difficult for the discriminator $D_N$ to differentiate from RIRs generated from DAS.

$$L_{CGAN} = \mathbb{E}_{\pi_A \sim p_{data}}[\log(1 - D_N(G_N(\pi_A)))]. \tag{3.2}$$

**Mean Square Error (MSE):** We compare each sample ($s$) of the RIR generated using our FAST-RIR ($R_N$) with RIR generated using DAS ($R_D$) for each $\pi_A$ to calculate the following MSE.

$$L_{MSE} = \mathbb{E}_{\pi_A \sim p_{data}}[\mathbb{E}[(R_N(\pi_A, s) - R_D(\pi_A, s))^2]]. \tag{3.3}$$

**$T_{60}$ Error:** We generate RIRs using our FAST-RIR and calculate their $T_{60}$ using a method based on ISO 3382-1:2009. We compare the $T_{60}$ of each generated RIRs with the $T_{60}$ given as input to the network in the embedding $\pi_A$ as follows:

$$\mathcal{L}_{T_{60}} = \mathbb{E}_{\pi_A \sim p_{data}}[|T_{60}(G_N(\pi_A)) - T_{60}(\pi_A)|]. \tag{3.4}$$

**Full Objective:** We train the $G_N$ and $D_N$ alternatingly to minimize the generator objective function $\mathcal{L}_{G_N}$ (Equation 8.5) and maximize the discriminator objective function $\mathcal{L}_{D_N}$ (Equation 8.6). We control the relative importance of the MSE ($\mathcal{L}_{MSE}$) and $T_{60}$ error ($\mathcal{L}_{T_{60}}$) using the weights $\lambda_{MSE}$ and $\lambda_{T_{60}}$, respectively.

$$\mathcal{L}_{G_N} = \mathcal{L}_{CGAN} + \lambda_{MSE}\mathcal{L}_{MSE} + \lambda_{T_{60}}\mathcal{L}_{T_{60}}. \tag{3.5}$$

$$\begin{aligned} \mathcal{L}_{D_N} = {} & \mathbb{E}_{(R_D,\pi_A) \sim p_{data}}[\log(D_N(R_D(\pi_A)))] \\ & + \mathbb{E}_{\pi_A \sim p_{data}}[\log(1 - D_N(G_N(\pi_A)))]. \end{aligned} \tag{3.6}$$

## 3.4 Implementation

### 3.4.1 Network Architecture

We adapt the generator network ($G_N$) and the discriminator network ($D_N$) proposed in Stage-I of StackGAN architecture [151] and modify the networks. StackGAN takes a text description and a noise vector as input and generates a photo-realistic two-dimensional (2D) image as output. Our FAST-RIR takes acoustic environment details as input and generates an RIR as a one-dimensional (1D) raw-waveform audio output. We flatten the 2D convolutions into 1D to process 1D RIR in both $G_N$ and $D_N$.

Unlike photo-realistic images, raw-waveform audio exhibits periodicity. Donahue et al. [152] suggest that filters with larger receptive fields are needed to process low frequencies (large wavelength signals) in the audio. We improve the receptive field of the original $G_N$ and the encoder in $D_N$ by increasing the kernel size (i.e., 3×3 2D convolution becomes length 41 1D convolution) and strides (i.e., stride 2×2 becomes stride 4×1). We also replace the upsampling layer and the following convolutional layer with a transposed convolutional layer.

### 3.4.2 Dataset

The sizes of the existing real-world RIR datasets [30, 153, 154] are insufficient to train our FAST-RIR. Therefore, we generate 75,000 medium-sized room impulse responses using a DAS [1] to create a training dataset. We choose 15 evenly spaced room lengths within the range 8m to 11m, 10 evenly spaced room widths between 6m and 8m, and 5 evenly spaced room heights between 2.5m and 3.5m to generate RIRs. We position the speaker and the listener at random positions within the room and generate 100 different RIRs for each combination of room dimensions (15×10×5). The $T_{60}$ values of our training dataset are between 0.2s and 0.7s.

### 3.4.3 Training

We iteratively train $G_N$ and $D_N$ using RMSprop optimizer with batch size 128 and learning rate $8\text{x}10^{-5}$. For every 40 epochs, we decay the learning rate by 0.7.

## 3.5 Experiment and Results

### 3.5.1 Baselines

We randomly select 30,000 different acoustic environments within the range of the training dataset (Section 3.4). We generate RIRs corresponding to the selected acoustic environments using image method [16], gpuRIR [28], DAS [1] and FAST-RIR to evaluate the performance of our proposed FAST-RIR. IR-GAN [48] does not have the capability to precisely generate RIRs for a given speaker and listener positions; therefore, we did not use IR-GAN in our experiments.

Table 3.1: The runtime for generating 30,000 RIRs using image method, gpuRIR, DAS, and our FAST-RIR. Our FAST-RIR significantly outperforms all other methods in runtime.

| **RIR Generator** | **Hardware** | **Total Time** | **Avg Time** |
|---|---|---|---|
| DAS [1] | CPU | $9.01 \times 10^5$s | 30.05s |
| Image Method [16] | CPU | $4.49 \times 10^3$s | 0.15s |
| **FAST-RIR(Batch Size 1)** | **CPU** | **$2.15 \times 10^3$s** | **0.07s** |
| gpuRIR [28] | GPU | 16.63s | $5.5 \times 10^{-4}$s |
| FAST-RIR(Batch Size 1) | GPU | 34.12s | $1.1 \times 10^{-3}$s |
| **FAST-RIR(Batch Size 64)** | **GPU** | **1.33s** | **$4.4 \times 10^{-5}$s** |
| FAST-RIR(Batch Size 128) | GPU | 1.77s | $5.9 \times 10^{-5}$s |

### 3.5.2 Runtime

We evaluate the runtime for generating 30,000 RIRs using image method, gpuRIR, DAS and FAST-RIR on an Intel(R) Xenon(R) CPU E52699 v4 @ 2.20 GHz and a GeForce RTX 2080 Ti GPU (Table 9.1). The gpuRIR is optimized to run on a GPU; therefore, we generate RIR using gpuRIR only on a GPU. For a fair comparison with CPU implementations of image-method and DAS, we also generate RIRs using our FAST-RIR with batch size 1 on a CPU.

From Table 9.1, we can see that our proposed FAST-RIR with batch size 1 is 400 times faster than DAS [1] on a CPU. Our FAST-RIR is optimized to run on a GPU. We compare the performance of our FAST-RIR with an existing GPU-based RIR generator gpuRIR [28]. We can see that gpuRIR performs better than our FAST-RIR with batch size 1, which is not the real use case of our generator. To our best knowledge, the gpuRIR does not leverage the batch parallelization while this was supported in our FAST-RIR. We can see that our proposed FAST-RIR with batch size 64 is 12 times faster than gpuRIR.

### 3.5.3 $T_{60}$ Error

Table 3.2 shows the $T_{60}$ error of the generated RIRs calculated using Equation 3.4. We can see that the testing $T_{60}$ error of our FAST-RIR is high for input $T_{60}$ below 0.25s (0.068s) when compared to the input $T_{60}$ greater than 0.25s (0.021s).

Our FAST-RIR is trained to generate RIRs with durations slightly above 0.25s. For the input $T_{60}$ below 0.25s, the generated RIR has a noisy output between $T_{60}$ and 0.25s. We notice that cropping the generated RIRs at $T_{60}$ improves the overall $T_{60}$ error from 0.029s to 0.023s.

Table 3.2: $T_{60}$ error of our FAST-RIR for 30,000 testing acoustic environments. We report the $T_{60}$ error for RIRs cropped at $T_{60}$ and full RIRs. We only crop RIRs with $T_{60}$ below 0.25s.

| $T_{60}$ Range | Crop RIR at $T_{60}$ | $T_{60}$ Error |
|---|---|---|
| 0.2s - 0.25s | No | 0.068s |
| **0.2s - 0.25s** | **Yes** | **0.033s** |
| 0.25s - 0.7s | - | 0.021s |
| 0.2s - 0.7s | No | 0.029s |
| **0.2s - 0.7s** | **Yes** | **0.023s** |

### 3.5.4 Simulated Speech Comparison

We simulate reverberant speech $x_r[t]$ by convolving clean speech $x_c[t]$ from the LibriSpeech test-clean dataset [29] with different RIRs $r[t]$ (Equation 3.7).

$$x_r[t] = x_c[t] \circledast r[t]. \tag{3.7}$$

We decode the simulated reverberant speech using Google Speech API[1] and Microsoft

[1]https://cloud.google.com/speech-to-text/

Speech API[2]. Table 3.3 shows the Word Error Rate (WER) of the decoded speech. No text normalization was applied in both cases, as only the relative WER differences between different RIR generators are concerned. For Google Speech API, we report WER for the clean and reverberant LibriSpeech test sets that are successfully decoded. The results of each speech API show that compared with the reverberant speech simulated using traditional RIR generators, reverberant speech simulated using our FAST-RIR is closer to the reverberant speech simulated using DAS [1]. We provide reverberant speech audio examples, spectrograms and the source code for reproducibility at github[3].

Table 3.3: Automatic speech recognition (ASR) results were obtained using Google Speech API and Microsoft Speech API. We simulate a reverberant speech testing dataset by convolving clean speech from the LibriSpeech dataset with different RIR datasets. We compare the reverberant speech simulated using the image method, gpuRIR and our FAST-RIR with the reverberant speech simulated using DAS. We show that the relative WER change from our method is the smallest.

| **Testing Dataset** | **Word Error Rate** [%] | |
|---|---|---|
| **Clean Speech ⊛ RIR** | **Google API** | **Microsoft API** |
| Libri ⊛ DAS (baseline) [1] | 6.56 | 2.63 |
| Libri ⊛ gpuRIR [28] | 9.39 (+43%) | 3.78 (+44%) |
| Libri ⊛ Image Method [16] | 9.03 (+38%) | 3.86 (+47%) |
| **Libri ⊛ FAST-RIR (ours)** | **7.14 (+9%)** | **2.76 (+5%)** |

### 3.5.5 Far-Field Automatic Speech Recognition

We want to ensure that our FAST-RIR generates RIRs that are better than or as good as existing RIR generators for ASR. We use the AMI corpus [155] for our far-field ASR experiments. AMI contains close-talk speech data recorded using Individual Headset Microphones (IHM) and distant speech data recorded using Single Distant Microphones (SDM).

[2] `https://azure.microsoft.com/en-us/services/cognitive-services/speech-services/`

[3] `https://anton-jeran.github.io/FRIR/`

Table 3.4: Far-field ASR results were obtained for far-field speech data recorded by single distance microphones (SDM) in the AMI corpus. The best results are shown in **bold**.

| **Training Dataset** | **Word Error Rate** [%] | |
|---|---|---|
| **Clean Speech ⊛ RIR** | **dev** | **eval** |
| IHM ⊛ None | 55.0 | 64.2 |
| IHM ⊛ Image Method [16] | 51.7 | 56.1 |
| IHM ⊛ gpuRIR [28] | 52.2 | 55.5 |
| IHM ⊛ DAS [1] | 47.9 | **52.5** |
| IHM ⊛ DAS-cropped [1] | 48.3 | **52.6** |
| **IHM ⊛ FAST-RIR (ours)** | 47.8 | **53.0** |

We use a modified Kaldi recipe [4] to evaluate our FAST-RIR. The modified Kaldi recipe takes IHM data as the training set and tests the model using SDM data. The IHM data can be considered clean speech because the echo effects in IHM data are negligible when compared to SDM data. We augment far-field speech data by reverberating the IHM data with different RIR sets using Equation 3.7. The 30,000 RIRs generated using the image method, gpuRIR, DAS, and FAST-RIR are used in our experiment.

The IHM data consists of 687 long recordings. Instead of reverberating a speech recording using a single RIR, we do segment-level speech reverberation, as proposed in [30]. We split each recording at the beginning of at least continuous 3 seconds of silence. We split at the beginning to avoid inter-segment reverberated speech overlapping. We can split IHM data into 17749 segments. We reverberate each segment using a randomly selected RIR from an RIR dataset (either image method, gpuRIR, DAS, DAS-cropped or our FAST-RIR).

Table 9.4 presents far-field ASR development and test WER for far-field SDM data. We can see that our FAST-RIR outperforms gpuRIR [28] by up to 2.5% absolute WER. The DAS [1] with full duration and the DAS cropped to have the same duration as our FAST-RIR (DAS-cropped) performs similarly in the far-field ASR experiment. We see that the performance of DAS and FAST-RIR has no significant difference.

[4] https://github.com/RoyJames/kaldi-reverb/

## 3.6 Discussion and Future Work

We propose a novel FAST-RIR architecture to generate a large RIR dataset on the fly. We show that our FAST-RIR performs similarly in ASR experiments when compared to the RIR generator (DAS [1]), which is used to generate a training dataset to train our FAST-RIR. Our FAST-RIR can be easily trained with RIR generated using any state-of-the-art accurate RIR generator to improve its performance in ASR experiments while keeping the speed of RIR generation the same.

Although we trained our FAST-RIR for limited room dimensions ranging from (8m,6m,2,5m) to (11m,8m,3.5m) using 75,000 RIRs, we believe that our FAST-RIR will give a similar performance when we train FAST-RIR for a larger room dimension range with a huge amount of RIRs. We would like to evaluate the performance of our FAST-RIR in the multi-channel ASR [156] and speech separation [157] tasks.

# Chapter 4: Fast Monaural Room Impulse Response Generator for AR/VR Application

## 4.1 Motivation

Rapid developments in interactive applications (e.g., games, virtual environments, speech recognition, etc.) demand realistic sound effects in complex dynamic indoor environments with multiple sound sources. Generating realistic sound effects is still a challenging problem because of the geometric and aural complexity of the real-world environment. The geometric complexity is a measure of the number of vertices and faces needed to represent the complex environment as a 3D mesh. The aural complexity depends on the number of sound sources, the number of static and dynamic objects in the environment and the acoustic effects (e.g., early reflection, late reverberation, diffraction, scattering, etc.). The way the sound propagates from a sound source to the listener can be modeled as an impulse response (IR) [10]. We can generate sound effects by convolving the IR with a dry sound. The IRs are used to generate plausible sound effects in many interactive applications used for games and AR/VR: Steam Audio [158], Project Acoustics [159], and Oculus Spatializer [160].

Recently, many machine learning-based algorithms are proposed to synthesize the sounds in musical instruments [161,162], estimate the acoustic material properties, and model the acoustic effects from finite objects [163–166]. Neural-network-based IR generators [2, 48] for simple rooms are proposed for speech processing applications (e.g., far-field speech recognition, speech enhancement, speech separation, etc.). FAST-RIR [2] is a GAN-based IR generator that takes shoe-box-shaped room dimensions, listener and source positions, and reverberation time as inputs and generates a large IR dataset on the fly. Gaming applications demand fast IR generators for complex scenes. Complex indoor 3D scenes with furniture can be represented in detail using mesh models. Traditional geometric IR simulators take a 3D scene mesh-model and listener and source positions as inputs and generate realistic environmental sound effects [35]. The complexity of geometric IR simulators increases exponentially with the reflection depth [167] and makes them impractical for interactive applications. No current simulation and learning methods can compute real-time IRs for unseen complex dynamic scenes.

Sound propagation can be accurately simulated using wave-based methods. The wave-based methods solve wave equations using different numerical solvers such as the boundary-element method [20], the finite-difference time-domain simulation [21], etc. The complexity of the wave-based approach grows as the fourth power of the frequency and is limited to static scenes. Geometric acoustic algorithms are a less complex alternative to the wave-based method. Geometric acoustic algorithms can handle complex environments with dynamic objects [22] and multiple sources [23]. However, geometric acoustic methods do not model the low-frequency components of the IRs accurately because of the ray assumption. The sound wave can be treated as a ray when the wavelength of the sound is smaller than the obstacles in the environment. At low frequencies under 500 Hz, the ray assumption is not valid for most scenes and results in significant simulation errors. Hybrid sound propagation algorithms [25, 26] combine IRs from wave-based and geometric techniques to generate accurate IRs in the human aural range for complex dynamic scenes. Generating IRs corresponding to thousands of sound sources at an interactive rate is not possible with the hybrid method because of its complexity.

## 4.2 Main Contributions

We propose a novel learning-based IR generator (MESH2IR) to generate realistic IRs for furnished 3D scenes with arbitrary topologies (i.e., meshes with 2000 faces to 3 million faces) not seen during the training. For a given complex scene, MESH2IR can generate IRs for any listener and source positions. We perform mesh simplification to even handle complex 3D scenes represented using a mesh with millions of triangles. Our mesh encoder network significantly reduces the input data size by transforming the 3D scene meshes into low-dimensional vectors of a latent space. Mesh simplification and its encoder network allow us to handle general 3D scenes meshes with a varying number of faces. We present an efficient approach to preprocess the IR training dataset and show that training MESH2IR on preprocessed dataset gives a significant improvement in the accuracy of IR generation. We also evaluate the contribution of energy decay relief in improving the IRs generated from MESH2IR. We train our MESH2IR on an IR dataset computed using a hybrid sound propagation algorithm [26]. Therefore, the predicted IRs from MESH2IR exhibit good accuracy for both low-frequency and high-frequency components. MESH2IR can generate more than 10,000 IRs for a given indoor 3D scene. We show that our MESH2IR can generate IRs 200 times faster than a geometric acoustic simulator [1] on a single CPU. We evaluate the accuracy of the predicted IRs from MESH2IR using power spectrum and acoustic metrics which characterize the acoustic environment. Our MESH2IR can predict the IRs for unseen 3D scenes during training with less than 10% error in three different acoustic metrics. We also show that far-field speech augmented using the IRs generated from MESH2IR significantly improves the performance in speech processing applications.

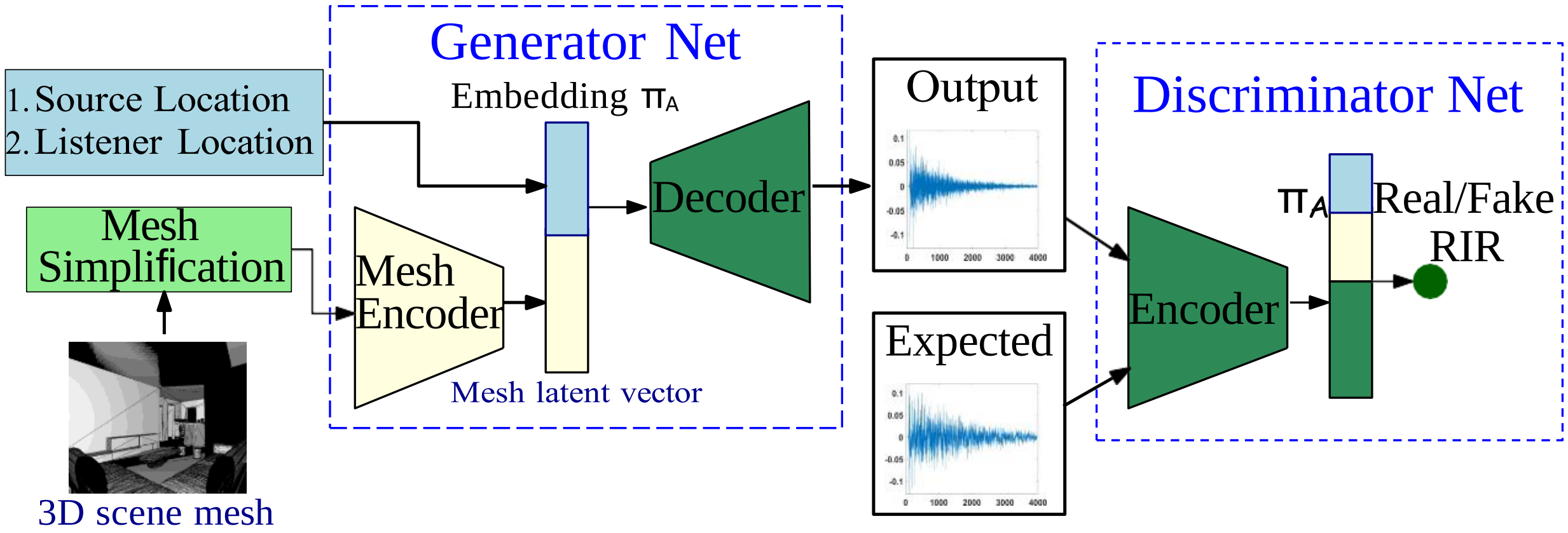

Figure 4.1: The architecture of our MESH2IR. Our mesh encoder network encodes an indoor 3D scene mesh to the latent space. The mesh latent vector and the source and listener locations are combined to produce a scene vector embedding ($\pi_A$). The generator network generates an IR corresponding to the input scene vector embedding. For the given scene vector embedding, the discriminator network discriminates between the generated IR and the ground truth IR during training.

## 4.3 MESH2IR: Our Approach

### 4.3.1 Overview

Our goal is to predict the IR for the given indoor 3D scene and the source and listener positions (Equation 4.1). The 3D scenes are represented as triangular meshes and we do not make any assumptions about the topology of the 3D scenes. The mesh format describes the shape of an object using vertices, edges, and faces consisting of triangles represented using 3D Cartesian coordinates (x,y,z). The source positions and listener positions are also represented using 3D Cartesian coordinates. For each scene material (e.g., furniture, floor, wall, ceiling, etc.), we do not explicitly control the characteristics of the scene materials (the amount of absorption or scattering of sound by the scene material). The IRs used to train our MESH2IR is computed by considering the characteristic of scene material in every indoor 3D scene. Therefore, our trained MESH2IR randomly assigns the characteristics of scene materials based on their shape.

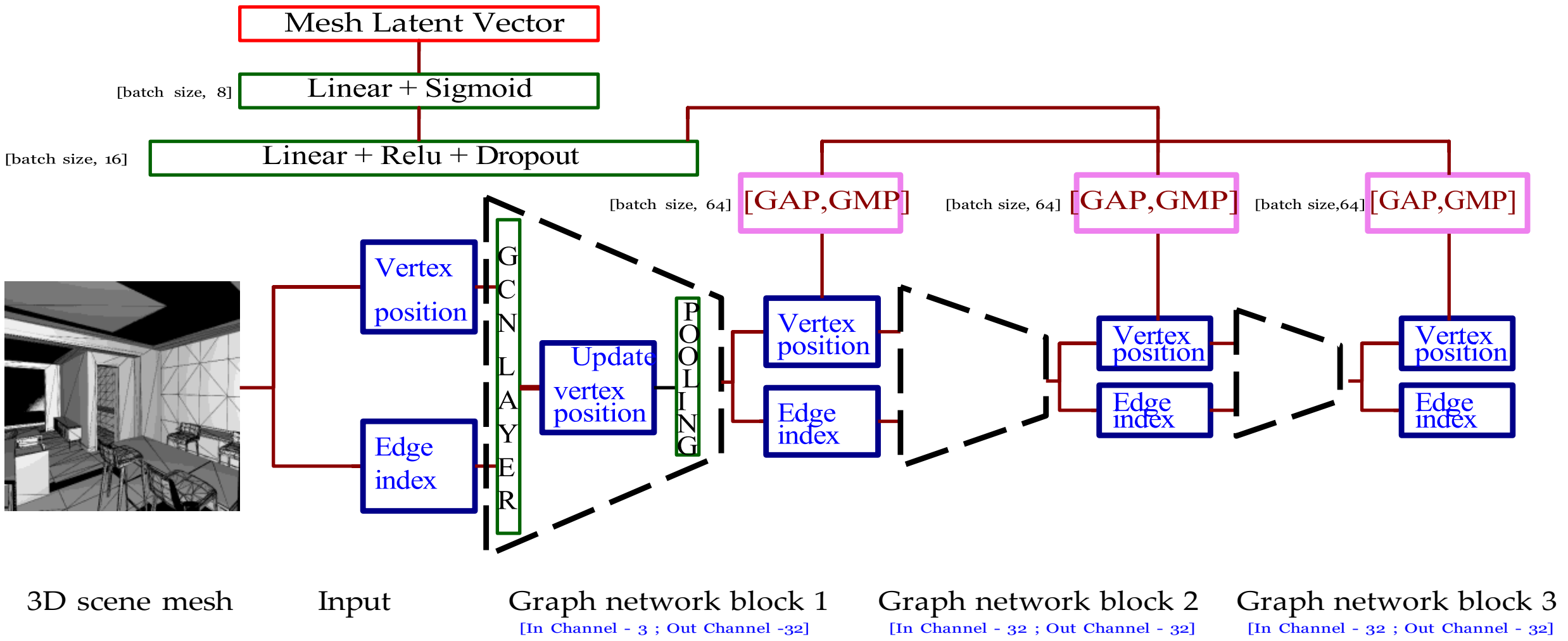

Figure 4.2: The expansion of our mesh encoder in Figure 4.1. Our encoder network transforms the indoor 3D scene mesh into a latent vector. The topology information (edge connectivity) and the node features (vertex coordinates) are extracted from the mesh and passed to our graph neural network.

The IR is the response of an impulse signal emitted in an environment. IR describes the relationship between a dry sound and the reflected sound from the boundaries in the scene. The reflected sound signal depends on the scene geometry, scene materials, and source and listener positions. IRs can be accurately simulated by solving the wave equation [21]. However, solving a wave equation is not practical for interactive applications because of its complexity. In our work, we propose a learning-based IR generator (MESH2IR) that is capable of approximating thousands of IRs per second for a given complex scene. Our network can be formally described as:

$$IR_n = \mathbf{N}_{\theta_1}(\mathbf{N}_{\theta_2}(M_n), SP_n, LP_n), \tag{4.1}$$

where $IR_n$ is the predicted IR for the given scene $n$, $M_n$ is the 3D mesh representation of the scene simplified to have around 2000 faces, $LP_n$ and $SP_n$ are listener location and source location, respectively. $LP_n$ and $SP_n$ are represented using a 3D vector. We simplify the 3D scene mesh with an arbitrary number of faces to have a constant number of faces (2000 faces). We use a mesh-based encoder network $\mathbf{N}_{\theta_2}$ to transform the simplified 3D mesh into a latent vector of a latent space (Figure 4.2). $\mathbf{N}_{\theta_1}$ is a generator network, and $\theta_1$ and $\theta_2$ are the trained network parameters. Our overall network architecture is shown in Figure 4.1.

In recent years, cross-modal translation neural networks have gained attention in computer vision. Several algorithms have been proposed for translating video to audio [168], image to audio [5], text to image [151, 169], image to mesh [170], etc. In our work, we translate complex scenes represented using meshes to acoustic IRs represented as audio signals.

### 4.3.2 Training Dataset

In our work, we train MESH2IR on the GWA IR dataset [26] simulated using a hybrid algorithm to generate realistic IRs on unseen complex environments. The IRs in GWA are created by automatically calibrating the ray energies simulated using the geometric acoustic method [1] with the wave effects simulated using finite-difference time-domain wave solver [21] to create high-quality low-frequency and high-frequency wave effects. The GWA dataset consists of 2 million IRs simulated on the indoor 3D environments represented as meshes in the 3D-FRONT dataset. Out of the 2 million IRs, we train MESH2IR on 200,000 IRs simulated in 5,000 different indoor environments from 3D-FRONT dataset [58].

### 4.3.3 Mesh Simplification

The number of faces in the 3D-FRONT dataset varies from about 2000 faces to 3 million faces. We initially simplify the meshes using a quadratic-based edge collapse algorithm in PyMeshLab [171] to have a fewer number of faces (i.e., 2000 faces). The edge collapse algorithm makes sure that the approximation error between the original mesh and the simplified mesh, in terms of Hausdorff distance is small. Therefore the acoustic characteristics do not change due to the simplification step.

Figure 4.3 depicts an example of the original indoor 3D scene mesh and the simplified mesh using the quadratic edge collapse algorithm. We can see that high-level details of the furniture such as bed, pillow, settee, table and cupboard are preserved in the simplified mesh, in addition to the scene geometry. In this example, the simplified mesh has only 2% of faces of the original mesh.

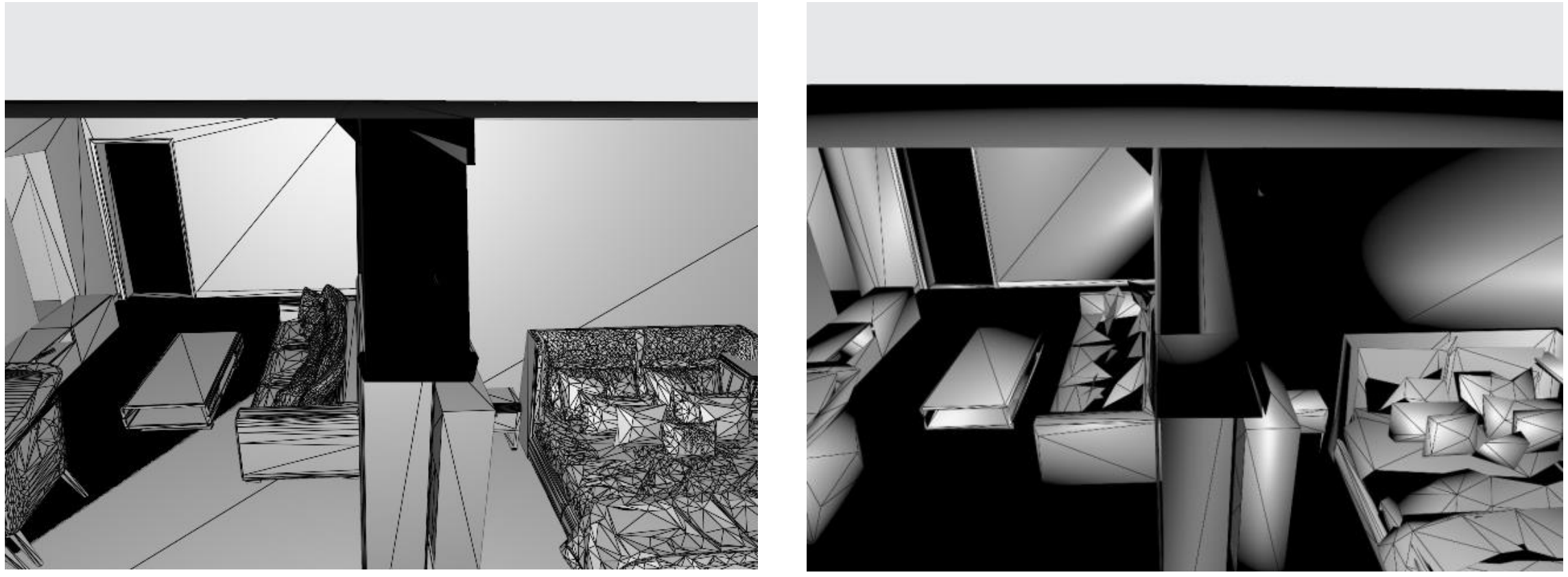

(a) Before mesh simplification. (b) After mesh simplification.

Figure 4.3: The indoor 3D scene is represented using an original mesh with around 100,000 faces and a simplified mesh with 2,000 faces. We can see that high-level details of the room geometry and furniture are preserved in the simplified mesh.

### 4.3.4 Mesh Encoder

Our goal is to transform the simplified indoor scene meshes into a low-dimensional latent space. The triangular meshes can be represented as graph data. Therefore, we represent the 3D scene meshes as a graph and use a graph network (Figure 4.2) to reduce the dimension. Our graph network uses graph convolution (GCN) layers [172] to encode the topology information (i.e., edge connectivity) and node features (i.e., vertex coordinates) of the graph. We use graph pooling (gpool) layers [173, 174] to reduce the topology information of the graph.

The layer-wise propagation rule of multi-layer GCN [172]:

$$X^{(l+1)} = \sigma(\hat{D}^{-\frac{1}{2}}\hat{A}\hat{D}^{-\frac{1}{2}}X^{(l)}W^{(l)}), \tag{4.2}$$

where $\hat{D}_{ii} = \sum_j \hat{A}_{ij}$, and $\hat{A} = A + I$. $A$ is the adjacency matrix, $I$ is the identity matrix and $W^{(l)}$ is a trainable weight matrix for layer $l$. The feature matrices at layers $l$ and $l+1$ are $X^{(l)}$ and $X^{(l+1)}$, respectively. The topology information is not modified in the GCN layer.

The gpool layer [173] creates a new graph with high-level feature encoding by choosing $K$ vertices from the original vertex set and discarding the other vertices. Some edges are removed when discarding vertices in the gpool layer. This results in some isolated vertices in the graph. To address this problem, the gpool algorithm increases the graph connectivity by calculating the square of the adjacency matrix $A^{(l)}$ at layer $l$ and uses the new adjacency matrix $A_n^{(l)}$ for gpool computation (Equation 4.3).

$$A_n^{(l)} = A^{(l)}A^{(l)}. \tag{4.3}$$

We calculate the channel-wise average (GAP) and the channel-wise maximum (GMP) of the node features after each gpool layer and aggregate the GAP and GMP values. We concatenate the aggregated GAP and aggregated GMP values, pass them to linear layers, and get the mesh latent vector ($\pi_M$) (dimension=8) as the output.

### 4.3.5 Scene Vector Embedding

We concatenate the mesh latent vector ($\pi_M$) with the source ($SP$) and listener positions ($LP$) in 3D Cartesian space to generate scene vector embedding $\pi_A$ of dimension 14 (Equation 4.4). The mesh latent vector will be learned during training.

$$\pi_A = [\pi_M \; SP \; LP]. \tag{4.4}$$

### 4.3.6 IR Representation & Preprocessing

The IRs in the GWA dataset [26] have a sampling rate of 48 kHz. We downsample the IRs in the dataset to 16 kHz and crop the IRs. Downsampling IRs allows us to maintain a longer duration of IRs in a fixed-length input IR vector (3968 samples). The IRs with full duration and the IRs cropped to have a duration of around 0.25 seconds give similar performance in speech applications [2].

**IR preprocessing:** The IRs in the GWA dataset have large variations in standard deviation (STD) of the magnitude values ($10^{-12}$ to $10^{-2}$). We noticed that it is hard for MESH2IR to learn from such datasets with high dynamic ranges. To overcome this issue, we divide the IRs by ten times the STD to create preprocessed IRs with the same STD of 0.1 (or any constant STD). We noticed that preprocessing IRs to have constant STD improves the accuracy of MESH2IR (see Section 4.4.1).

To recover the IR with the original magnitude, we duplicate the STD of the IRs 128 times and concatenate them at the end of the preprocessed IRs. The concatenated IRs will have a length of 4096 (3968+128). The convolution layers in MESH2IR calculate an average of 41 neighboring sample values for each sample of the concatenated IRs and pass them to the next layer. Therefore, concatenated STD values near the end of preprocessed IR magnitudes in a concatenated IR will be corrupted when we calculate the average and pass it to the next layers. We can recover the STD values from the tail-end of the concatenated IR where all the 41 neighboring samples are STD.

### 4.3.7 Energy Decay Relief

The energy decay relief (EDR) obtained from the physics-based algorithm contains enough information to be converted into an "equivalent impulse response" [175]. Therefore, we use EDR in the cost function of our MESH2IR. Constraining the Generator network with additional information (i.e., EDR) helps the model to converge easily.

The energy decay curve (EDC) describes the energy remaining in the IR with respect to time [176]. When compared to the IR itself, the EDC decays more smoothly, and we can use it to measure IR acoustic metrics. The generalized EDC for multiple frequency bands is called EDR [177]. EDR is the total amount of energy remaining in the IR at time $t_n$ seconds in a frequency band centered at $f_k$ Hz:

$$EDR(t_n, f_k) = \sum_{m=n}^{M} |H(m, k)|^2. \tag{4.5}$$

In Equation 4.5, bin $k$ of the short-time Fourier transform at time-frame $m$ is denoted by $H(m, k)$. $M$ is the total number of time frames.

### 4.3.8 Modified Conditional GAN

Our MESH2IR passes the scene vector embedding $\pi_A$ (Equation 4.4) to a one-dimensional modified conditional GAN (CGAN) architecture proposed in FAST-RIR [2] to generate a single precise IR for the given indoor 3D scene. CGAN [149, 150] is conditioned on a random noise $z$ and on a condition $y$ to generate multiple different outputs that satisfy the condition $y$. The modified CGAN is only conditioned on $y$ to generate a single output.

MESH2IR consists of a generator network ($G_n$) and a discriminator network ($D_n$), which are alternately trained at each iteration. We train $G_n$ and $D_n$ using the IR samples from the data distribution $p_{data}$. The objective function of our generator network consists of modified CGAN error, EDR error, and mean square error (MSE). We train the discriminator network with a modified CGAN objective function. For each $\pi_A$, we use the corresponding IRs in the GWA dataset [26] as the ground truth when training our network.

**Modified CGAN Error (Generator Network):** The CGAN error is used to generate IRs that are hard for the $D_n$ to differentiate from the ground truth IRs.

$$L_{CGAN} = \mathbb{E}_{\pi_A \sim p_{data}}[\log(1 - D_N(G_N(\pi_A)))]. \tag{4.6}$$

**EDR Error:** For each $\pi_A$, we calculate the EDR of the generated IR using our MESH2IR ($E_N$) and the ground truth IR ($E_G$). We calculate EDR at a set of frequency bands with center frequencies at 125 Hz, 250 Hz, 500 Hz, 1000 Hz, 2000 Hz, and 4000 Hz ($B$). We compare the

$E_N$ and $E_G$ for each sample ($s$) as follows:

$$\mathcal{L}_{EDR} = \mathbb{E}_{\pi_A \sim p_{data}}[\mathbb{E}_{b \sim B}[\mathbb{E}[(E_N(\pi_A, b, s) - E_G(\pi_A, b, s))^2]]]. \tag{4.7}$$

The signal energy in high-frequency components of EDR is high when compared with the signal energy in low-frequency components of EDR in the training IR dataset [26]. Therefore, we give more weight to low-frequency components of EDR.

**MSE Error:** For each $\pi_A$, we compare the IR generated from MESH2IR ($I_N$) with the ground truth IR ($I_G$). We calculate the squared difference of each sample ($s$) in $I_N$ and $I_G$ as follows:

$$\mathcal{L}_{MSE} = \mathbb{E}_{\pi_A \sim p_{data}}[\mathbb{E}[(I_N(\pi_A, s) - I_G(\pi_A, s))^2]]. \tag{4.8}$$

The generator network ($G_N$) and the discriminator network ($D_N$) are trained to compete each other by minimizing the generator objective function $\mathcal{L}_{G_N}$ (Equation 4.9) and maximizing the discriminator objective function $\mathcal{L}_{D_N}$ (Equation 4.10 ). In Equation 4.9, the relative importance of the EDR error and MSE error are controlled using $\lambda_{EDR}$ and $\lambda_{MSE}$ respectively.

$$\mathcal{L}_{G_N} = \mathcal{L}_{CGAN} + \lambda_{EDR}\,\mathcal{L}_{EDR} + \lambda_{MSE}\,\mathcal{L}_{MSE}. \tag{4.9}$$

$$\begin{aligned} \mathcal{L}_{D_N} = {} & \mathbb{E}_{(I_G, \pi_A) \sim p_{data}}[\log(D_N(I_G(\pi_A)))] \\ & + \mathbb{E}_{\pi_A \sim p_{data}}[\log(1 - D_N(G_N(\pi_A)))]. \end{aligned} \tag{4.10}$$

### 4.3.9 Implementation Details

**Network Architecture:** We adapt the graph encoder network in the PyG official repository [178] and modify the network to encode indoor 3D scene meshes to the latent space (Figure 4.2). Our gpool layer keeps 60% of the original number of vertices in each layer. We use the Generator ($G_N$) network and the Discriminator network ($D_N$) architectures proposed in FAST-RIR [2] and modify their cost functions. We extend the $G_N$ architecture proposed in FAST-RIR by adding our mesh encoder network (Figure 4.1).

**Training:** We trained $G_N$ and $D_N$ using the RMSprop optimizer with a batch size of 256. The $G_N$ is iterated 3 times for every $D_N$ iteration. Initially, we start with the learning rate of $8\text{x}10^{-5}$ and decay the learning rate by 85% of the original value for every 7 epochs. We trained our network for 150 epochs. We published our code for reproducibility at github [179].

## 4.4 Ablation Experiments

We perform an ablation study to analyze the importance of our IR preprocessing approach (Section 4.3.6) and to find an efficient way of adding EDR to the cost function. We quantitatively evaluate different variations of our network by calculating the mean absolute difference of different acoustic metrics of the generated IRs and the ground truth IRs. The acoustic characteristics of the 3D indoor environment are described using acoustic metrics [180]. We also measure the mean square error (MSE) of the generated IRs and the ground truth IRs (Equation 4.8). We train all the variations of our network with 200K IRs from 5000 different indoor 3D scene meshes and generate 11K testing IRs from 600 scene meshes not used during training. We use commonly-used acoustic metrics such as reverberation time ($T_{60}$), early-decay-time (EDT), and direct-to-reverberant ratio (DRR) in our experiment. The time taken to decay the sound pressure by 60 decibels (dB) is called $T_{60}$. DRR is the ratio of the sound pressure level of the direct sound to the sound pressure level of reflected sound [128]. EDT is six times the time taken for the sound source to decay by 10 dB, and it depends on the type and location of the sound source.

Table 4.1: The accuracy of IRs generated using MESH2IR on 5 different furnished indoor 3D scenes not used for training. We plot the time-domain representation of ground truth IRs and the predicted IRs, and evaluate the accuracy of predicted IRs using relative reverberation time (**$T_{60}$**) error, relative direct-reverberant-ratio (DRR) error, and relative early-decay-time (EDT) error. Our MESH2IR can predict the IRs with less than 10% **$T_{60}$** and DRR errors, and less than 3% EDT error. We also can see that the envelope and magnitude of the predicted IRs matches the ground truth IRs.

| | | | | | |
|---|---|---|---|---|---|
| Benchmark Scene | 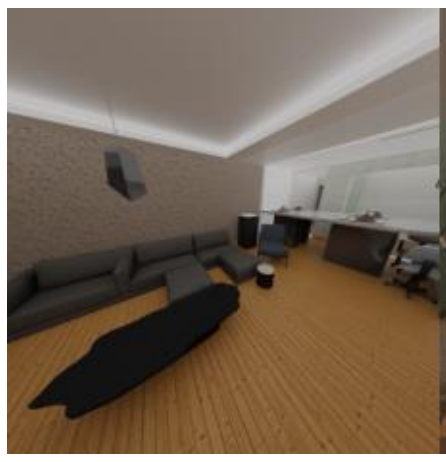 | 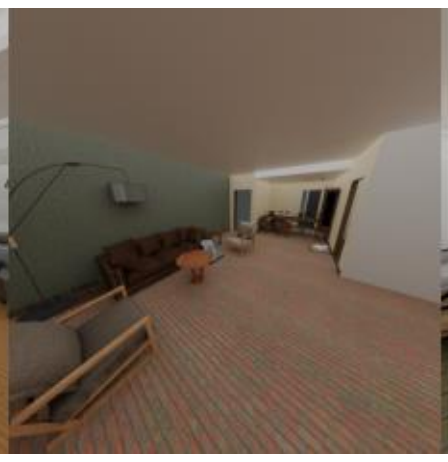 | 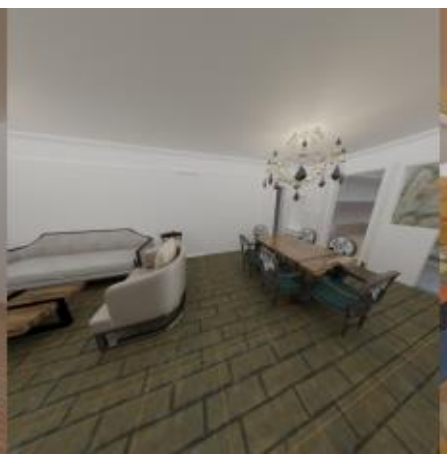 | 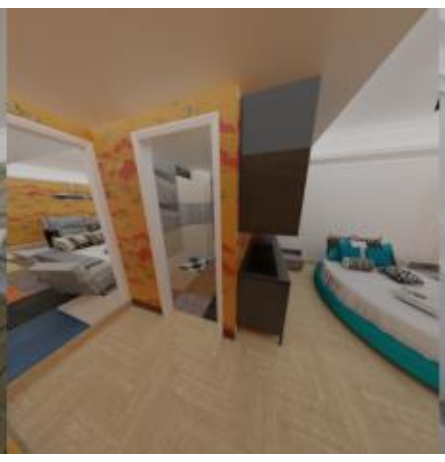 | 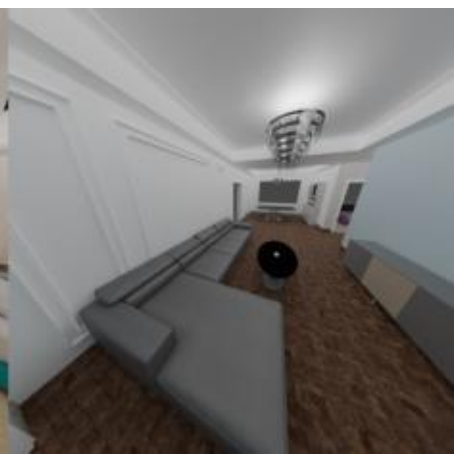 |
| # faces | 185,319 | 92,422 | 181,536 | 253,684 | 282,375 |
| Ground truth IR | 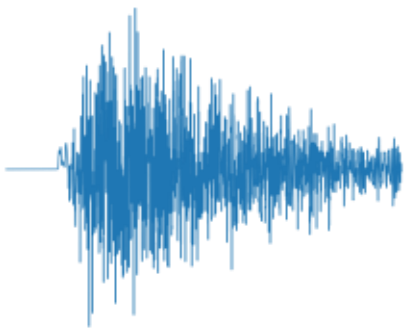 | 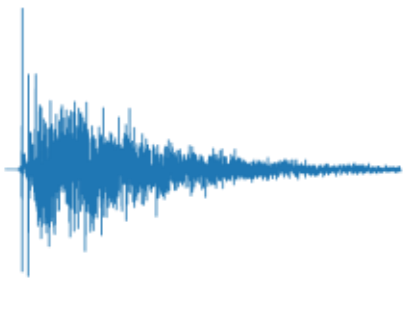 | 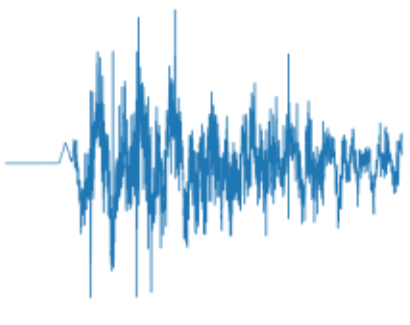 | 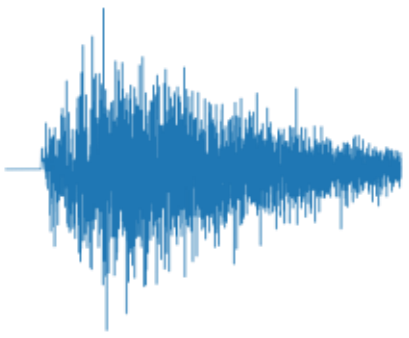 | 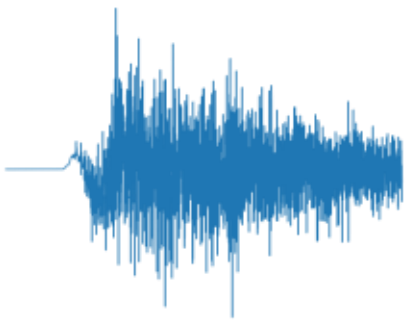 |
| Our Mesh2IR | 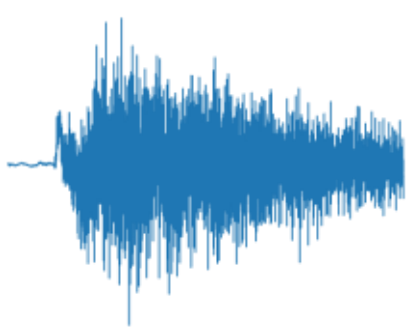 | 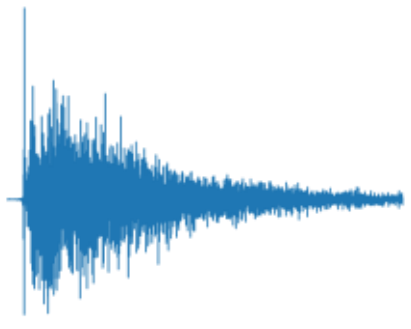 | 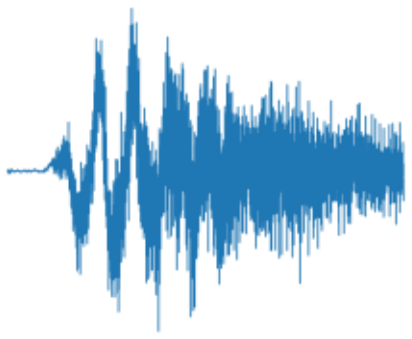 | 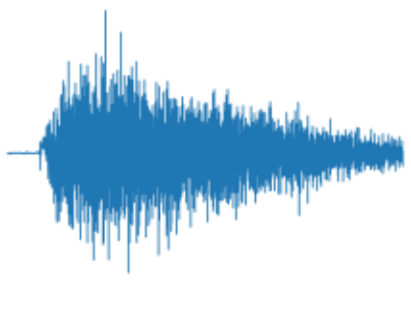 | 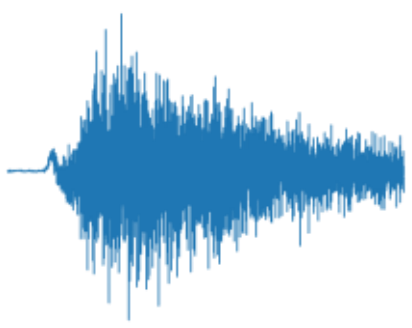 |
| $T_{60}$ error | 2.7% | 5.9% | 8.2% | 0.6% | 7.5% |
| DRR error | 2.6% | 9.4% | 1.4% | 1.9% | 9.3% |
| EDT error | 2.9% | 2.1% | 2.1% | 1.7% | 0.9% |

### 4.4.1 IR Preprocessing

We evaluate the contribution of our IR preprocessing approach discussed in Section 4.3.6. We train our MESH2IR on IRs without IR preprocessing (MESH2IR-UNPROCESSED) and generate IRs from our trained network. Figure 4.4 shows a ground truth IR and the IR generated from MESH2IR-UNPROCESSED. We can see that MESH2IR-UNPROCESSED generates noisy output. In Table 4.1 we can see that MESH2IR estimates the IRs for the given 3D scene to a greater extent.

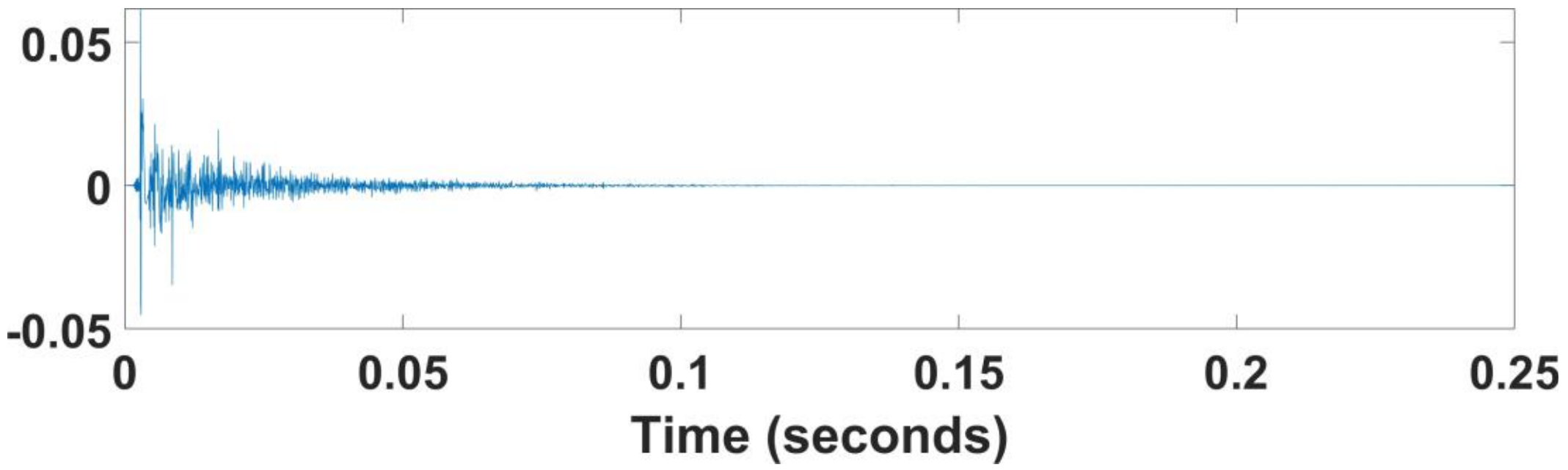

(a) Ground truth IR.

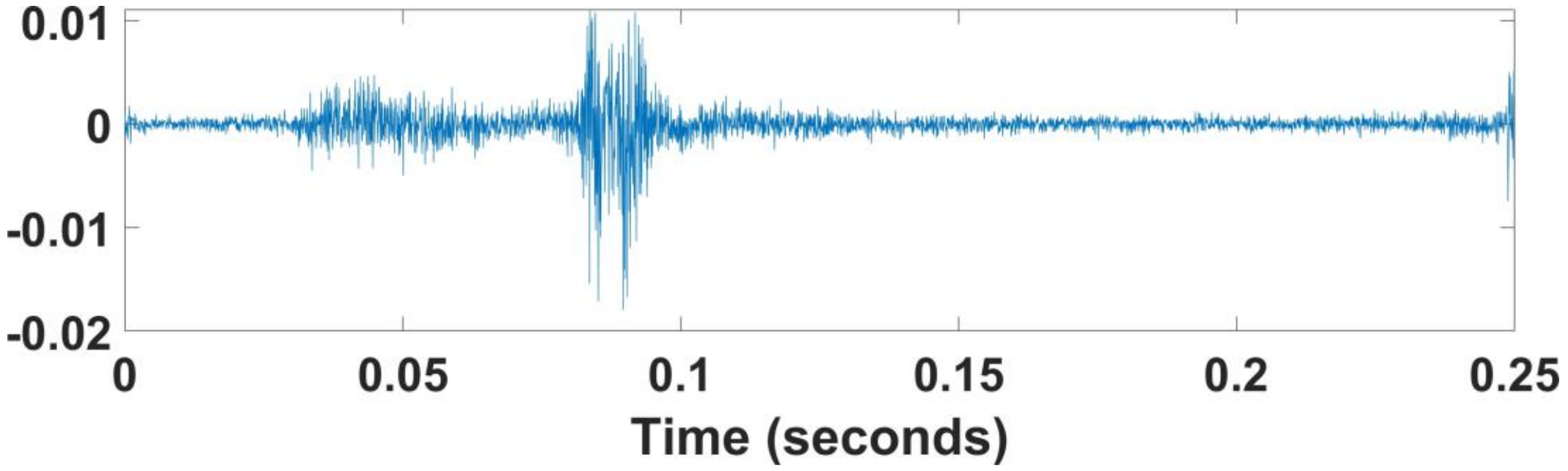

(b) Predicted IR.

Figure 4.4: The ground truth IR, and the IR predicted by MESH2IR trained with an unprocessed IR training set (MESH2IR-UNPROCESSED). We can see that MESH2IR-UNPROCESSED gives noisy output.

### 4.4.2 Energy Decay Relief (EDR)

To evaluate the importance and the efficient way of adding EDR to the cost functions, we train and compare two different variations of our MESH2IR network.

**Variation 1 (MESH2IR-NO-EDR) :** We evaluate the role played by the EDR loss in the generator network by training a variant of MESH2IR without the EDR loss (Equation 4.7).

**Variation 2 (MESH2IR-D-EDR) :** Instead of training the Discriminator network to discriminate between generated IRs and ground truth IRs of dimension 3968 x 1, we train the Discriminator network to discriminate between the EDR of the generated IRs and the ground truth IRs. EDR is calculated at six octave bands with center frequencies at 125 Hz, 250 Hz, 500 Hz, 1000 Hz, 2000 Hz and 4000 Hz. The EDR of an IR has a dimension of 3986x6. We also train the Generator network with the EDR loss.

From Table 4.2, we can see that adding EDR in the generator loss function improves the $T_{60}$, EDT, and MSE. Training the Discriminator network with IRs (MESH2IR) gives a similar performance to passing EDR of the IRs (MESH2IR-D-EDR).

### 4.4.3 Power Spectrum

The power spectrum describes the energy distribution of the IR in the frequency components that compose the waveform. In Figure 4.5, we compare the power spectrum of ground truth IRs with the power spectrum of the predicted IR using MESH2IR, MESH2IR-D-EDR, and MESH2IR-NO-EDR. In this example, we placed the listener 1m and 8m away from the source. We can see that in both cases the power spectrum of MESH2IR is closer to GWA when compared with other variants of our approach.

Table 4.2: Mean absolute error of the acoustic metrics and mean square error of the generated IRs from MESH2IR-NO-EDR, MESH2IR-D-EDR and MESH2IR. The acoustic metrics used in our experiments are reverberation time (**$T_{60}$**) measured in seconds, direct-to-reverberant ratio (DRR) measured in decibels and early-decay-time (EDT) measured in seconds. We can see that overall MESH2IR gives the least error. The best results of each metric are ”bolded”.

| **IR Dataset** | **Mean Absolute Error ↓** | | | **MSE (x $10^{-4}$) ↓** |
|---|---|---|---|---|
| | **$T_{60}$** | **DRR** | **EDT** | |
| MESH2IR-NO-EDR | 0.16 | **2.68** | 0.23 | 1.44 |
| MESH2IR-D-EDR | **0.13** | 2.74 | **0.22** | **1.23** |
| **MESH2IR** | **0.13** | 2.72 | **0.22** | **1.23** |

## 4.5 Acoustic Evaluation

In this section, we evaluate the accuracy, power spectrum and runtime of our MESH2IR network. We evaluate the accuracy of our MESH2IR network by comparing the relative acoustic metric error values of the IRs predicted from MESH2IR on indoor 3D scenes not used during training and the ground truth IRs from the GWA dataset [26]. We compare the run-time of MESH2IR network with a state-of-the-art geometric acoustic simulator [1].

### 4.5.1 Accuracy Analysis

To evaluate the robustness of our MESH2IR, we selected ground truth IRs from the GWA dataset [26] from 5 different indoor 3D scenes with different levels of complexity (the number of faces in a 3D scene mesh). The selected IRs have a large variation in magnitudes and shapes. The distance between the listener and the source position varies from 3.6m to 10.5m. We predicted

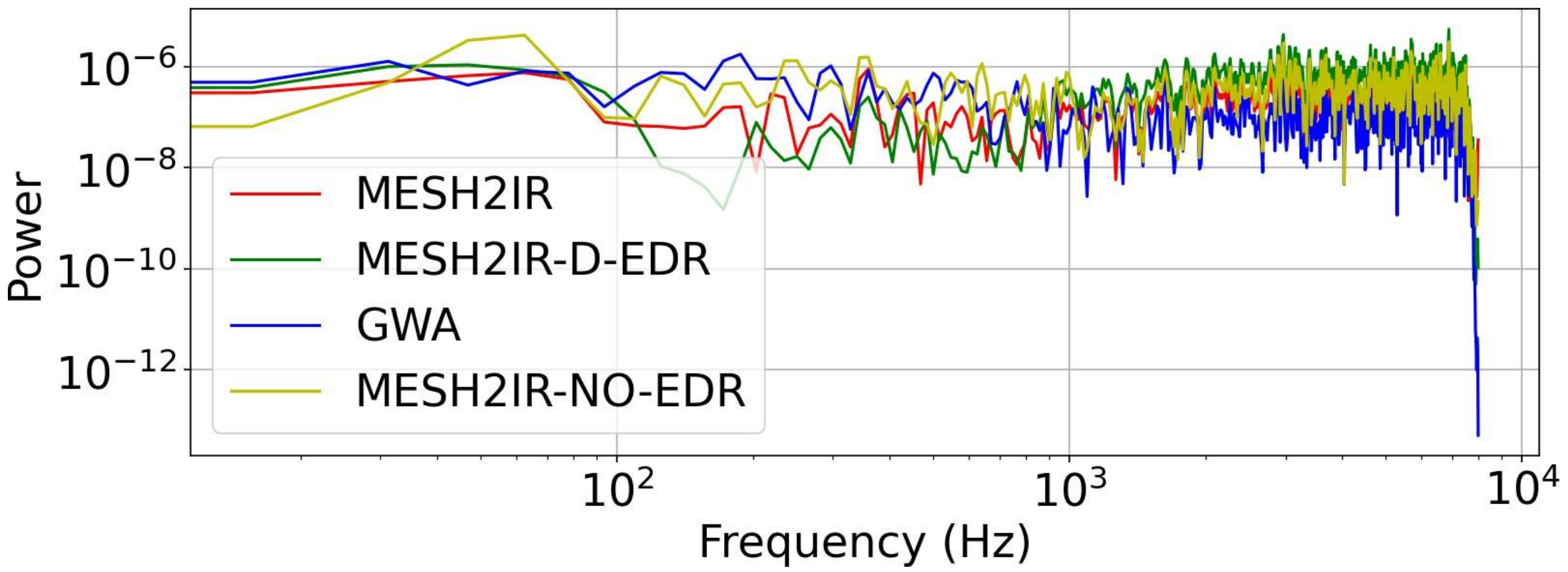

(a) The source is placed around 1m away from the listener.

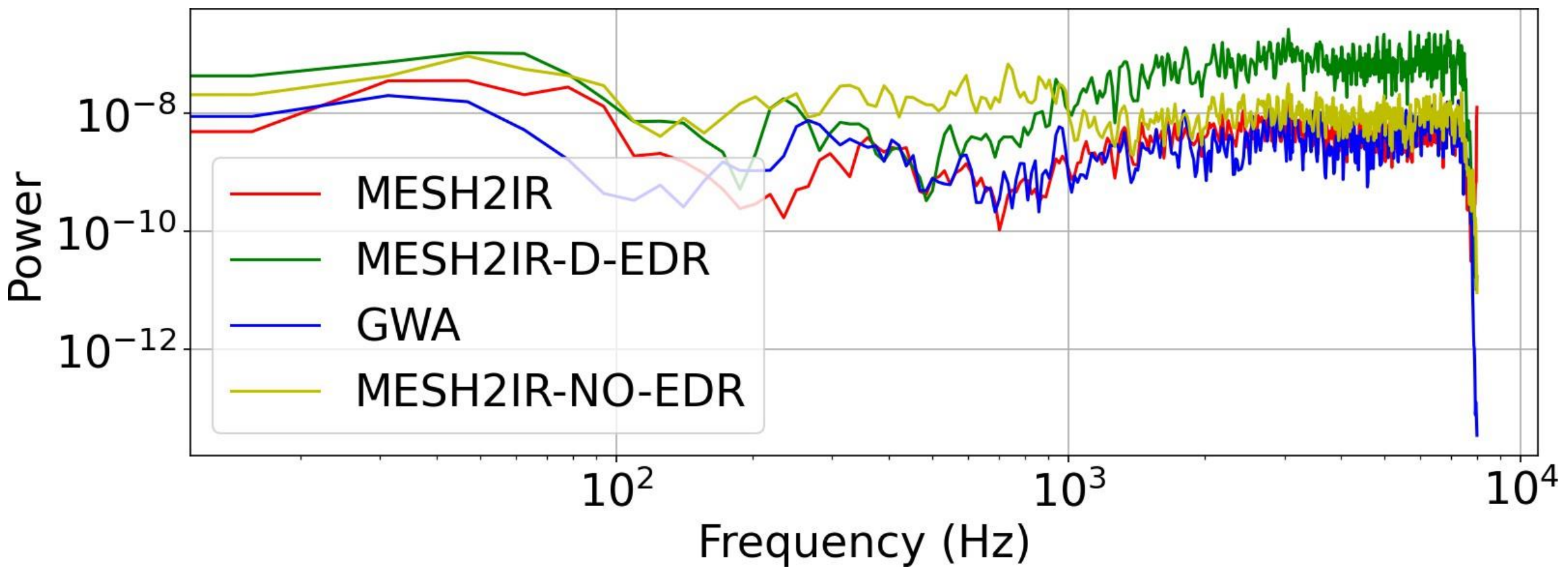

(b) The source is placed around 8m away from the listener.

Figure 4.5: The power spectrum of IRs generated using our proposed MESH2IR, MESH2IR-D-EDR, MESH2IR-NO-EDR, and the ground truth IRs from the GWA dataset. We can see that the power spectrum of MESH2IR is closest to the power spectrum of GWA.

IRs corresponding to the ground truth IRs in 3D scenes using our MESH2IR. We evaluate the accuracy of our prediction using relative $T_{60}$ error, relative DRR error, and relative EDT error. Table 4.1 shows the ground truth IRs, predicted IRs and the accuracy of our prediction. We can see that our approach can predict the IRs with less than 10% $T_{60}$ and DRR errors, and less than 3% EDT error, which are significantly lower than their just-noticeable-differences (JNDs) [181–183].

### 4.5.2 Runtime

We compare the runtime for generating 30,000 IRs from our MESH2IR with the time required to generate these samples from a geometric-based acoustic simulator (GA) [1] and FAST-RIR [2]. The runtime of GA depends on the complexity of the scene, and FAST-RIR is only capable of generating IRs for empty shoe-box-shaped medium-sized rooms (i.e., room dimensions varying from [8m,6m,2.5m] to [11m,8m,3.5m]). We compare the runtime of GA and FAST-RIR given in Ratnarajah et al. [2] with the runtime of MESH2IR for generating IRs for furnished indoor 3D scene meshes in the 3D-Front dataset [58]. We use an Intel(R) Xenon(R) CPU E52699 v4 @ 2.20 GHz and a GeForce RTX 2080 Ti GPU for our evaluation.

For a fair comparison with GA and FAST-RIR methods, which take 3D room dimensions as inputs, we did not consider the time taken for mesh simplification and mesh-to-graph conversion in Table 4.3. On average, mesh simplification takes around 7.5 seconds and mesh-to-graph conversion takes 0.04 seconds. From Table 4.3, we can see that MESH2IR is more than 200 times faster than GA on a CPU. MESH2IR is optimized to run on GPUs and supports batch parallelization. MESH2IR is slower than FAST-RIR because we encode a 3D scene mesh represented using a graph to a mesh latent vector using our graph neural network (Figure 4.2). To generate thousands of IRs for a given furnished indoor 3D scene, we perform mesh simplification, mesh-to-graph conversion, and mesh encoding only once.

We have shown the average time taken for mesh encoding (MESH2IR[Mesh Encoder]) and IR generation using the encoded mesh (MESH2IR[IR Generator]) separately in Table 4.3. Our MESH2IR can generate more than 10,000 IRs per second on a single GPU for a given furnished indoor 3D scene, and the runtime of MESH2IR is stable irrespective of the scene.

Table 4.3: The runtime of a geometric acoustic simulator (GA) [1], FAST-RIR [2], and our MESH2IR. MESH2IR is an extension of FAST-RIR to generate IRs for complex indoor scenes. We can see that the runtime of MESH2IR is higher than FAST-RIR because we use a graph-based network to process the mesh. MESH2IR still outperforms GA on a single CPU.

| IR Generator | Hardware | Avg time | Scene Type |
|---|---|---|---|
| GA [1] | CPU | 30.05s | Simple |
| **MESH2IR** | **CPU** | **0.13s** | **Complex** |
| MESH2IR(Batch Size 1) | GPU | $1.32 \times 10^{-2}$s | Complex |
| FAST-RIR(Batch Size 128) | GPU | $5.9 \times 10^{-5}$s | Simple |
| **MESH2IR(Batch Size 128)** | **GPU** | **$2.6 \times 10^{-3}$s** | **Complex** |
| **MESH2IR[Mesh Encoder]** | **GPU** | **$2.57 \times 10^{-3}$s** | **Complex** |
| **MESH2IR[IR Generator]** | **GPU** | **$7.4 \times 10^{-5}$s** | **Complex** |

## 4.6 Applications

We demonstrate the benefit of our MESH2IR for several downstream speech applications - neural sound rendering, speech dereverberation and reverberant speech separation. For a fair comparison between different methods, we generate 11K IRs from 600 scene meshes not used during training our MESH2IR using GA [1], GWA [26], MESH2IR-D-EDR, and MESH2IR. We generate reverberant speech using the 11K IRs, and train speech dereverberation and speech separation methods. Reverberant speech can be generated from a dry sound source and an IR via a convolution:

$$x(t) = s(t) * r(t) \tag{4.11}$$

where $x(t)$ is the reverberant speech signal, $r(t)$ is the IR, $s(t)$ is the dry speech signal.

### 4.6.1 Neural Sound Rendering

Our MESH2IR is the first neural-network-based approach to predict IRs from a given 3D scene mesh at interactive rates. Given this advantage, MESH2IR can be applied to general 3D scenes to enable real-time neural sound rendering, without pre-computation on new scenes. For single IR updates in dynamic scenes, MESH2IR can operate at more than 100 frames per second (fps), which is significantly beyond the requirement for interactive applications (e.g., 10 fps). We demonstrate its sound rendering quality in our supplemental video [1].

### 4.6.2 Speech Dereverberation

Speech dereverberation is the process of obtaining reverb-free speech from reverberant speech. We train speech dereverberation models using data generated from different synthetic IR generation methods and compare their performance on data generated from recorded IRs. We test the models on IRs from the MIT IR dataset [184], the BUTReverb dataset [185], and the RWCP Aachen IR dataset [186, 187]. For all experiments, we train the SkipConvNet [188] model with its default parameters. We use the metric speech-to-reverberation modulation energy ratio (SRMR) [189] to measure the improvement obtained by the dereverberation process. Higher SRMR implies higher speech quality and lower reverberation effects. We also report the SRMR of the baseline where we do not apply any dereverberation (Reverb).

In Table 4.4, we test the models on reverberant data generated from recorded IRs. Across all IR datasets, we see that the SRMR of our MESH2IR model is similar to the SRMR of the GWA IRs [26]. We also see that MESH2IR outperforms the GA [1] method, which only generates

[1] `https://anton-jeran.github.io/M2IR/`

Table 4.4: Speech dereverberation results are obtained when training data is generated by different synthetic IR generation methods. Testing is done on reverberant data synthesized from IRs present in three different datasets containing recorded IRs collected in a variety of environments. Higher SRMR is better.

| **Training Dataset** | **SRMR** | | |
|---|---|---|---|
| | **MIT** | **BUTReverb** | **RWCP Aachen** |
| Reverb | 7.35 | 3.14 | 5.16 |
| GA [1] | 6.39 | 3.74 | 4.83 |
| GWA [26] | **7.67** | **4.6** | **6.14** |
| **MESH2IR-D-EDR** | 6.18 | 3.29 | 4.32 |
| **MESH2IR (ours)** | **7.82** | **4.27** | **5.88** |

IRs using geometric simulations (less accurate than IRs generated from the GWA dataset). The SRMR of MESH2IR is closer to GWA IRs when compared with MESH2IR-D-EDR. Therefore, MESH2IR generates more accurate IRs when compared with MESH2IR-D-EDR.

### 4.6.3 Speech Separation

Speech separation, also referred to as the cocktail party problem, is the process of separating a mixture of speech signals into its constituent speech signals. We check the performance of different synthetic IR generation methods on the task of reverberant speech separation - separating a reverberant mixture into its constituent reverberant speech signals. The dry speech signals and mixtures from the Libri2Mix [190] dataset are convolved with IRs to generate the reverberant data. We train the DPRNN-TasNet [191] model for all speech separation experiments. We utilize the default implementation provided by the Asteroid [192] framework. The model is tested on reverberant mixtures created from real recordings obtained from four different room configurations in the VOiCES [193] dataset. We clearly see that our MESH2IR performs similar to GWA [26], and outperforms GA method [1] and MESH2IR-D-EDR.

Table 4.5: Speech separation results in the presence of reverberation are shown below. We report the improvement in the Scale-Invariant Signal Distortion Ratio (SI-SDRi) over the reverberant mixture. Higher Si-SDRi is better. We report performance on reverberant mixtures generated in four different room configurations present in the VOiCES dataset.

| **Training Dataset** | **SI-SDRi** | | | |
|---|---|---|---|---|
| | **Room** 1 | **Room 2** | **Room 3** | **Room 4** |
| GA [1] | 2.26 | 2.22 | 1.33 | 2.35 |
| GWA [26] | **4.75** | **4.75** | **2.41** | **4.91** |
| **MESH2IR-D-EDR** | 4.68 | 4.35 | 1.87 | 4.72 |
| **MESH2IR (ours)** | **4.91** | **4.89** | **2.54** | **5.13** |

## 4.7 Conclusion and Future Work

We present a novel neural-network-based MESH2IR architecture to generate thousands of IRs for a given furnished indoor 3D scene on the fly. The IR generation speed of our MESH2IR is constant within a given complex 3D scene, irrespective of the complexity of the scene. Our MESH2IR can generate thousands of IRs per second for a given 3D scene. We show that the IRs predicted by our MESH2IR in unseen indoor 3D scenes are highly similar to the ground truth IRs generated from the GWA dataset, which is used to train our MESH2IR.

Our work has some limitations. One is we cannot control the characteristics of the scene materials such as the amount of sound absorption and scattering, which can affect the overall amplitude of the IR. Efficiently inputting the characteristics of scene material to our MESH2IR may improve the accuracy of IR generation. In addition, while MESH2IR can handle moving sound by simulating many IRs according to a trajectory, when objects in the scene moves, the scene representation changes, which incurs additional encoding time for MESH2IR, making it less efficient in highly dynamic scenes. In the future, we would like to integrate our MESH2IR into game engines and other interactive applications, and evaluate its benefits.

# Chapter 5: Fast Binaural Room Impulse Response Generator for AR/VR Application

## 5.1 Motivation

Recent advances in computer vision and 3D reconstruction algorithms have made it possible to generate 3D models of real scenes in real-time [56, 194]. These reconstructed 3D models are used for ray-tracing simulation [195], surveying [196], visual analysis or interactive walkthroughs of buildings [197]. Furthermore, many tools or systems are available to transform real-life spaces into digital models [198], which offer higher visual fidelity than panoramic scans. The resulting static 3D models are used to generate immersive 3D experiences for VR or AR applications.

Many reconstructed models corresponding to apartments, houses, offices, public places, malls, or tourist attractions consist of multiple sound sources (e.g., human speaker, dishwasher, telephone, music). In order to improve the sense of the presence for a user, it is important to augment the visual realism with acoustic effects generated by these sources. It is well known that a user's sense of presence in VR or AR environments can be improved by generating plausible sounds [199]. The resulting acoustic effects vary based on the location of each source, the listener and the environment characteristics [200]. In practice, the acoustic effects in VR or AR environments can be modeled using impulse responses (IRs), which capture how sound propagates from a source location to the position of the receiver in a given scene. IRs contain the necessary information for acoustic scene analysis such as the early reflections, late reverberation, arrival time, energy of direct and indirect sound, etc. The IR can be convolved with any dry sound (real or virtual) to apply the desired acoustic effects. Binaural IR characterizes the sound propagation from the sound source to the left and right ears of the listener. Unlike monaural IRs, binaural IRs have sufficient spatial information to locate the sound source accurately. Therefore binaural impulse responses (BIRs) give an immersive experience in AR and VR applications. It turns out that recording the BIRs in real scenes can be challenging and needs special capturing hardware. Furthermore, these BIRs need to be recaptured if the source or listener position changes.

In synthetic scenes, the IRs can be computed in real-time using sound propagation algorithms [200, 201]. However, current propagation algorithms are limited to synthetic scenes where an exact geometric representation of the scene and acoustic material properties are known as apriori. On the other hand, generating a large number of high-quality IRs for complex 3D real scenes in real-time remains a challenging problem [27].

Recently, neural-network-based sound propagation methods to generate IRs have been proposed for interactive audio rendering applications [2, 3, 53, 202]. After training, the network can be used to generate a large number of IRs for 3D scenes. However, current learning methods have some limitations. They only deal with the mesh geometry, compute monaural IRs, and do not consider the acoustic material properties of the objects in the 3D scene. The material acoustic properties depend on the surface roughness, thickness and acoustic impedance [203, 204]. The materials in the 3D scene strongly influence the overall accuracy of the IR by controlling the amount of sound absorption and scattering when propagating sound waves interact with each surface in the scene. Moreover, current methods may not be directly applied to reconstructed 3D scenes with significant holes.

## 5.2 Main Contributions

We present a novel neural-network-based sound propagation method to render audio for real indoor 3D scenes in real-time. Our approach is general and can generate BIRs for arbitrary topologies and material properties in the 3D scenes, based on the source and listener locations. Our sound propagation network comprises a graph neural network to encode the 3D scene materials and the topology, and a conditional generative adversarial network (CGAN) conditioned on the encoded 3D scene to generate the BIRs. The CGAN consists of a generator and a discriminator network. Some of the novel components of our work include:

**1. Material-aware learning-based method:** We represent the material's acoustic properties using the frequency-dependent absorption and scattering coefficients. We calculate these material properties using average sound absorption and scattering coefficients for each vertex in a 3D scene from the input semantic labels of the 3D model and acoustic material databases. We propose an efficient approach to incorporate material properties in our Listen2Scene architecture. Our method results in 48% better accuracy over prior learning methods in terms of acoustic characteristics of the IRs.

**2. Binaural Impulse Response (BIR) Generation:** We present a simple and efficient cost function to the generator network in our CGAN to incorporate spatial acoustic effects such as the difference in the time-of-arrival of sound arriving in left and right ears (interaural time difference) [205] and sound level difference in both ears caused by the barrier created by the head when the sound is arriving (interaural level difference).

**3. Perceptual evaluation:** We performed a user study to evaluate the benefits of our proposed audio rendering approach. We rendered audio for 5 real environments with different levels of complexity with the number of vertices in the selected environments varying from 0.5 million to 2.5 million (Fig. 5.8) and asked the participants to choose between our proposed approach and the baseline methods. More than 67% and 45% of the participants observed that the audio rendered from Listen2Scene is more plausible than the prior learning-based approach MESH2IR and interactive geometric-based sound propagation algorithm respectively. We also compared the audio rendering using our approach with recorded IRs [206] where the materials are an independent variable.

**4. Novel Dataset:** We generate 1 million high-quality BIRs using the geometric-based sound propagation method [207] for around 1500 3D real scenes in the ScanNet dataset [56]. Among 1 million BIRs, we randomly sampled 200,000 BIRs to train our network. We release the full BIR dataset in the wav format [1].

## 5.3 Model Representation and Dataset Generation

Our approach is designed for real scenes. We use 3D reconstructed scenes from the RGB-D data captured using commodity devices (e.g., iPad and Microsoft Kinect). These reconstructed 3D scenes are segmented and the objects in the 3D scene can be annotated by crowdsourcing [56, 194]. Our goal is to use these mesh representations and semantic information to generate plausible

[1] https://drive.usercontent.google.com/download?id=1FnBadVRQvtV9jMrCz_F-U_YwjvxkK8s0&authuser=0

acoustic effects. An overview of our approach is given in Figure 5.1.

We preprocess the annotated 3D scene to close the holes in the reconstructed 3D scene and simplify the 3D scene by reducing the number of faces. We perform mesh simplification using graph processing to reduce the complexity of the 3D scene input into our network. We represent the simplified 3D scene as a graph $GN$ and input $GN$ to our graph neural network $Net_{GR}$ (Figure 5.3) to encode the input 3D scene as an 8-dimensional latent vector. Then we pass the encoded 3D scene latent vector along with the listener position $LP$ and the source position $SP$ to our generator network $Net_{GN}$ (Figure 5.1) to generate binaural impulse response $BIR$ (Equation 5.1).

$$BIR = Net_{GN}(Net_{GR}(GN), LP, SP). \tag{5.1}$$

We rendered audio $S_R$ for the given spatial locations of the receiver and listener in a given 3D scene at time $t$ by convolving the corresponding $BIR$ with any clean or dry audio signal $S_C$ (Equation 5.2).

$$S_R[t] = S_C[t] \circledast BIR[t]. \tag{5.2}$$

### 5.3.1 Dataset Creation

There aren't real-world and synthetic BIR datasets for a wide range of real 3D scenes captured using commodity hardware available to train our Listen2Scene. Therefore we create synthetic BIRs using a geometric simulator [1] for 3D reconstructed real-world scenes in the ScanNet dataset [56] to train our Listen2Scene. We preprocess the 3D meshes and assign meaningful acoustic material properties to each object and surface in the 3D scene (Section 5.3.1.1). Next, we sample source and listener positions and simulate BIRs using the geometric simulator (Section 5.3.1.2).

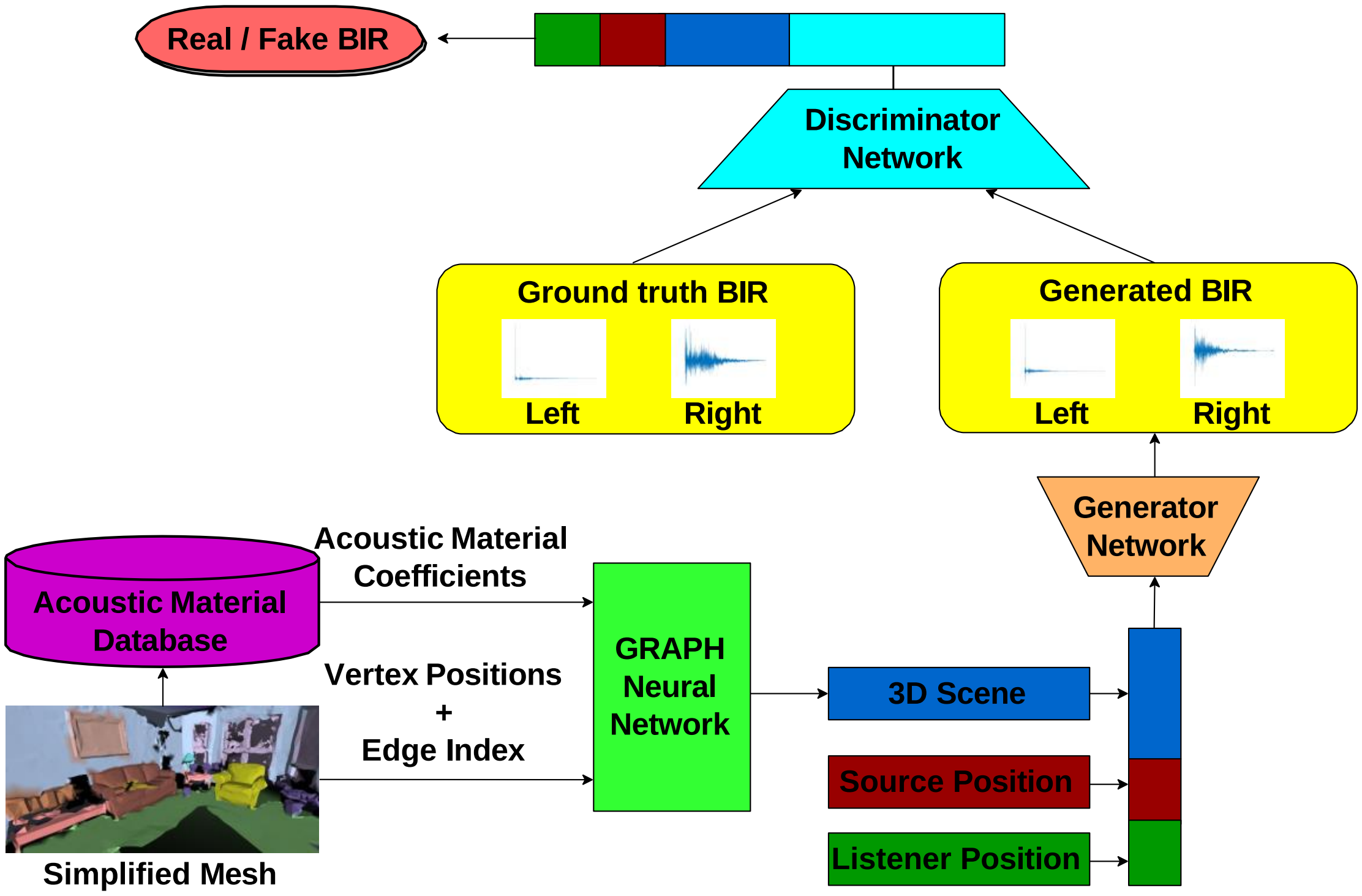

Figure 5.1: The overall sound propagation architecture of our Listen2Scene method: The simplified 3D scene mesh with material annotations is passed to the acoustic material database to estimate the acoustic material coefficients (absorption and scattering coefficient). We pass the acoustic material coefficients, vertex positions, and edge index to our graph neural network (Figure 5.3) to encode the 3D scene into a latent vector. Our generator network takes the 3D Scene and listener and source positions as input and generates a corresponding BIR. The discriminator network discriminates between the generated BIR and the ground truth BIR during training.

#### 5.3.1.1 Mesh Preprocessing and Material Assignment

The ScanNet dataset contains vertex-level segmented mesh. To make the dataset compatible with a geometric-based sound propagation system, we convert vertex-level segmentation of the 3D scene to face-level segmentation of the 3D scene. Face-level segmentation is used to assign material acoustic coefficients to each surface in the 3D scene. Many of the meshes in the ScanNet dataset have holes in the surface boundary and the ceiling is not present. The holes can prevent some of the sound rays from reflecting back to the listener and result in generating unrealistic acoustic effects using the ray tracing-based geometric sound propagation algorithm. We compute the convex hull of the overall 3D scene mesh and merge it with the original mesh to close the holes in the outer surface boundaries. We fill small holes on internally separated spaces using fill_holes() function in the trimesh library [208].

The ScanNet dataset also contains the semantic annotation (i.e., instance-level object category labels such as dish rack, wall, laundry basket etc.) for every 3D scene. We use the absorption coefficient acoustic database with more than 2000 materials [209] to get the absorption coefficient of each material in the ScanNet 3D reconstructions. We do not always find exact ScanNet object labels in the acoustic database. Therefore, we use the natural language processing (NLP) technique to find the closest matching material in the acoustic database for every ScanNet object label and assign its absorption coefficient to the ScanNet object label. To find the closest matching material, we encode the object labels in ScanNet and material names in the acoustic database into fixed-length sentence embeddings [210]. Transformer-based sentence embedding vectors are close in cosine similarity distance for sentences with similar meaning and outperform in many NLP tasks [211]. We use the Microsoft pre-trained sentence transformer model to encode materials into 768-dimensional sentence embedding. We use the cosine similarity of the ScanNet object labels and materials in the acoustic database and assign the closest materials absorption coefficients to the objects in the ScanNet.

In addition to absorption coefficients, we need scattering coefficients for geometric sound propagation. The scattering coefficients are not available in the acoustic database [209]. Therefore, we adapt the sampling approach proposed in GWA [26]. We fit a Gaussian distribution by calculating the mean and standard deviation of 37 sets of scattering coefficients collected from the BRAS benchmark [206] and we sample randomly from the distribution for every 3D scene.

#### 5.3.1.2 Geometric Sound Propagation

For every 3D scene, we perform grid sampling with 1m spacing in all three dimensions. We also ensure that there is a minimum gap of 0.2 m between the sampled position and objects in the scene to prevent collisions. The number of grid samples varies with the dimension of the 3D scene. We randomly place 10 sources in the grid sampled locations and the rest of the samples are assigned to listener locations. We perform geometric simulations for every combination of listener and source positions. We use 20, 000 rays for geometric propagation and the simulation stops when the maximum depth of specular or diffuse reflection is 2000 or the ray energy is below the hearing threshold.

## 5.4 Our Learning Approach

In this section, we present the details of our learning method. Our approach learns to generate BIRs for 3D reconstructed real scenes, which may have noise or holes. We first present our approach to representing the topology and material details of the 3D scene using our graph neural network (Section 5.4.1). Next, we present our overall architecture, which takes the 3D scene and generates plausible BIRs and training details (Section 5.4.2).

### 5.4.1 3D Scene Representation

The ScanNet dataset represents the RGB-D data collected from the 3D scene in the form of a 3D mesh. The shapes of the objects in the 3D scene are represented using the vertices and triangular faces in the 3D Cartesian coordinates. The ScanNet dataset also provides object category labels at the vertex level. We perform the mesh pre-processing and material assignment approach as mentioned in Section 5.3.1.1. To reduce the size and complexity of the data passed to the neural network while preserving high-level object details, we adapt and modify prior work [3] by performing mesh simplifications using PyMeshlab's implementation of the quadratic-based edge collapse mesh simplification algorithm [171]. We simplify the meshes to have only 2.5% of the initial number of faces. The mesh simplification algorithm can simplify the mesh while preserving the vertex-level segmentation of the mesh (Figure 5.2). The simplified meshes typically have around 10,000 faces.

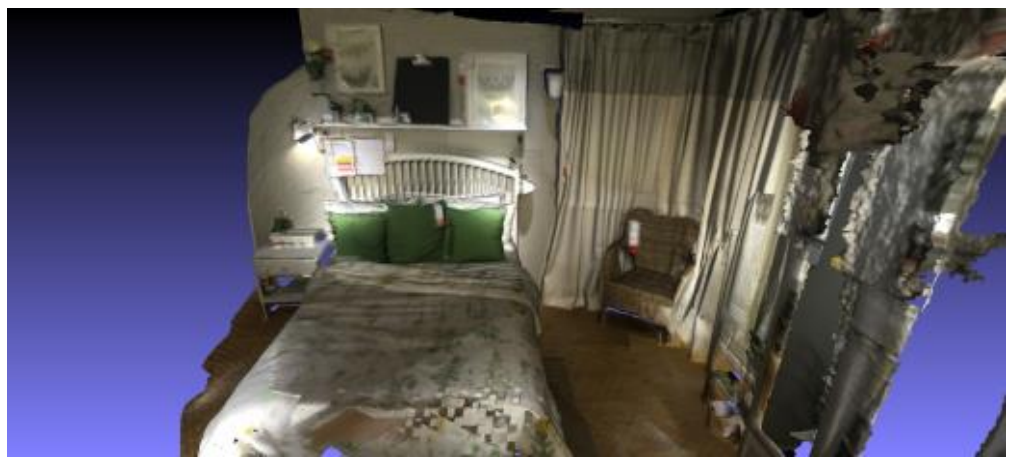

(a) 3D reconstructed mesh.

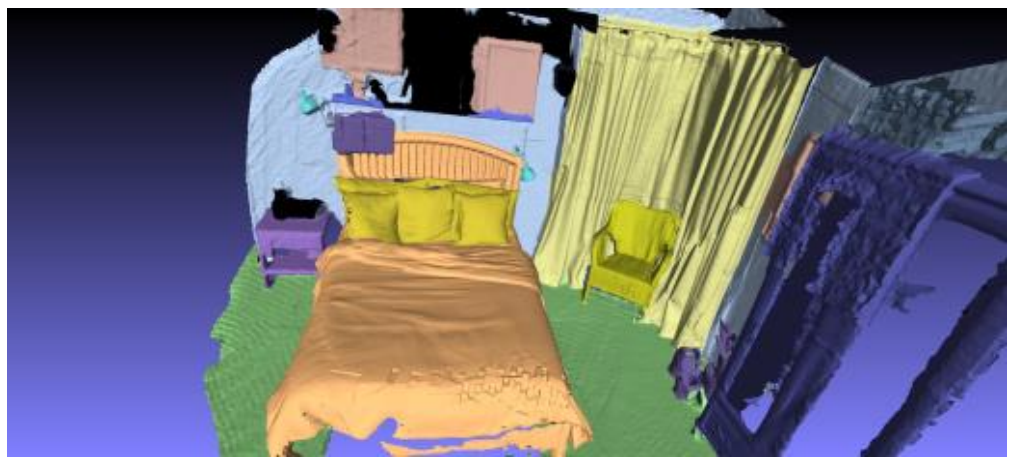

(b) Segmented mesh.

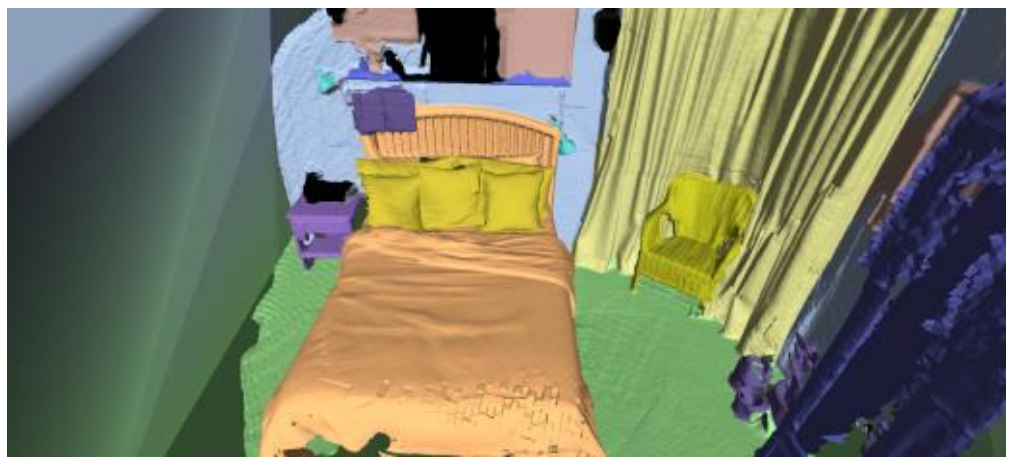

(c) Closed segmented mesh.

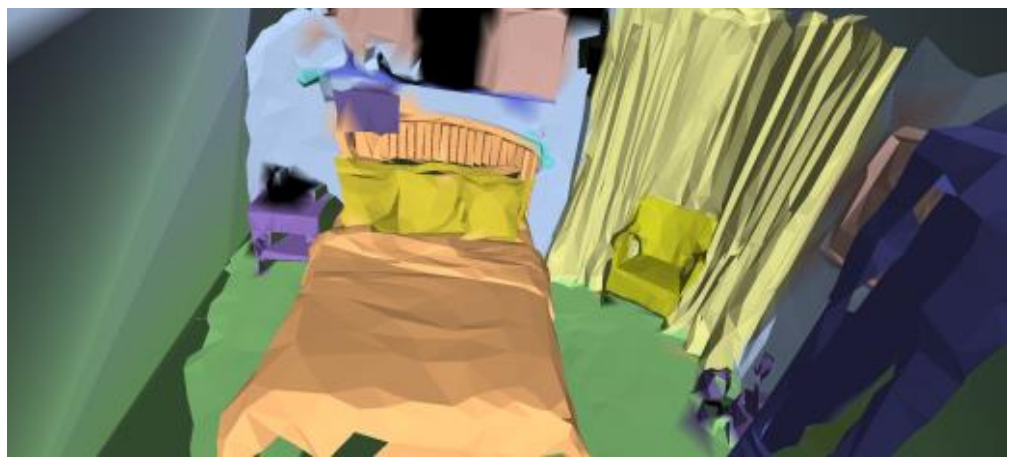

(d) Mesh after simplification.

Figure 5.2: The 3D reconstruction of the real scene from the ScanNet (a); object category-level segmentation of the 3D scene with each category is represented by a different color (b); the modified mesh after closing the holes using convex hull (c); the simplified mesh with object-level segmentation information preserved (d); we observe that high-level object shapes (e.g., bed, office chair, wooden table, etc.) and materials are preserved even after simplifying the mesh to 2.5% of the original size.

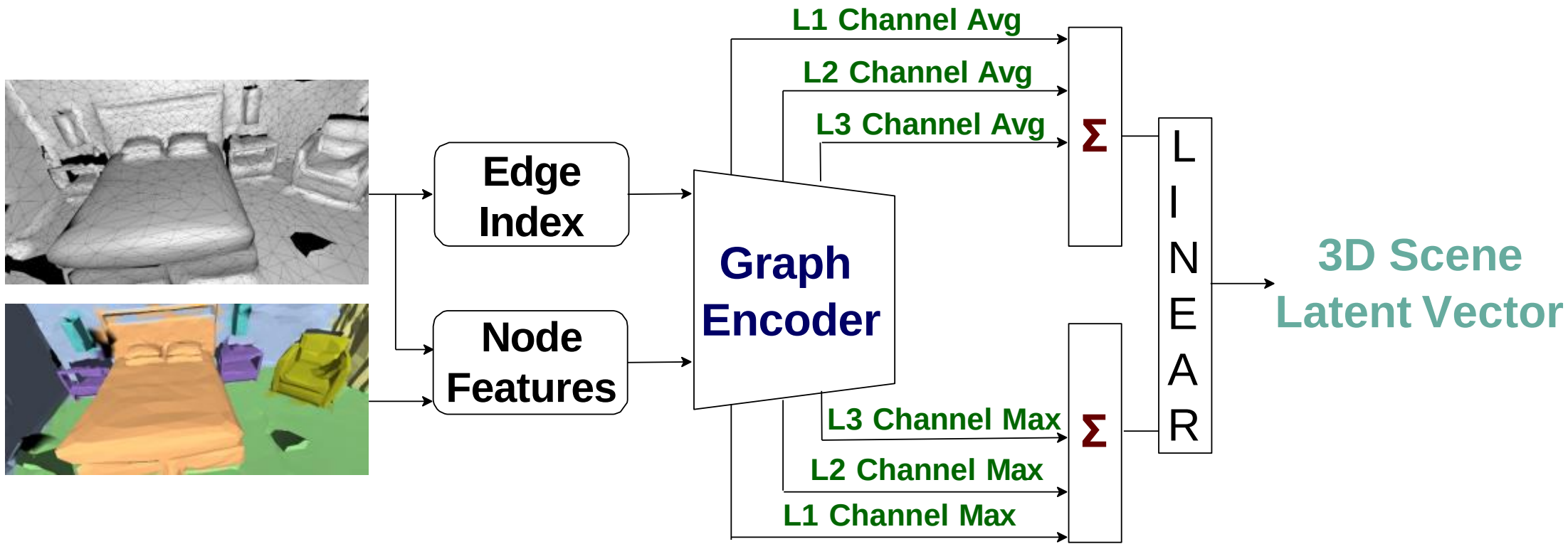

Figure 5.3: Our network architecture represents a 3D scene as an 8-dimensional latent vector. The vertex positions and material properties are combined to produce the node features. We pass the edge index and node features from the 3D scene as input to the graph encoder. The graph encoder consists of 3 graph layers (L1, L2, and L3). The channel-wise average and the channel-wise maximum of the node features in each layer are aggregated and passed to linear layers. Linear layers output a 3D scene latent vector.

In Figure 5.2, we observe that segmented mesh interpolates the nearby materials to the closed holes (e.g., holes near the floor are assigned to materials of the floor and the material is represented in green). We observe that even after mesh simplification to 2.5% of the original size, high-level object structures are preserved.

The triangular mesh of the 3D scene can be represented using graph $G = \langle V, E \rangle$, where $V$ represents the 3D Cartesian coordinates of the set of vertices/nodes and $E$ is the connectivity of each node (edge index). The vertex coordinates of three dimensions are features of the node in a graph. To add the material properties of the 3D scene, we increase the node feature dimension to five. The material properties can be represented using the material's absorption coefficient and scattering coefficient. The absorption coefficient represents how much sound can be absorbed by the material. Metal absorbs the least sound and has a very low coefficient. A cushion is a sound-absorbing material and has a high coefficient. The scattering coefficient represents the roughness of the material's surface. When the surface is rough, the sound will be scattered in all directions and has a high coefficient; smooth surfaces have a low coefficient value.

The absorbing and scattering coefficients are frequency-dependent coefficients. The coefficients are defined for the 8-octave bands between 62.5 Hz and 8000 Hz. To reduce the dimensionality of the coefficients, we calculate the average coefficients by taking the coefficients at 500 Hz and 1000 Hz. We show the benefit of our approach of calculating average coefficients at 500 Hz and 1000 Hz in Section 5.6.2. In many practical applications, the average value of room acoustics parameters like reverberation time is used for analysis instead of all the values at different octave bands [27, 47, 68]. We increase the node features $V$ by combining $(x, y, z)$ Cartesian coordinates of the vertex with the average absorption coefficient $ab$ and average scattering coefficient $sc$ ($V = [x, y, z, ab, sc]$).

We input node features and edge index to the graph encoder network to encode the 3D scene to a low dimensional space. The encoder network has 3 layers. In each layer, the graph convolution layer [172] is used to encode the node features (Equation 5.3). We gradually reduce the size of the graph by dropping the number of node features to 0.6 times the original number of node features in each layer using the graph pooling layer.

In Equation 5.3, the adjacency matrix representing the edge index of the 3D scene ($A$) and the identity matrix $I$ are aggregated to calculate $\hat{A}$ ($\hat{A} = A + I$). Each column of $\hat{A}$ is summed to get diagonal matrix $\hat{D}$ ($\hat{D}_{ii} = \sum_j \hat{A}_{ij}$). $W^{(n)}$ is a trainable weight matrix for layer $n$. Node features at layers $n$ and $n + 1$ are $N_F^{(n)}$ and $N_F^{(n+1)}$, respectively.

$$N_F^{(n+1)} = \sigma(\hat{D}^{-\frac{1}{2}} \hat{A} \hat{D}^{-\frac{1}{2}} N_F^{(n)} W^{(n)}), \tag{5.3}$$

The output of the graph convolution layer is passed to the graph pooling layers [173, 174] to simplify the graph by reducing the node features and edge index. The graph pooling layer initially calculates the square of the adjacency matrix ($A_{new}^{(n)} = A^{(n)}A^{(n)}$) to increase the graph connectivity and is used to choose the top N node features. The adjacency matrix $A_{new}^{(n)}$ prevents isolated edges in the graph encoded 3D scene when choosing top N node features from the input graph and discarding other features.

We calculate the channel-wise average and channel-wise maximum of the output node features in each graph layer in the graph encoder network. We aggregate the channel-wise average and channel-wise maximum separately over the 3 layers. We concatenate the aggregated maximum and aggregated average values and pass them as input to a set of linear layers. We concatenate the learned features in each layer to ensure that the linear layers use all the learned features to construct an accurate 3D scene latent vector of dimension 8 as an output from the linear layer.

### 5.4.2 BIR Generation

We use a one-dimensional modified conditional generative adversarial network (CGAN) to generate BIRs. The standard CGAN architectures [149, 150] generate multiple different samples corresponding to input condition $y$ by changing the input random noise vector $z$. In our CGAN architecture, we only input the condition $y$ to generate a single precise output. Our CGAN network takes a 3D scene latent vector as the input condition and generates a single precise BIR. We propose a novel cost function to trigger the network to generate binaural effects such as interaural level difference (ILD) and interaural time difference (ITD) accurately.

We extend the IR preprocessing approach proposed in MESH2IR to make the network learn to generate BIRs with large variations of standard deviation (SD) efficiently. In Section 5.3.1, we generate high-fidelity BIRs with a sampling rate of 48,000 Hz. We initially downsample the BIRs to 16,000 Hz to represent a longer duration of BIRs. We train our network to generate around 0.25 seconds (3968 samples) of BIR to reduce the complexity of the network. Our architecture can be easily modified to train the network to generate any duration of BIRs. The complexity of our network changes linearly with the duration of generated BIRs. We calculate the SD of the BIR and divide the BIR with SD to have fewer variations over training samples. We replicate the SD 128 times and concatenate it towards the end of the BIR. Therefore, each channel of the preprocessed BIR will have 4096 samples (3968+128). We train our network to generate preprocessed BIRs. Later, we can recover the original BIR by removing SD represented in the last 128 samples, getting the average of SD values, and multiplying the first 3968 samples by the average SD value. We get the average SD over 128 samples to reduce the error of the recovered SD.

Our CGAN architecture consists of a generator network ($G$) and a discriminator network ($D$) (Figure 5.1). We pass the 3D scene information $\Gamma_S$ consisting of mesh topology and materials of the 3D scenes represented using a latent vector, and the listener and source position as an input to $G$. We train the $G$ and the $D$ in our CGAN architecture using our created BIRs (Section 5.3.1) and $\Gamma_S$ in the data distribution $p_{data}$. We train $G$ to minimize the objective function $L_G$ and the $D$ to maximize the objective function $L_D$ alternatingly.

**Generator Objective Function ($L_G$) :** The $L_G$ is minimized during training to generate accurate BIRs for the given condition $\Gamma_S$. The $L_G$ (Equation 5.4) consists of modified CGAN error ($L_{CGAN}$), BIR error ($L_{BIR}$), ED error ($L_{ED}$), and mean square error ($L_{MSE}$). The contribution of each individual error is controlled using the weights $\lambda_{BIR}$, $\lambda_{ED}$ and $\lambda_{MSE}$:

$$L_G = L_{CGAN} + \lambda_{BIR}\, L_{BIR} + \lambda_{ED}\, L_{ED} + \lambda_{MSE}\, L_{MSE}. \tag{5.4}$$

The modified CGAN error is minimized when the BIRs generated using $G$ are difficult to differentiate from the ground truth BIRs by $D$ for each 3D scene $\Gamma_S$:

$$L_{CGAN} = \mathbb{E}_{\Gamma_S \sim p_{data}}[\log(1 - D(G(\Gamma_S), \Gamma_S))]. \quad (5.5)$$

The time of arrival of the direct signal and the magnitude levels of the left and right channels of the BIRs vary significantly with the direction of the sound source. To make sure the network captures the relative variation of the IRs in the left and right channels, we propose the BIR error formulation.

$$L_{BIR} = \mathbb{E}_{(B_G, \Gamma_S) \sim p_{data}}[\mathbb{E}[((B_{LN}(\Gamma_S, s) - B_{RN}(\Gamma_S, s)) - (B_{LG}(\Gamma_S, s) - B_{RG}(\Gamma_S, s)))^2]], \quad (5.6)$$

where $B_{LN}$ and $B_{RN}$ are the left and right channels of the BIRs generated using our network and $B_{LG}$ and $B_{RG}$ are the left and right channels of the ground truth BIRs.

The energy remaining in the BIR ($b$) with respect to the time $t_i$ seconds and at frequency band with center frequency $f_c$ Hz (Equation 5.7) is described using energy decay relief (ED) [176, 177]. In Equation 5.7, the bin $c$ of the short-time Fourier transform of $b$ at time $t$ is defined as $H(b, t, c)$. The ED curves decay smoothly over time and they can be converted into an ”equivalent IR” [175]. In previous works [3, 6], it is observed that ED helps the model to converge.

$$ED(b, t_i, f_c) = \sum_{t=i}^{T} |H(b, t, c)|^2. \quad (5.7)$$

The ED curves reduce exponentially over time. In previous works [3], the mean square error

(MSE) between the ED curves of the ground truth BIR ($B_G$) and the generated BIR ($B_N$) is calculated. This approach does not capture the latter part of ED curves accurately. Therefore we compare the log of the ED curves between ground truth and generated BIRs for each sample ($s$) as follows:

$$\mathcal{L}_{ED} = \mathbb{E}_{(B_G, \Gamma_S) \sim p_{data}}[\mathbb{E}_{c \sim C}[\mathbb{E}[(\log(ED(B_G(\Gamma_S), c, s)) - \log(ED(B_N(\Gamma_S), c, s)))^2]]]. \quad (5.8)$$

To capture the structures of the BIR, we also calculate MSE error in the time domain. For each 3D scene $\Gamma_S$ we compare $B_G$ and $B_N$ over the samples ($s$) of BIR as follows:

$$\mathcal{L}_{MSE} = \mathbb{E}_{(B_G, \Gamma_S) \sim p_{data}}[\mathbb{E}[(B_G(\Gamma_S, s) - B_N(\Gamma_S, s))^2]]. \quad (5.9)$$

**Discriminator Objective Function ($\mathcal{L}_D$) :** The discriminator ($D$) is trained to maximize the objective function $\mathcal{L}_D$ (Equation 5.10) to differentiate the ground truth BIR ($B_G$) and the BIR generated using the generator ($G$) during training for each 3D scene $\Gamma_S$.

$$\mathcal{L}_D = \mathbb{E}_{(B_G, \Gamma_S) \sim p_{data}}[\log(D(B_G(\Gamma_S), \Gamma_S))] + \mathbb{E}_{\Gamma_S \sim p_{data}}[\log(1 - D(G(\Gamma_S), \Gamma_S))]. \quad (5.10)$$

**Network Architecture and Training:** We extend the standard time domain Generator ($G$) and Discriminator ($D$) architectures proposed for monaural IR generation [2, 3]. We modify $G$ to take our 3D scene latent vector of 8 dimensions (Figure 5.3) and the source and listener positions in 3D Cartesian coordinates. Our $G$ takes 14-dimensional conditional vectors and generates 4096x2 preprocessed BIR as output. We also modify our $D$ to differentiate between two channel ground truth and generated BIRs. We train all networks with a batch size of 96 using an RMSprop optimizer. The hyperparameter is chosen manually by looking at how the network converges at the initial epochs. We initially started with a learning rate of 8 x $10^{-5}$, and the learning rate decayed to 0.7 of its previous value every 7 epochs. We trained our network for 100 epochs.

## 5.5 Ablation Experiments

We perform ablation experiments to analyze the contribution of our proposed BIR error (Equation 5.6) and Energy Decay (ED) error (Equation 5.8) in training our network. We also analyze the performance of the network with and without closing holes in the 3D mesh. We generated 900 BIRs for 20 real testing environments for our ablation study.

### 5.5.1 BIR Error

Our BIR error (Equation 5.6) helps to generate binaural acoustic effects by incorporating magnitude level differences between the left and right channels of BIRs. In Figure 5.4, we plot the difference between the left and right channels of the ED curve of BIRs generated using a geometric-based approach, Listen2Scene and Listen2Scene approach trained without BIR error (Listen2Scene-No-BIR). We can observe that incorporating BIR error reduces the gap between the geometric approach (ground truth) and our Listen2Scene.

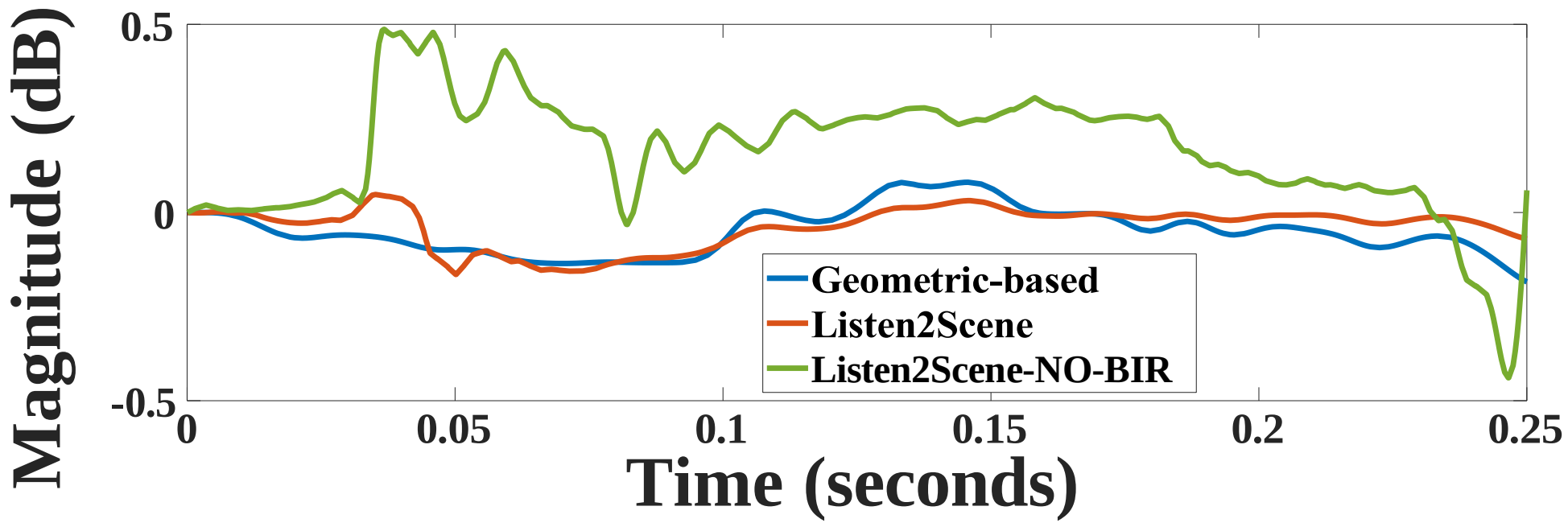

Figure 5.4: The normalized difference in energy decay (ED) curves of left and right channels of BIR. The BIRs are generated using the geometric method, Listen2Scene and Listen2Scene-No-BIR (Listen2Scene trained without BIR error). We observe that the ED curve difference of Listen2Scene closely matches the geometric method.

### 5.5.2 ED Error

We trained our Listen2Scene network with the ED error proposed in MESH2IR [3] (Listen2Scene-ED) and our proposed ED error (Equation 5.8). We calculated the MSE between the normalized ED curves of the ground truth BIRs from the geometric-based approach and the generated BIRs over the center frequencies 125Hz, 500Hz, 1000Hz, 2000Hz and 4000 Hz covering voice frequency and reported in Table 5.1. We can see that MSE of the normalized ED curves in the testing environment is low for our proposed ED error (Listen2Scene). Figure 5.5, shows the normalized ED curves of the left channel BIR from the geometric-based method, Listen2Scene and Listen2Scene-ED at 2000Hz. We can see that the ED curve of Listen2Scene-ED diverges from the geometric-based method after 0.1 seconds.

Table 5.1: The MSE error between the normalized energy decay (ED) curves of the ground truth BIRs from the geometric sound propagation algorithm and the generated BIRs from our Listen2Scene and Listen2Scene trained with ED error proposed in MESH2IR [3] (Listen2Scene-ED). We calculate the MSE over the center frequencies 125Hz, 500Hz, 1000Hz, 2000Hz and 4000 Hz. The best results are shown in **bold**

| **Method** | **Frequency** | | | | |
|---|---|---|---|---|---|
| | **125Hz** | **500Hz** | **1000Hz** | **2000Hz** | **4000Hz** |
| Listen2Scene-ED | 2.58 | 3.28 | 3.99 | 4.16 | 4.23 |
| **Listen2Scene** | **2.50** | **2.93** | **3.54** | **3.56** | **3.56** |

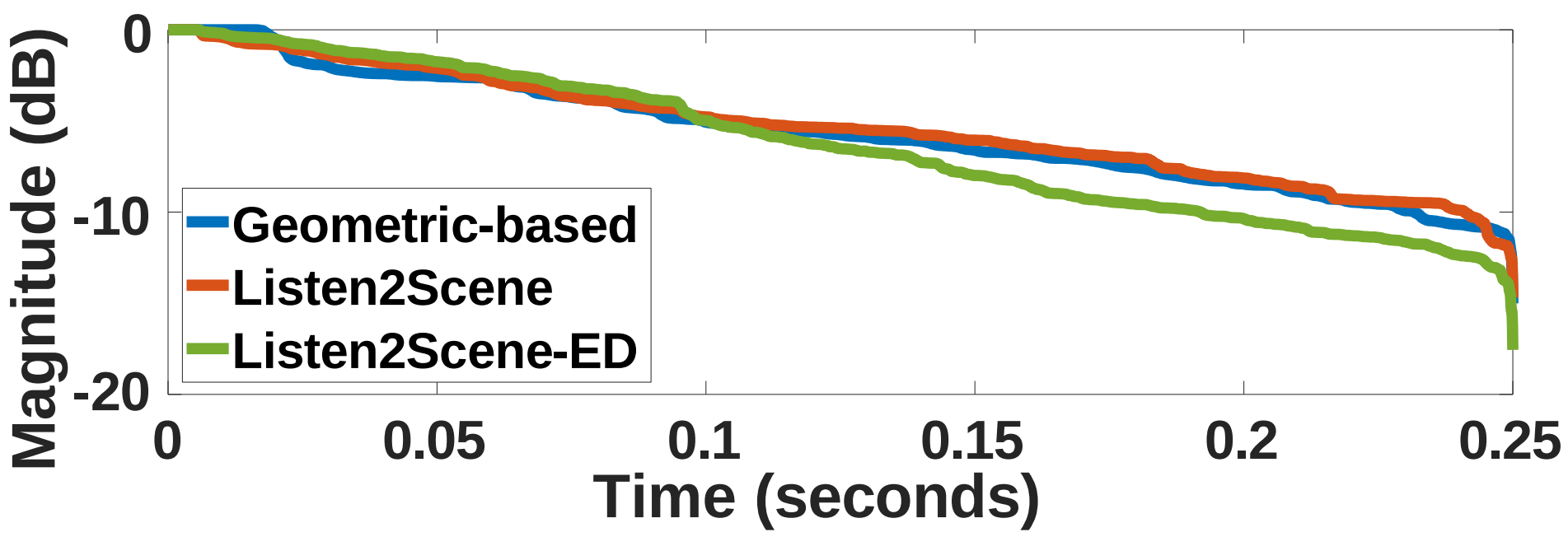

Figure 5.5: The normalized energy decay (ED) curve of the BIRs (left channel) generated using the geometric-based method, Listen2Scene and Listen2Scene-ED (Listen2Scene trained with ED error proposed in MESH2IR [3]) at 2000 Hz. We can see that the ED curve of Listen2Scene matches the geometric method for the entire duration while the ED curve of Listen2Scene-ED starts diverging after 0.1 seconds.

### 5.5.3 Closed and Open Mesh Models

We trained and evaluated our Listen2Scene network using the default 3D mesh with holes (Listen2Scene-Hole) and a closed mesh using our proposed approach (Section 5.3.1.1). We can see in Table 5.2 that the BIRs generated using Listen2Scene match the geometric-based sound propagation algorithm. Figure 5.6, shows the left channel of the BIR from the geometric-based approach and the corresponding BIR from Listen2Scene-Hole. We can see that the BIRs from the geometric-based approach and Listen2Scene-Hole are significantly different.

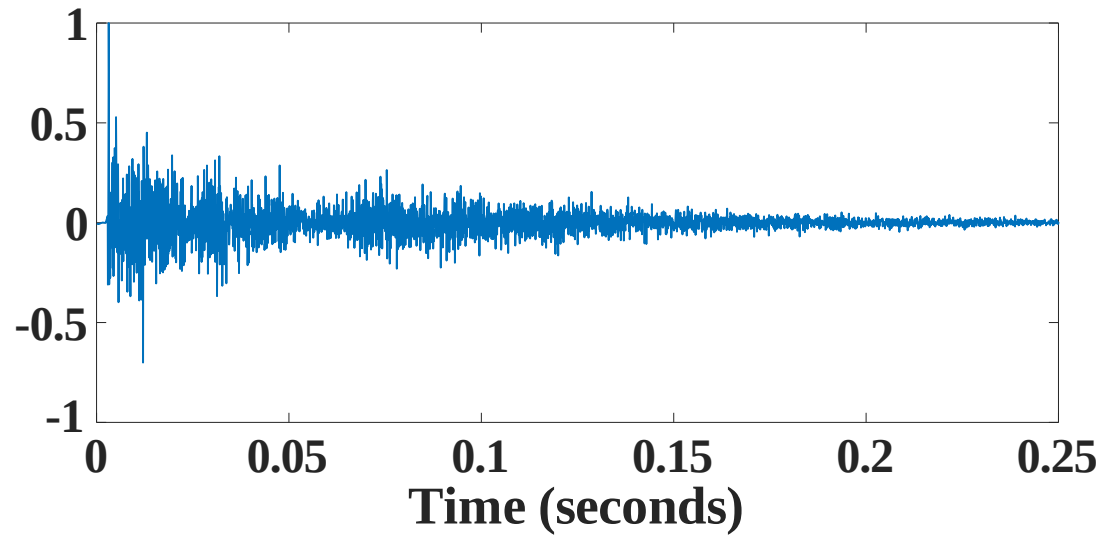

(a) Geometric-based.

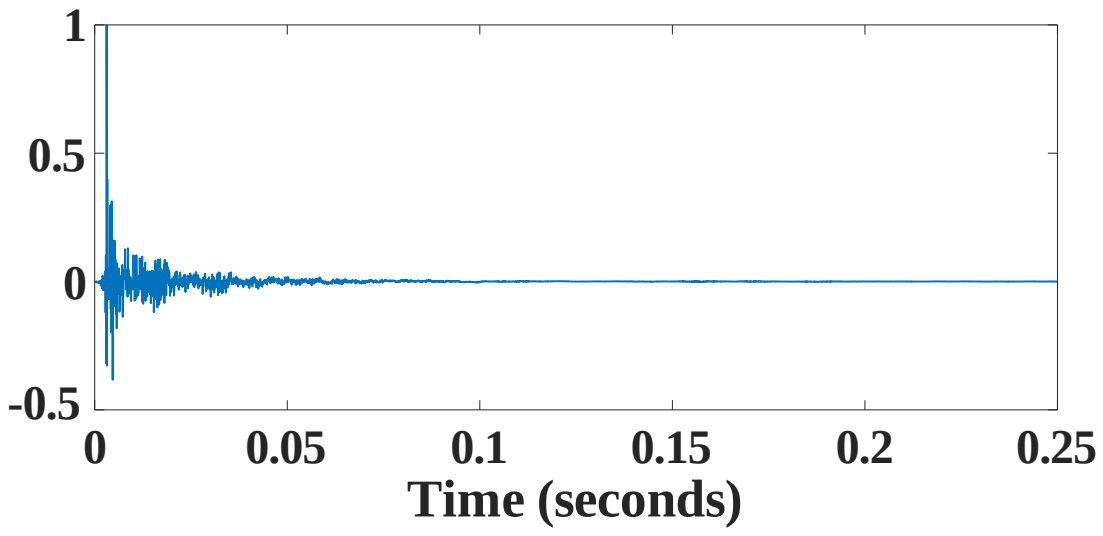

(b) Listen2Scene-Hole.

Figure 5.6: The left channel of the BIR generated using a geometric-based sound propagation algorithm and our Listen2Scene approach without closing the holes (Listen2Scene-Hole). We can see that the BIR from Listen2Scene-Hole significantly varies from the geometric-based approach.

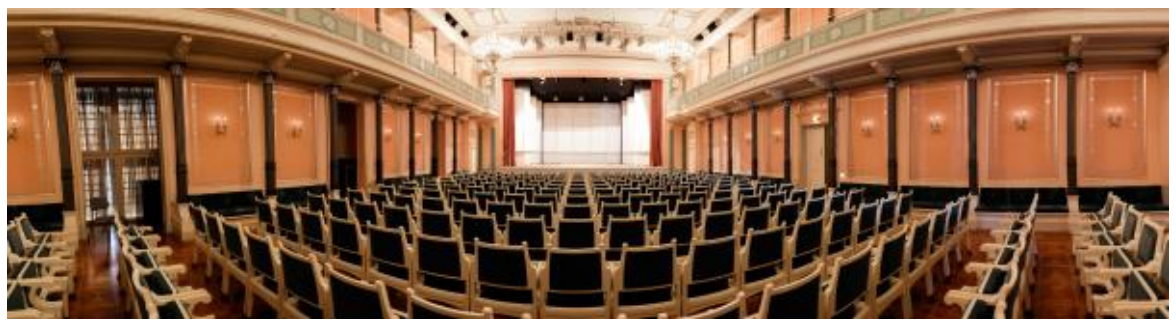

(a) Chamber music hall (Medium).

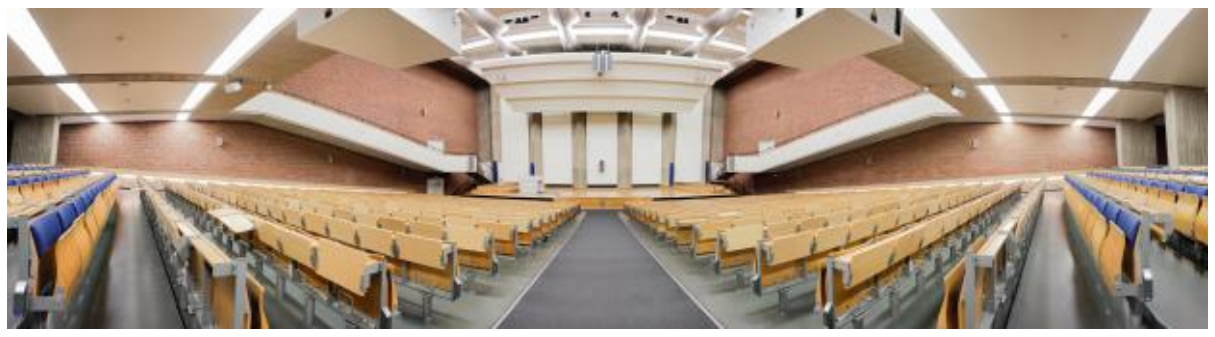

(b) Auditorium (Large).

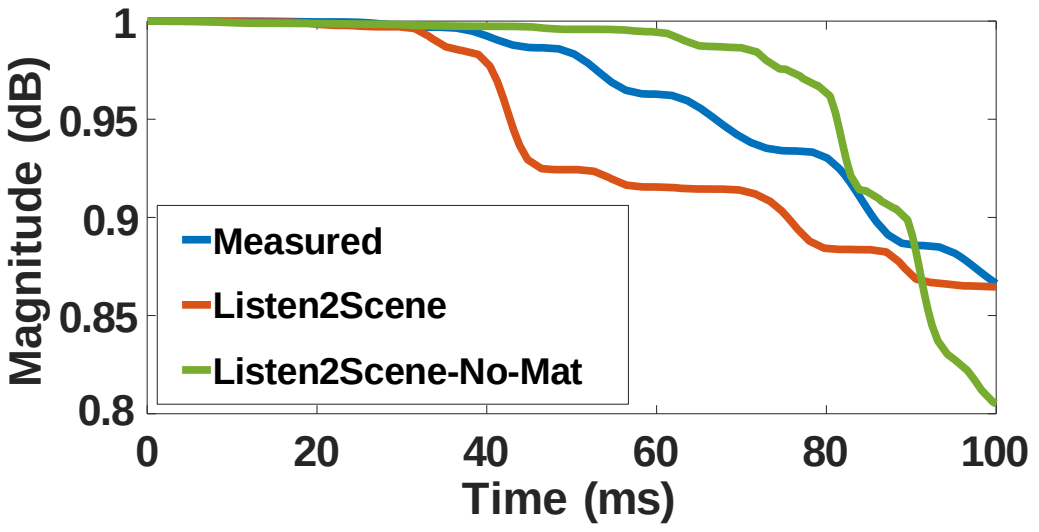

(c) Left channel.

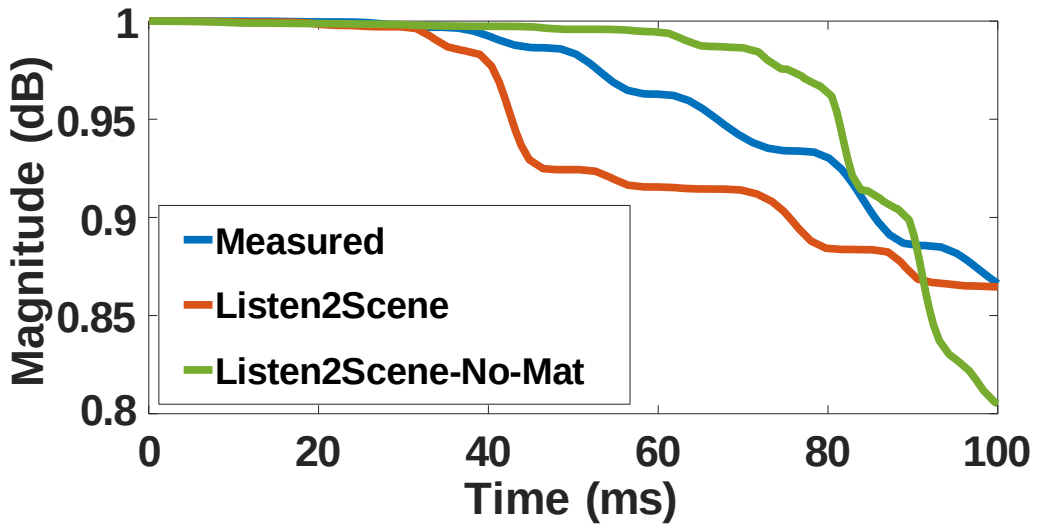

(d) Right channel.

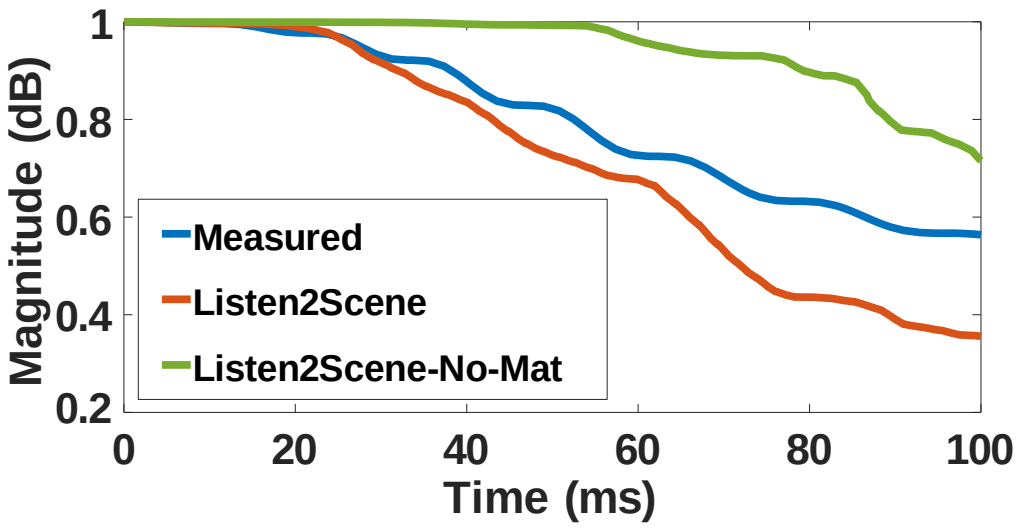

(e) Left channel.

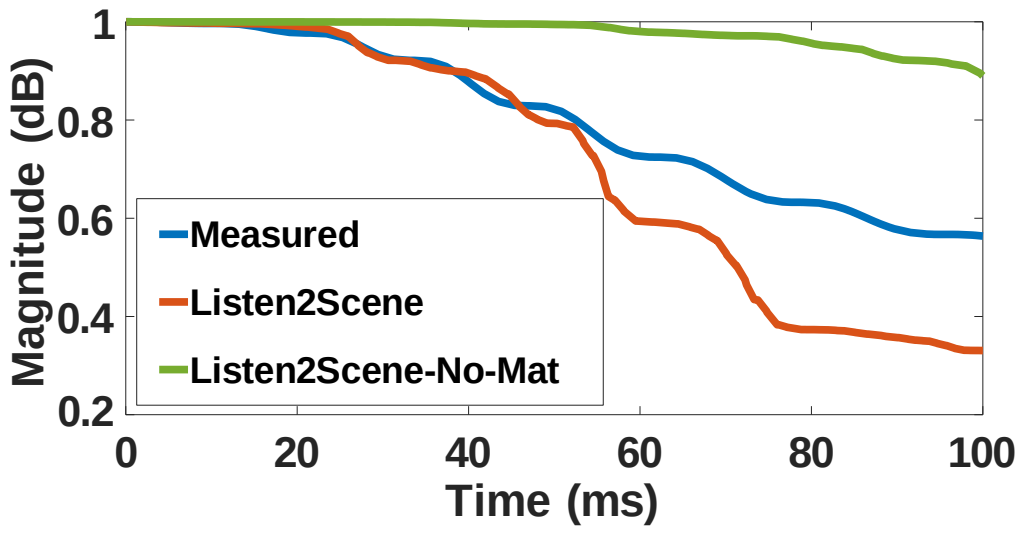

(f) Right channel.

Figure 5.7: The normalized energy decay curves (EDC) of the captured BIRs and the BIRs generated using our approach with material (Listen2Scene) and without material (Listen2Scene-No-Mat) for the 3D scenes in BRAS ((a),(b)). We plot the EDC for the BIRs from the chamber music hall ((c),(d)) and auditorium ((e),(f)). We observe that the EDC of Listen2Scene is closer to the EDC of captured BIRs.

Table 5.2: The binaural impulse responses (BIR) synthesized for real-world 3D scenes. We compare the accuracy of our learning-based sound propagation method (Listen2Scene) with geometric sound propagation algorithms. These 3D reconstructed scenes were not used in the training data for Listen2Scene. Our Listen2Scene can synthesize BIRs corresponding to left and right channels by considering interaural level differences (ILD) and interaural time differences (ITD). We can see high-level structures of BIRs from our Lisen2Scene is similar to the geometric-based method. The mean absolute error of the normalized BIRs (MAE) is less than 0.5 x $10^{-2}$.

| | **Real-world environment 1** | | **Real-world environment 2** | |
|---|---|---|---|---|
| **3D Scene** | 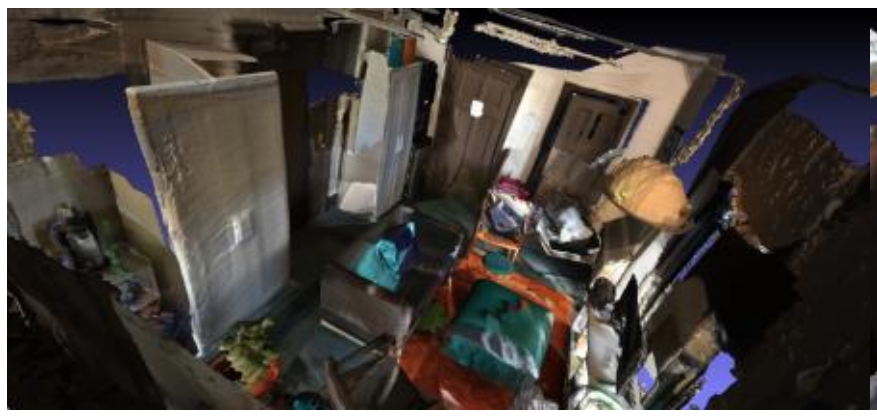 | | 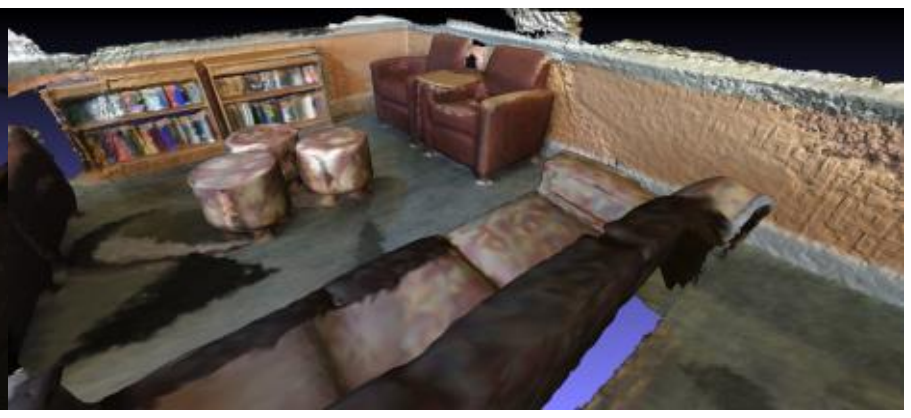 | |
| **Channel** | **Left** | **Right** | **Left** | **Right** |
| **MAE ($10^{-2}$)** | **0.50** | **0.49** | **0.19** | **0.21** |
| **Geometric-based BIR** | 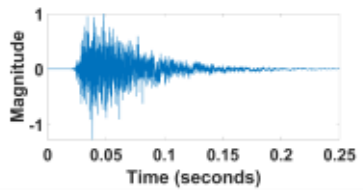 | 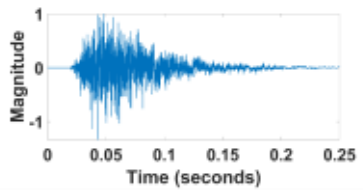 | 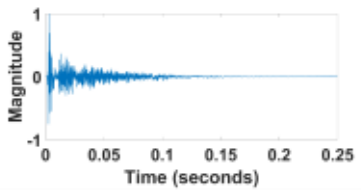 | 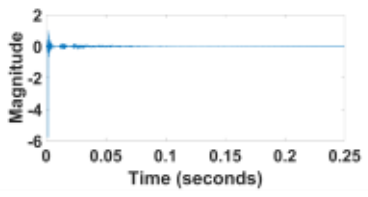 |
| **Our Listen2Scene** | 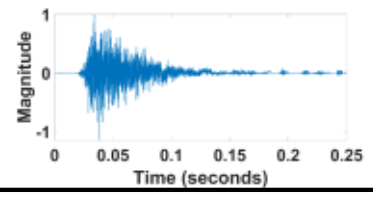 | 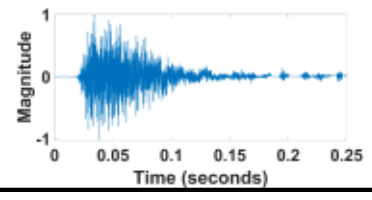 | 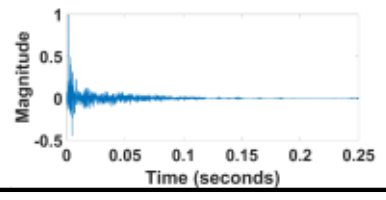 | 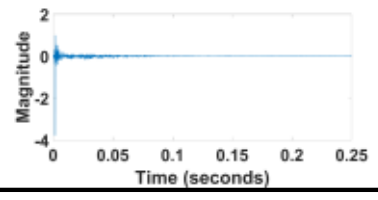 |
| | **Real-world environment 1** | | **Real-world environment 2** | |
| **3D Scene** | 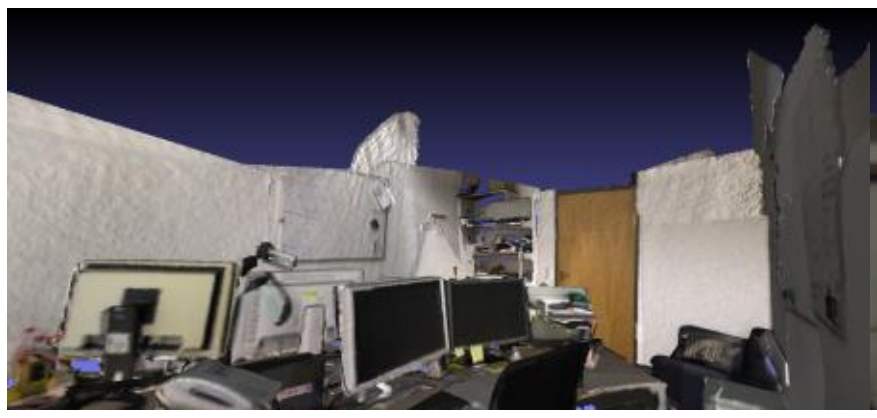 | | 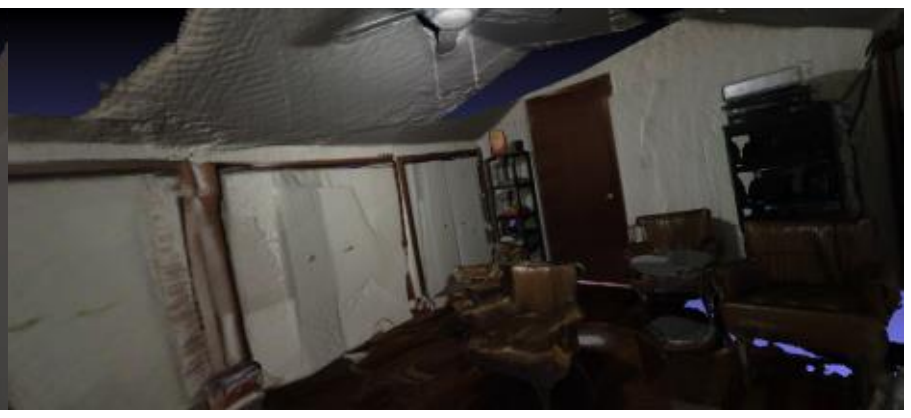 | |
| **Channel** | **Left** | **Right** | **Left** | **Right** |
| **MAE ($10^{-2}$)** | **0.44** | **0.40** | **0.34** | **0.37** |
| **Geometric-based BIR** | 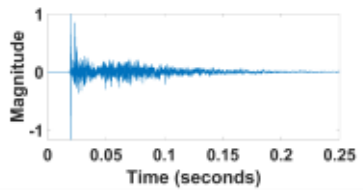 | 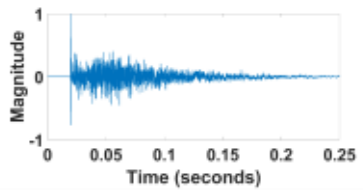 | 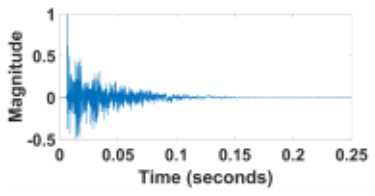 | 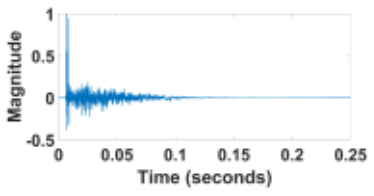 |
| **Our Listen2Scene** | 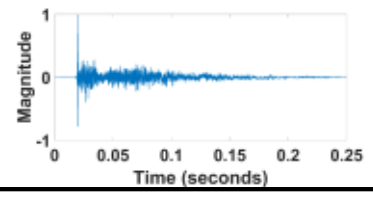 | 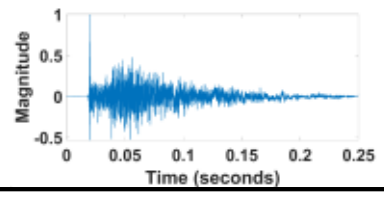 | 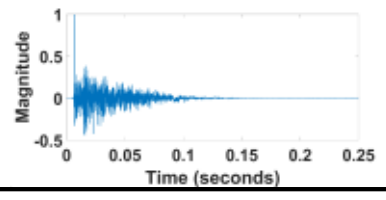 | 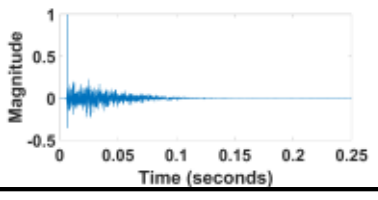 |

## 5.6 Acoustic Evaluation

### 5.6.1 BRAS Benchmark

We use the BRAS benchmark [206] to evaluate the contribution of material properties to the accuracy of the BIR generated using our Listen2Scene method. The BRAS contains a complete scene description, including the captured BIRs (i.e. ground truth) and the 3D models with semantic annotations for a wide range of scenes. We trained our approach without including the material properties (Listen2Scene-No-Mat) and including material properties (Listen2Scene). We evaluate our approach using recorded BIRs from the chamber music hall and auditorium (Figure 5.7). We generated BIRs corresponding to the source and listener positions in the same 3D models and compared the accuracy. We plot the normalized early reflection energy decay curves (EDC) of the captured BIRs and the BIRs generated using our models (Figure 5.7). The EDC describes the amount of energy remaining in the BIR with respect to time [176]. We observe that in 2 different scenarios, adding material improves the energy decay pattern of the BIRs. We calculated the mean absolute error (MAE) between the EDC of captured BIRs and generated BIRs. MAE decreases by 3.6% for the medium room and 6.6% for the large room.

Table 5.3: The responses from the acoustic experts and AMT participants on the plausibility of the sounds in each video created using 3D scenes in the ScanNet. We report the response from each age-category separately and the standard deviation (SD) of the combined results. We shifted our rating scale from -2 - 2 to 1 - 5 and calculated the SD (Section 5.7.3). We compare video auralized using our Listen2Scene approach with the videos auralized using clean speech, MESH2IR, Listen2Scene-No-Mat and geometric-method. We compare Listen2Scene-No-Mat using a single source in medium (M) and large (L) 3D scenes. We observe that 67% of total participants prefer Listen2Scene when we play video generated using Listen2Scene and MESH2IR with a single source. The highest comparative percentage is **bolded**.

| **Participants** | | **Acoustic Experts (13 participants) [%]** | | | **AMT (57 participants) [%]** | | | **Combined (70 participants) [%]** | | | |
|---|---|---|---|---|---|---|---|---|---|---|---|
| **Baseline Method** | **No of Sources** | **Baseline** | **No Preference** | **Listen2Scene** | **Baseline** | **No Preference** | **Listen2Scene** | **Baseline** | **No Preference** | **Listen2Scene** | **SD** |
| Clean | 1 | 15.38 | 0.00 | **84.62** | 29.82 | 1.76 | **68.42** | 27.14 | 1.43 | **71.43** | 1.44 |
| | 2 | 7.69 | 0.00 | **92.31** | 19.30 | 3.51 | **77.19** | 17.14 | 2.86 | **80** | 1.39 |
| Mesh2IR [3] | 1 | 23.08 | 0.00 | **76.92** | 31.58 | 3.51 | **64.91** | 30 | 2.86 | **67.14** | 1.51 |
| | 2 | 15.38 | 7.69 | **76.92** | 15.79 | 5.26 | **78.95** | 15.71 | 5.71 | **78.57** | 1.26 |
| Listen2Scene-No-Mat | 1 (M) | 30.77 | 23.08 | **46.15** | 29.82 | 19.30 | **50.88** | 30 | 20 | **50** | 1.29 |
| | 1 (L) | 7.69 | 30.77 | **61.54** | 26.31 | 7.02 | **66.66** | 22.85 | 11.43 | **65.71** | 1.15 |
| | 2 | 23.08 | 23.08 | **53.85** | 22.81 | 15.79 | **61.40** | 22.86 | 17.14 | **60** | 1.30 |
| Geometric-Method | 2 | 23.08 | 46.15 | **30.77** | 38.60 | 12.28 | **49.12** | 35.71 | 18.57 | **45.71** | 1.43 |
| **Age Category** | | **18 - 24 (16 participants) [%]** | | | **25 - 34 (47 participants) [%]** | | | **35 or older (7 participants) [%]** | | | |
| **Baseline Method** | **No of Sources** | **Baseline** | **No Preference** | **Listen2Scene** | **Baseline** | **No Preference** | **Listen2Scene** | **Baseline** | **No Preference** | **Listen2Scene** | |
| Clean | 1 | 31.25 | 0.00 | **68.75** | 19.15 | 2.13 | **78.72** | **71.43** | 0.00 | 25.57 | |
| | 2 | 12.5 | 0.00 | **87.5** | 14.89 | 2.13 | **82.98** | 42.86 | 14.28 | **42.86** | |
| Mesh2IR [3] | 1 | 18.75 | 6.25 | **75.00** | 34.04 | 0.00 | **65.96** | 28.57 | 14.28 | **57.14** | |
| | 2 | 12.5 | 6.25 | **81.25** | 14.89 | 4.26 | **80.85** | 28.57 | 14.28 | **57.14** | |
| Listen2Scene-No-Mat | 1 (M) | 18.75 | 0.00 | **81.25** | 36.17 | 25.53 | **38.30** | 14.28 | 28.57 | **57.14** | |
| | 1 (L) | 12.5 | 12.5 | **75.00** | 21.28 | 12.77 | **65.96** | **57.14** | 0.00 | 42.86 | |
| | 2 | 6.25 | 18.75 | **75.00** | 27.66 | 17.02 | **55.32** | 28.57 | 14.29 | **57.14** | |
| Geometric-Method | 1 (L) | 31.25 | 25.00 | **43.75** | 27.66 | 23.40 | **48.94** | 42.86 | 14.23 | **42.86** | |

Table 5.4: We calculate the mean absolute reverberation time ($T_{60}$) error, direct-to-reverberant ratio (DRR) error and early-decay-time (EDT) error for monaural IRs generated using MESH2IR and BIRs generated using our approach with materials (Listen2Scene) and without material (Listen2Scene-No-Mat), Listen2Scene-Full, Listen2Scene-No-BIR, and Listen2Scene-ED. We compare them with BIRs computed using the geometric method (Section 5.3.1). We compare the monaural IRs generated using MESH2IR with each channel in BIRs separately and compute the average. The best results of each metric are shown in **bold**.

| **IR Dataset** | **Mean Absolute Error ↓** | | |
|---|---|---|---|
| | **$T_{60}$ (s)** | **DRR (dB)** | **EDT (s)** |
| MESH2IR [3] | 0.16 | 5.06 | 0.25 |
| Listen2Scene-No-Mat | 0.10 | 3.15 | 0.14 |
| Listen2Scene-Full | 0.10 | 3.18 | 0.16 |
| Listen2Scene-Fix | 0.11 | 2.56 | 0.17 |
| Listen2Scene-No-BIR | **0.08** | 4.21 | 0.21 |
| Listen2Scene-ED | 0.10 | 3.49 | 0.16 |
| **Listen2Scene** | **0.08** | **1.7** | **0.13** |

### 5.6.2 Accuracy Analysis

We quantitatively evaluate the accuracy of our proposed approach using standard acoustic metrics such as reverberation time ($T_{60}$), direct-to-reverberant ratio (DRR), and early-decay-time (EDT). $T_{60}$ measures the time taken for the sound pressure to decay by 60 decibels (dB). The ratio of the sound pressure level of the direct sound to the sound arriving after surface reflections is DRR [128]. The six times the time taken for the sound pressure to decay by 10 dB corresponds to EDT. We generate 2000 high-quality BIRs using many rays with the geometric method [207] for 166 real scenes not used to train our networks in the ScanNet dataset. We compare the accuracy of Listen2Scene with the BIRs computed using the geometric method on these scenes.

In our Listen2Scene network, we pass the average sound absorption and reflection coefficients at 500 Hz and 1000 Hz as input. In our Listen2Scene-Full variant, the average coefficient over the 8-octave bands between 62.5 Hz and 8000 Hz is given as input. Also, in our Listen2Scene, we simplify the mesh to 2.5% of the original size. Our Listen2Scene-Fix variant simplifies all the meshes to have a constant number of faces (2000 faces). The motivation behind our approach is that we empirically observed that instead of having a fixed size if we simplify the meshes to 2.5% of the original size, the contextual information is preserved better. We calculate the mean absolute acoustic metrics error of the BIRs generated using our approach with materials (Listen2Scene) and without materials (Listen2Scene-No-Mat), Listen2Scene-ED, Listen2Scene-No-BIR, Listen2Scene-Fix and Listen2Scene-Full. We report the average error from two channels in our generated BIRs (Table 5.4). Many prior learning-based approaches are not capable of generating IRs for new scenes not used during training [202] or generating BIRs for standard inputs taken by physics-based BIR simulators [46]. MESH2IR [3] can generate monaural IRs from 3D mesh models. Therefore, we compare the acoustic metrics of MESH2IR separately with the left and right channels and report the average error. We highlight the accuracy improvements in Table 5.4. We can see that our Listen2Scene outperforms MESH2IR and other variants of the Listen2Scene network.

### 5.6.3 Time-domain comparison

We plot additional time-domain representation of BIRs generated using a geometric-based sound propagation approach [207] and our proposed Listen2Scene (Table 5.2) for two different 3D scenes. We can see that the amount of reverberation and the high-level structures of the BIRs generated using our approach match BIRs generated using the geometric-based method. Also, we can see that the ILD and ITD in our generated BIRs match the BIRs from a geometric method. The mean absolute error of the normalized BIRs generated using Listen2Scene is less than 0.5 x $10^{-2}$.

Table 5.5: The total participants' (acoustic experts and AMT participants) responses on which synthetic speech sample is closer to real-world speech created using captured IRs in the BRAS dataset. We created synthetic speech samples using Listen2Scene and Listen2Scene-No-Material for 2 different real-world environments. The highest comparative percentage is **bolded**.

| **Environment** | **Listen2Scene-No-Material** | **Listen2Scene** |
|---|---|---|
| Chamber music hall | 44.29% | **55.71%** |
| Auditorium | 21.43% | **78.57%** |

### 5.6.4 Run Time

We generated 2500 BIRs for a given 3D scene to calculate the run time. Our network comprises a graph neural network (GNN) and a BIR generator network. For a given 3D scene, we perform mesh encoding using GNN only once, and we generate BIRs by varying source and listener positions. On average, our network takes 0.21 seconds to encode the scene using GNN and 0.023 milliseconds to generate a BIR. Therefore, on average, our network takes 0.1 milliseconds per BIR to generate 2500 BIRs for a given 3D scene. On average, interactive image-based geometric sound propagation algorithm [167] takes around 0.15 seconds to generate an impulse response [2]. Therefore, our Listen2Scene is more than two orders of magnitude faster than image-based sound propagation methods [167].

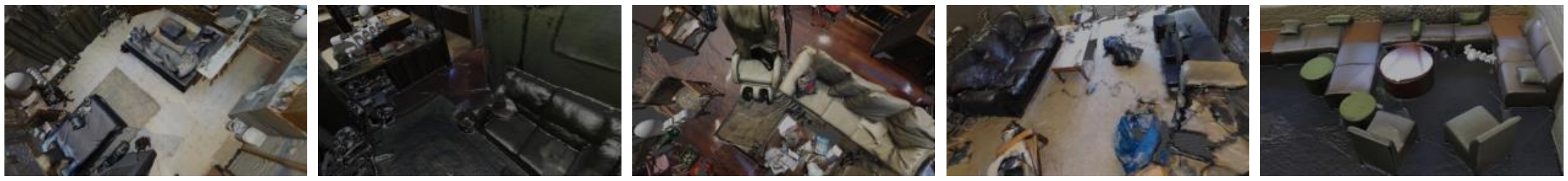

Figure 5.8: We show the 3D reconstructed real-world environments with various levels of complexity that are used to evaluate our learning-based real-time sound propagation method and our audio rendering quality. In practice, our Listen2Scene approach is two orders of magnitude faster than the interactive geometric sound propagation algorithm. In our supplementary demo video, we show that the overall sound quality of Listen2Scene is very similar to the interactive geometric sound propagation algorithm.

## 5.7 Perceptual Evaluation

We perceptually evaluate the audio rendered using Listen2Scene and compare them with prior learning-based and geometric-based sound propagation algorithms. Our study aims to verify whether the audio rendered using our Listen2Scene is plausible (with left and right channels). We auralized three scenes with a single sound source and two scenes with two sound sources from the ScanNet test dataset (more details in the video). Figure 5.8 shows the snapshot of 5 scenes used to evaluate the quality of our proposed audio rendering method. We created a 40-second video of each scene by moving the listener around the scene. Figure 5.9 shows the listener path in a 3D scene with two sound sources. We evaluate our approach by adding sounds synthesized using different methods to the 3D scene walkthrough: clean or dry sound (Clean), sound propagation effects created using MESH2IR, Listen2Scene-No-Material, geometric-based method and Listen2Scene. We also compared the reverberant speech created using Listen2Scene-No-Material and Listen2Scene with the captured / real-world IRs from two different scenes in the BRAS dataset (Figure 5.7).

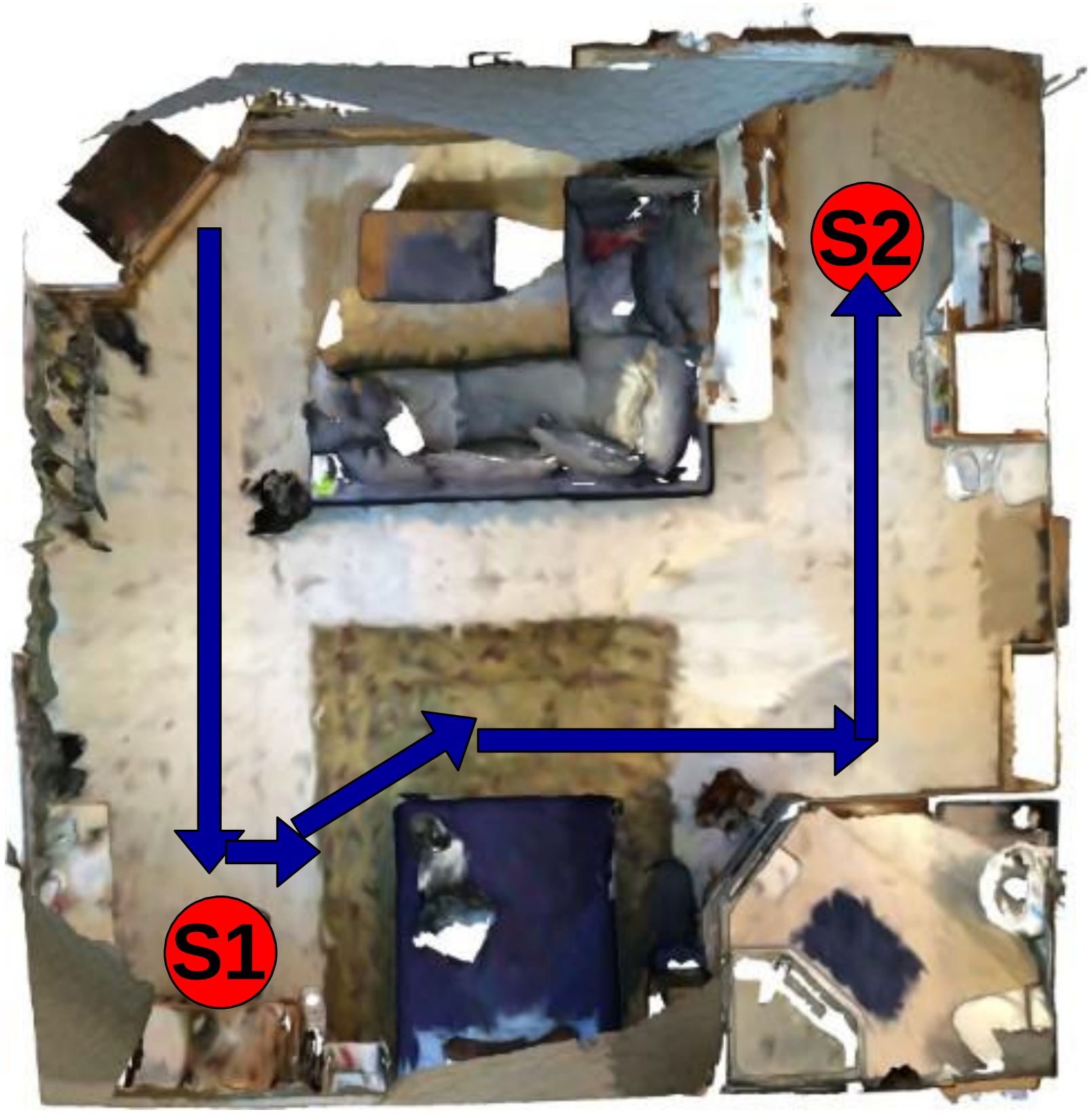

Figure 5.9: The path covered by the listener in a real-world 3D scene (studio apartment) with two sound sources. The listener path is shown in blue arrows. The red circle represent the two sound sources in the 3D scene. The source S1 is a speech signal from a speaker, and the source S2 is water pouring from the kitchen.

### 5.7.1 Participants

We conducted our user study among the acoustic experts (13 participants) and the participants from Amazon Mechanical Turk (AMT) (57 participants), an online crowdsourcing platform that can be used to collect data from diverse participants. Since we have a limited number of acoustic experts to evaluate our approach, we also evaluated using AMT. We conducted our user study on 70 participants (47 males and 23 females), of which 16 participants were between 18 and 24 years of age, 48 participants were between 25 and 34 years of age, and 7 participants were above 35 years of age. We ensured the quality of our evaluation by pre-screening the participants. As our pre-screening questions, we asked the online participants whether they use headphones with a laptop/desktop and only allowed them to proceed with the survey if they answered yes. The average completion time of our user study is 20 minutes for each user. The just-noticeable-difference (JND) relative reverberation time change is 5% - 25% [212]. On average on each rendered 40-second video, the reverberation time changes by 30%. Therefore, under normal conditions, we expect the listeners to identify the relative changes in the audio correctly.

### 5.7.2 Benchmarks

We performed the following five benchmark comparisons in perceptual evaluation. Our first four benchmarks compare 40-second-long audio-rendered 3D environment walkthrough videos from our Listen2Scene with baseline methods. In our last benchmark, we compare the real speech with speech rendered using our Listen2Scene and Listen2Scene-No-Material.

**Clean vs. Listen2Scene:** We compared audio-rendered 3D scenes with and without acoustic effects from Listen2Scene. We created two different 3D scene walkthrough videos for our experiment with a single sound source and two sound sources. For a single sound source, we evaluate whether our approach creates continuous and smooth acoustic effects when moving around the scene and whether the user can perceive the indirect acoustic effects. In the two sound sources walkthrough video, we evaluate whether the relative distance between the two sound sources in the rendered audio using our Listen2Scene matches the video.

**MESH2IR vs. Listen2Scene:** We auralized a 3D walkthrough video each for a single source and two sources. We use the prior monaural audio rendering method MESH2IR and our proposed binaural audio rendering approach, Listen2Scene, for our comparison. We aim to investigate whether the participants feel that the acoustic effects in the left and right ears change smoothly and synchronously as the user walks into the real-world 3D scene. In addition to distance, we investigate whether our acoustic effects change smoothly with the direction of the source. We also evaluated whether our approach is plausible even when there is more than one source in the 3D scene.

**Listen2Scene-No-Material vs. Listen2Scene:** We auralized two real-world 3D scenes with a single source from a medium-sized and a large 3D scene, and another 3D scene with two sources. In this experiment, we evaluate whether the reverberation effects from Listen2Scene match closely with the environment when compared with Listen2Scene-No-Material. Our goal is to evaluate the perceptual benefits of adding material characteristics to our learning method. The amount of reverberation varies with the size of the 3D scene, therefore we compare the contribution of material to the plausibility of auralized medium and large 3D scenes. In real environments, the listener hears audio from multiple sound sources. Therefore, we evaluate the plausibility of our approach when more than one source is played in the 3D walkthrough video.

**Geometric-method vs. Listen2Scene:** We auralized one real-world 3D scene with two sources. In this experiment, we evaluate whether the participants feel the Listen2Scene or the geometric-based sound propagation [207] is more plausible for the corresponding 3D scene walkthrough video.

**BRAS benchmark:** We played reverberant speech created using captured left channel

IRs from the BRAS and left channel impulse responses synthesized using our Listen2Scene and Listen2Scene-No-Material in two different 3D scenes (Figure 5.7). We use single-channel IRs to remove acoustic effects from ITD and ILD and make the participants focus on reverberation effects corresponding to the complexity and shape of the environment. We asked the participants to choose which speech sampled auralized using our BIRs is closer to the real speech from the BRAS.

### 5.7.3 Experiment and Results

In our experiment, we randomly choose the location of two videos (left or right) used for the comparison to eliminate bias from collected data and ask the participants to rate from -2 to +2 based on which video sounds more plausible, i.e. the way the sound varies in both ears when the listener moves towards and away from the sound source. The participants rate -2 if the left video sounds more plausible and vice versa. If the participants have no preference, they rate 0. We group the negative scores (-1 and -2) and positive scores (1 and 2) to choose the participants' preferences.

Table 5.3 summarises all the participants' responses. We observe that 67% - 79% of the total participants find that the auralized scenes with 1-2 sources using Listen2Scene are more plausible than MESH2IR. Interestingly, 17% - 27% of total participants find that just adding clean sound to a 3D scene video is more plausible. When we further break down our results based on age, we observe that 42.86% to 71.43% of 35 or older participants prefer adding just clean sound to the video. We believe that this might be caused by an increase in volume from our approach when the listener moves too close to the speaker. All of our participants older than 35 are from AMT, therefore we were not able to get feedback from the participants after the studies. We also observed that when there is more than one source in the 3D scene, the relative sound variation of the sources based on their location is more plausible with Listen2Scene, as compared to using dry sound or MESH2IR. In large 3D models, where the $T_{60}$ tends to be higher, 66% of participants feel Listen2Scene is more plausible than Listen2Scene-No-Material. We also can see that 10% more participants feel our learning-based approach is more plausible than the geometric-based method. The BIRs generated using the learning-based method smoothly change with the distance and listeners can feel a smooth transition in audio when they move to different positions in the 3D scene. From Table 5.5, we can see that audio/speech rendered using our Listens2Scene approach is closer to the real-world speech. Overall, we notice that our approach creates plausible acoustic effects when there are one or more sound sources in the 3D scene.

## 5.8 Conclusion Limitations and Future Work

We present a material-aware learning-based sound propagation approach to render thousands of audio samples on the fly for a given real 3D scene. We propose a novel approach to handle material properties in our network. Moreover, we show that adding material information significantly improves the accuracy of BIR generation using our Listen2Scene approach and is comparable to geometric propagation methods or captured BIRs in terms of acoustic characteristics and perceptual evaluation. Overall, our algorithm offers two orders of magnitude performance improvement over interactive geometric sound propagation methods.

Our approach has some limitations. The performance of our network depends on the training data. We can train our network with real captured BIRs, though it is challenging and expensive to capture a large number of such BIRs. Currently, we use BIRs generated using geometric algorithms for medium-sized 3D scenes in the ScanNet dataset for training, and the overall accuracy of Listen2Scene is also a function of the accuracy of the training data.

Our approach is limited to static real scenes.Our material classification methods assume that accurate semantic labels for each object in the scene are known.It is possible to consider sub-band acoustic material coefficients to further improve the accuracy. However, the complexity of the graph representation of the 3D scene drastically increases, and we are limited by the GPU memory in handling such complex graphs. Due to the limitation of the training dataset used for training, the performance of our network has been currently evaluated on small and medium-sized scenes. In future work, we like to train and evaluate our approach on very large scenes. Since the ScanNet dataset does not have the same 3D environment with different structural changes, we are not able to train and evaluate different structural detail resolutions. As part of future work, it would be useful to analyze our learning-based sound propagation approach on different structural detail resolutions.

# Chapter 6: Room Impulse Response Augmentation using Real-World Room Impulse Response

## 6.1 Motivation

Physically-based acoustic simulators have been used over the decades to generate synthetic RIRs for far-field speech research. Wave-based methods and geometric methods are widely used to model RIRs for different acoustic environments. The wave-based approach [15] solves the wave equation using numerical methods. Although wave-based methods accurately compute the RIRs, these methods are computationally expensive and only feasible for low frequencies and less complex scenes. The image source method [16] and path tracing methods [1, 213] are common geometric acoustic simulation-based methods. The image source method only models specular reflections in simple rectangular rooms while path tracing-based geometric acoustic simulators model occlusion and specular and diffuse reflections. Geometric acoustic simulators treat sound waves in the form of a ray [44]. Although this assumption holds for high-frequency waves, ray assumption causes visible irregularities at low frequencies [214]. In recent works, TS-RIRGAN [47] is used to transfer low-frequency wave effects learned from real RIRs to synthetic RIRs generated using geometric acoustic simulators. In many scenarios, the target room environment (i.e., the exact geometric shape and material parameters) is unknown or too complex for the geometric acoustic simulators. As a result, their ability to generate RIRs for all kinds of scenarios can be limited.

To overcome the limitations of synthetic RIRs, real RIRs are recorded in a controlled environment. The maximum length sequence method [69], the time-stretched pulses method [11], and the exponential sine sweep method [12] are common methods to measure real RIRs. Among these approaches, the exponential sine sweep method is robust to changing loudspeaker output volume and performs well in automatic speech recognition tasks. The real RIRs in BUT ReverbDB [30] are collected using the exponential sine sweep method. Since collecting real RIRs is time-consuming and technically difficult, only a limited number of real RIRs is available to augment far-field speech.

GANs have made steady progress over the years in image generation [215], image inpainting [216], and domain adaptation [217]. The success of GAN in computer vision motivated researchers to use it in other fields. Recently, GANs have been becoming popular in the audio generation. GANs have shown progress from music generation [218] to any short audio clip generation [152, 219]. In this work, we aim to augment high-quality RIRs using existing real RIRs. We use GANs for RIR generation to complement the prior works.

## 6.2 Main Contributions

We present a novel GAN-based RIR generator (IR-GAN) that is trained on real-world RIRs. IR-GAN can parametrically control different acoustic parameters (e.g., reverberation time, direct-to-reverberant ratio, etc.) learned from real RIRs and generate synthetic RIRs that can imitate new or different acoustic environments. Moreover, we propose a constrained RIR generation approach that can avoid synthesizing RIRs with noisy artifacts to a greater extent.

## 6.3 Our Proposed Approach: IR-GAN

### 6.3.1 Room Impulse Response Statistics

Room impulse response acoustic parameters are used to characterize the acoustic environment [220] and control RIR generation using GAN. Reverberation time ($T_{60}$), direct-to-reverberant ratio (DRR), early-decay-time (EDT), and early-to-late index (CTE) are four acoustic parameters that can be estimated from RIRs. We use these acoustic parameters to constrain IR-GAN-based RIR augmentation. Reverberation time measures the amount of time taken to decay the sound pressure by 60 decibels (dB). The $T_{60}$ value depends on room size and the characteristics of the material (e.g., floor, walls, furniture, etc.). DRR is calculated by dividing the sound pressure level of a direct sound source by the sound pressure level of the sound arriving after one or more surface reflections [128]. DRR is measured in dB. Time taken for sound pressure to decay by 10 dB is multiplied by a factor of 6 to get early-decay-time. EDT depends on the type and location of the sound source. CTE measures the proportion of the total sound energy received in the first 50$ms$ to the energy received during the rest of the period [221].

### 6.3.2 Room Impulse Response Representation

The representation of the input data and the data generated by the neural network is important for synthesizing high-quality RIRs. Therefore, lossy representations of RIRs like Mel-frequency cepstral coefficients (MFCCs) are less favorable. Audio samples are a lossless representation that can be easily converted to an audio signal. As different datasets store RIRs with different sampling rates, we re-sample all the RIRs to 16 kHz. We pass audio samples as a 32-bit floating-point vector of length 16384 to the GAN. The vector length is sufficient to represent RIRs because most of the real RIRs are less than one second in duration. We can represent slightly more than one second with 16384 samples with a sampling rate of 16 kHz.

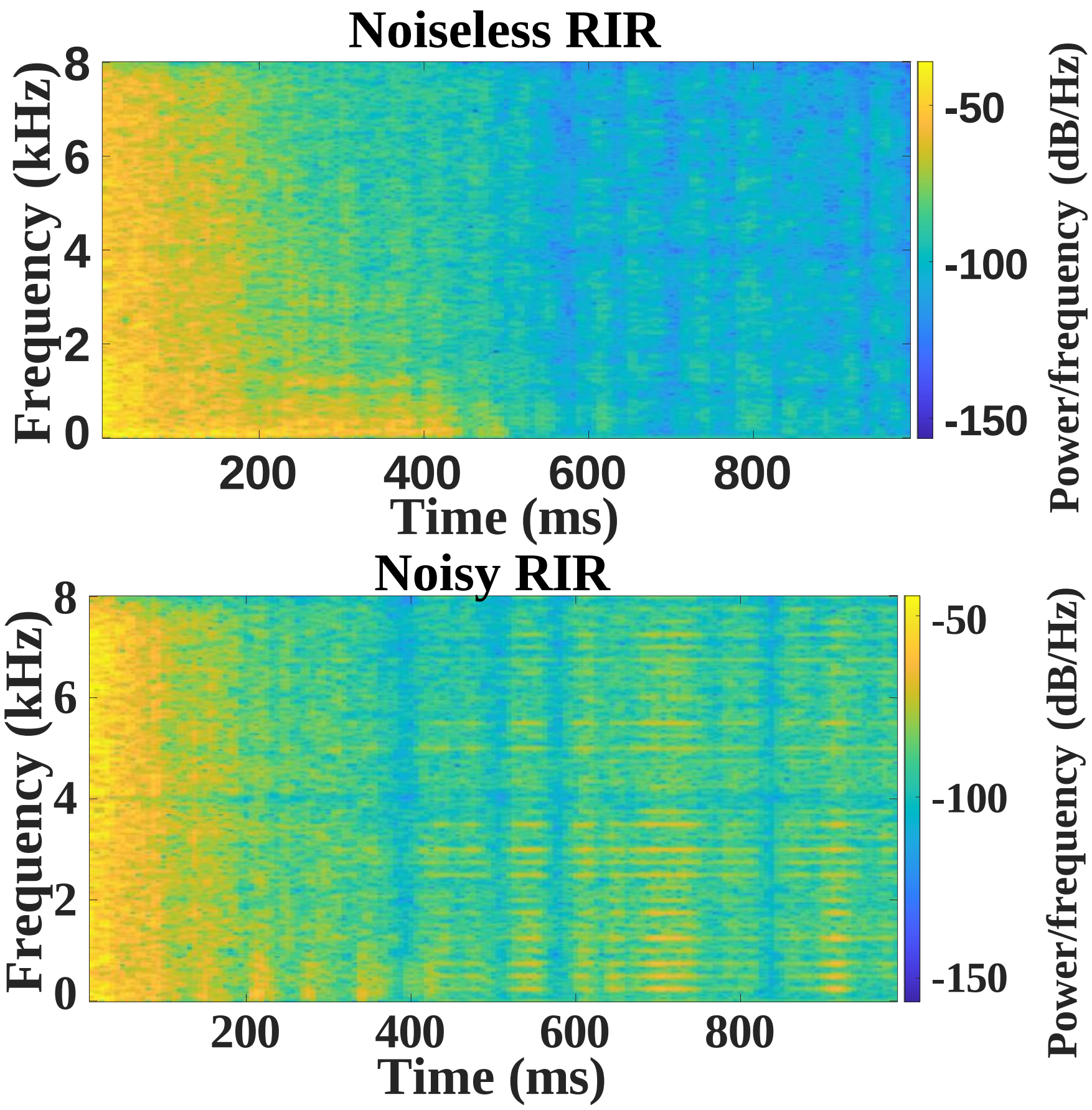

Figure 6.1: Spectrogram of noiseless RIR and noisy RIR. The noiseless RIR has a $T_{60}$ value of around 1, and the noisy RIR has a $T_{60}$ value of around 3. In the noisy spectrogram, we can see many horizontal artifacts around 700ms.

### 6.3.3 GAN

GAN is a generative model that learns a mapping from a low-dimensional vector space to a high-dimensional space where the data is represented. We adapt the WaveGAN architecture proposed in [152] to generate high-quality RIRs. WaveGAN is a one-dimensional version of DCGAN [222] where two-dimensional filters are replaced by one-dimensional filters.

GANs trained using the value function proposed in the original GAN paper [148] are often unstable, and mode collapse can occur when the generator architecture is varied. Therefore, we use a stable cost function introduced in WGAN [223]. In this cost function, we minimize the Wasserstein-1 distance between data distribution $p_{data}$(x) and model distribution (Equation 6.1). Model distribution is implicit in the second part of the equation because $G$(z) represents the mapping from a latent vector $z$ with distribution $p_z$(z) to the data space. In WGAN, the discriminator network $D_{WGAN}$ gives a score based on the realness of the given image instead of predicting the probability that $x$ comes from the real distribution. In Equation 6.1, $E$ represents expectation.

$$V_{WGAN}(D_{WGAN}, G) = E_{xSectionimp_{data}(x)}[\log D_{WGAN}(x)] - E_{zSectionimp_{z}(z)}[\log D_{WGAN}(G(z))]. \quad (6.1)$$

### 6.3.4 Constrained RIR Generation

In our approach, we train a GAN to learn the mapping from the 100-dimensional latent vector $z$ drawn from a Gaussian distribution to the RIR in data space. As the number of real-world RIR datasets is limited, we propose a constrained generation of RIRs from the generator network.

There is an infinite possibility to generate a 100-dimensional vector where each dimension can take any floating-point number between -1 and 1. Since we train GAN with a limited number of RIRs in real RIR datasets (BUT ReverbDB [30] contains less than 2000 RIRS.), there is a chance that some of the latent vectors map to noisy RIRs. For example, GAN may generate RIRs with unrealistically large $T_{60}$ values. In Figure 6.1, we can see a noisy RIR generated without any constraint. The generated noisy RIR with a $T_{60}$ value of around 3 has many horizontal artifacts around 700ms.

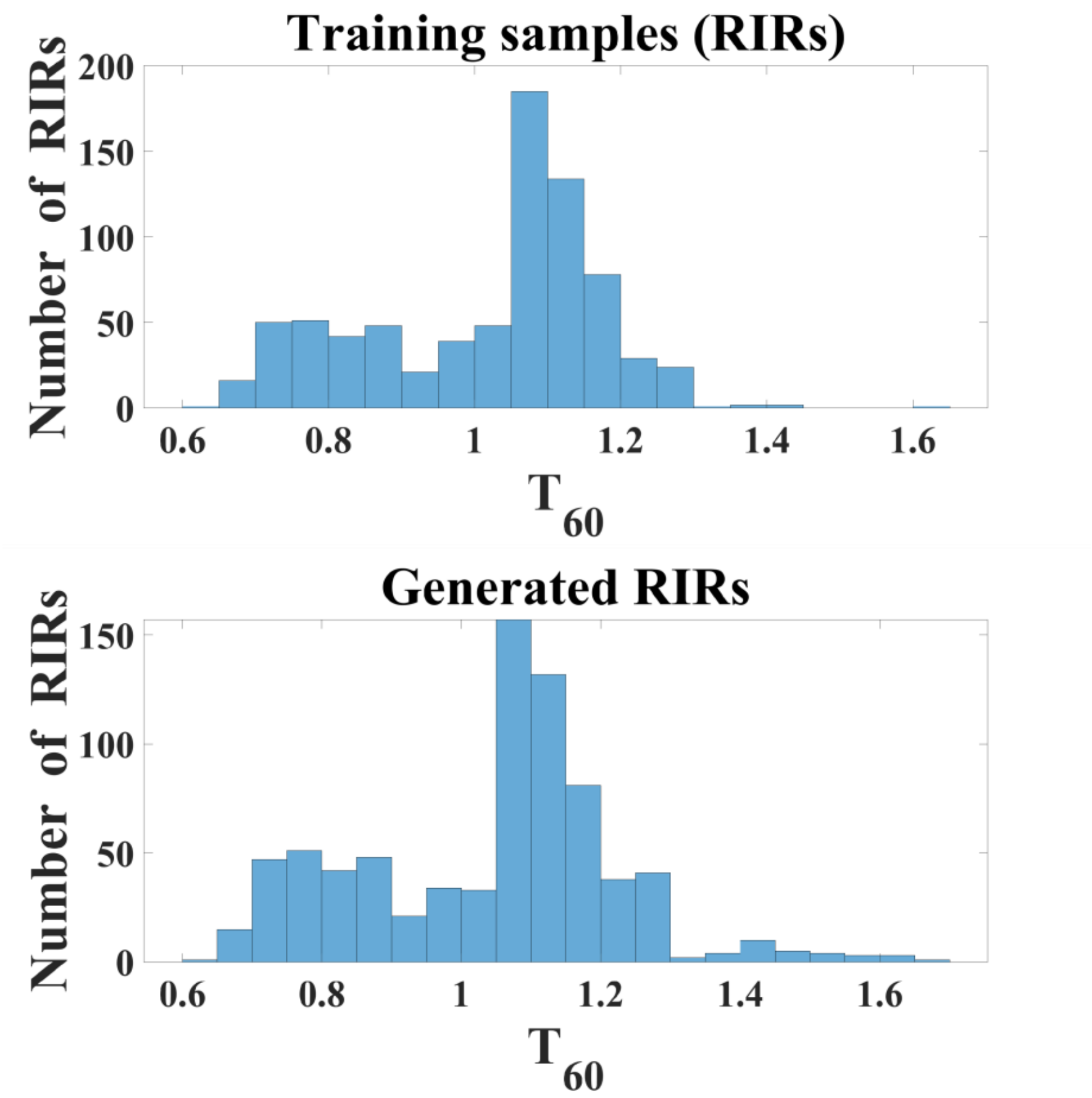

Figure 6.2: $T_{60}$ distribution of training samples and $T_{60}$ distribution of RIRs generated using our IR-GAN with the constraint.

To prevent such mappings, we calculate the key acoustic parameters of the training samples ($T_{60}$, DRR, CTE, and EDT) and use them to generate histograms. Later, we generate RIRs by constraining them to fit the distribution of key acoustic parameters as the training samples. In this way, we can avoid noisy mapping to a greater extent. Since it is difficult to match the exact distributions of the training samples, we relax GAN to generate samples closer to the distribution with low probability. Figure 6.2 depicts the $T_{60}$ distribution of the training samples and the $T_{60}$

Table 6.1: Different RIRs used in our experiment.

| RIR | Description |
|---|---|
| BUT | Real-world RIRS from the BUT ReverbDB dataset [30]. |
| AIR | Real-world RIRS from the AIR [31] dataset. |
| GAS | Simulated RIRs using the acoustic simulator [1]. |
| GAN.C | RIRs generated using our IR-GAN with constraint (Section 6.3.4). |
| GAN.U | RIRs generated using our IR-GAN without any constraint. |

distribution of RIRs generated with the constraint.

## 6.4 Experiments and Results

We evaluate the effectiveness of our proposed approach by conducting far-field automatic speech recognition (ASR) experiments using the modified Kaldi LibriSpeech ASR recipe[1]. We use augmented far-field speech to train and test the Kaldi LibriSpeech ASR model and evaluate the benefits in the following manner. First, we compare the performance of our proposed IR-GAN with the state-of-the-art synthetic RIR generator [1]. Second, we evaluate the robustness of our IR-GAN when we train the GAN on one dataset [30] and test the GAN on another dataset [31] from different acoustic environments. We use word error rate to evaluate the performance.

### 6.4.1 Data Preparation

As proposed in [214], we generate far-field speech from clean LibriSpeech by convolving it with RIRs and adding environmental noise. Since we mainly focus on the quality of the synthesized RIRs, we use the same environmental noise from the BUT ReverbDB [30] for training and test set generation.

We use real RIRs from the BUT ReverbDB dataset [30] and the AIR [31] dataset to conduct our experiments. BUT ReverbDB consists of 1891 RIRs and 9114 environmental noises covering nine different rooms. To make fair comparisons, we use the 1209 BUT ReverbDB RIRs picked in [214]. The AIR dataset consists of 344 real RIRs from 6 different rooms. From these, we select 68 RIRs from the 4 rooms (studio booth, office room, lecture room, and meeting room) mentioned in [31]. We split 1209 real-world RIRs from the BUT ReverbDB dataset into subsets of {773,194,242} to generate training, development, and test far-field speech datasets. To test the robustness of our proposed approach, we use 68 RIRs from the AIR dataset.

[1]https://github.com/RoyJames/kaldi.

Table 6.2: Detailed information about the augmented dataset generated using different RIRs (Table 6.1). GAN.C+GAS indicates an equal mixture of GAS and GAN.C synthesized RIRs. 2*GAN.C contains twice the number of RIRs when compared to GAN.C.

| Dataset | RIR | Hours | #RIRs | LibriSpeech Dataset |
|---|---|---|---|---|
| **Test Dataset** | BUT | 5.4 | 242 | test-clean |
| | AIR | 5.4 | 68 | test-clean |
| **Training Dataset** | BUT | 460 | 773 | train-clean-{100,360} |
| | GAS | 460 | 773 | train-clean-{100,360} |
| | GAN.C | 460 | 773 | train-clean-{100,360} |
| | GAN.U | 460 | 773 | train-clean-{100,360} |
| | GAN.C+GAS | 460 | 1546 | train-clean-{100,360} |
| | 2*GAN.C | 460 | 1546 | train-clean-{100,360} |

Table 6.1 describes different RIRs used to create far-field speech datasets in our experiment. Table 6.2 shows the detailed composition of the augmented datasets. The augmentation process does not change the overall duration of the original LibriSpeech dataset.

### 6.4.2 Synthetic RIR generation

For a fair comparison, we use the same synthetic RIRs generated using the state-of-the-art geometric acoustic simulator [1] as the previous benchmark. The geometric acoustic simulator synthesizes RIR using the meta-info provided in BUT ReverbDB. The meta-info includes room dimensions and loudspeaker and microphone locations. Therefore, the synthesized RIRs for the training and development sets mimic real RIR training and development sets to some extent.

We train our IR-GAN with 967 real RIRs from the BUT ReverbDB dataset. They are composed of real-world RIRs allocated for training and development. We generate synthetic RIRs with and without the constrained RIR generation process in Section 6.3.4.

### 6.4.3 ASR Experiment

We use the modified Kaldi LibriSpeech ASR recipe to conduct our ASR experiments. We train time-delay neural networks [224] for each of our augmented far-field training sets. We extract the identity vectors [225] (i-vectors) of the real-world far-field test set and decode them using large four-gram (fglarge), large tri-gram (tglarge), medium tri-gram (tgmed), and small tri-gram (tgsmall) phone language models. We also do online decoding on the tgsmall phone language model. In online decoding, extracted features are passed in real-time instead of waiting until the entire audio is captured. Word error rate (WER) of each of the language model is used to evaluate the synthesized RIRs.

All the training and testing is done on 32 Intel(R) Xeon(R) Silver 4208 CPUs @ 2.10 GHz and 2 GeForce RTX 2080 Ti GPUs. For a fair comparison, we generate all the results from the same environment. It takes around four days to prepare the dataset and conduct each experiment on the Kaldi toolkit.

Table 6.3: Far-field automatic speech recognition results obtained from the far-field LibriSpeech test set. In this table, *BUT and *AIR represent far-field test sets generated using real RIRs from the BUT ReverbDB and AIR datasets, respectively. clean* represents clean speech. WER is reported for the tri-gram phone (tglarge, tgmed, tgsmall) and four-gram phone (fglarge) language models, and online decoding using tgsmall. Best results in each comparison are marked in **bold**.

| Experiment Setup (training set)(test set) | Test Word Error Rate (WER) [%] | | | | |
|---|---|---|---|---|---|
| | fglarge | tglarge | tgmed | tgsmall | online |
| cleanBUT (Baseline) | 77.15 | 77.37 | 78.00 | 78.94 | 79.00 |
| BUT BUT (Oracle) | 12.40 | 13.19 | 15.62 | 16.92 | 16.88 |
| GASBUT [1] | 16.53 | 17.26 | 20.24 | 21.91 | 21.83 |
| GAN.UBUT | 19.71 | 20.74 | 24.27 | 25.93 | 25.90 |
| GAN.CBUT | **14.99** | **15.93** | **18.81** | **20.28** | **20.24** |
| 2*GAN.CBUT | 14.86 | 15.69 | 18.50 | 20.25 | 20.17 |
| GAN.C+GASBUT | **14.16** | **14.99** | **17.56** | **19.21** | **19.21** |
| cleanAIR | 26.79 | 27.40 | 29.64 | 30.88 | 31.15 |
| GAN.CAIR | **7.71** | **8.03** | **9.88** | **11.11** | **11.08** |

### 6.4.4 Results

Table 6.3 presents the ASR test WER for far-field speech generated using the BUT ReverbDB [30] and AIR [31] datasets. WER is calculated for four different phone language models (fglarge, tglarge, tgmed, and tgsmall) in Kaldi as well as for online decoding using a tgsmall phone language model.

We use WER to measure the robustness of the trained model. A lower WER indicates that the trained model shows superior accuracy in test conditions. Robustness depends on the model architecture and the input data used to train the model. In our experiments, we keep the model constant while we train with different datasets. Different training datasets are created by convolving the same LibriSpeech speech corpus with different RIRs. Therefore, the robustness of the model is only affected by the RIRs being used to generate the training datasets. As expected, a significantly high WER is reported when we train our baseline model on clean speech and test on the real RIRs. The lowest test WER is reported when we train and test on the real RIRs.

In Table 6.3, we can see that the proposed IR-GAN (GAN.C) gives a lower WER than the state-of-the-art geometric acoustic simulator (GAS). Lower WER indicates that the RIRs synthesized using IR-GAN are more realistically synthesized than the RIRs computed using the physically-based acoustic simulator. When we look at the WER for the fglarge model, we can see that our proposed IR-GAN gives an 8.95% lower error rate than the GAS. We can see that the RIRs generated using the unconstrained IR-GAN (GAN.U) performs poorly in our far-field speech recognition experiment. Therefore our constrained RIR generation approach is important to eliminate noisy RIR generation.

**Hybrid Combination:** Because the IR-GAN and GAS try to mimic real RIRs using two different approaches, we evaluate the WER when trained using a combination of synthetic RIRs generated using the IR-GAN and GAS. We observe that there is a further drop of up to 5% in WER compared to doubling the synthetic RIRs from the IR-GAN. The drop in WER indicates that we can boost the robustness of ASR systems by combining RIRs generated from our IR-GAN and GAS.

In practical scenarios, we do not have real RIRs from the acoustic environment where we need the capabilities for far-field ASR. Therefore, we augment RIRs using our IR-GAN trained with real RIRs from BUT ReverbDB [30]. Then we train the Kaldi LibriSpeech ASR model on far-field speech generated using the augmented RIRs and test this ASR model on the far-field speech augmented using the AIR dataset [31]. We can observe around a 19% absolute reduction in error when compared to training an ASR model with clean speech.

## 6.5 Vector arithmetic on RIR

The existing real RIR datasets are small and imbalanced. We can use our IR-GAN to overcome this issue by parametrically controlling RIR augmentation and generating samples covering a wide range.

Because the low-dimensional latent vectors $z$ encode all the acoustic parameters, we can parametrically control all the acoustic parameters by performing simple vector arithmetic on the latent vectors.

To illustrate, we parametrically control DRR by simple vector arithmetic. We randomly selected vector $z_1$ which maps to an RIR with low DRR of -16 dB and vector $z_2$ which maps to an RIR with high DRR of 0.56 dB. We then perform simple vector arithmetic to find nine intermediate vectors $z_i'$ in the linear path between $z_1$ and $z_2$ using the following formula:

$$z'_n = \frac{(10-n)}{10} z_1 + \frac{n}{10} z_2. \tag{6.2}$$

The intermediate vectors generate RIRs with DRR between -16 dB to 0.56 dB. Figure 6.3 shows RIRs generated from $z_1$ and $z_2$, and intermediate vectors $z'_3$ and $z'_6$. When the DRR increases, the direct signal becomes stronger than the reverberation. We can see in Figure 6.3 that the direct response becomes stronger than the reverberation when we move from $z_1$ to $z_2$. The calculated DRR using a method based on ISO 3382-1:2009 for RIRs generated from $z'_3$ and $z'_6$ are -12 dB and -6 dB respectively. We consider DRR because it is easier to visualize. Similar to DRR, we can parametrically control different acoustic parameters using the latent vectors, and we can generate desired RIRs.

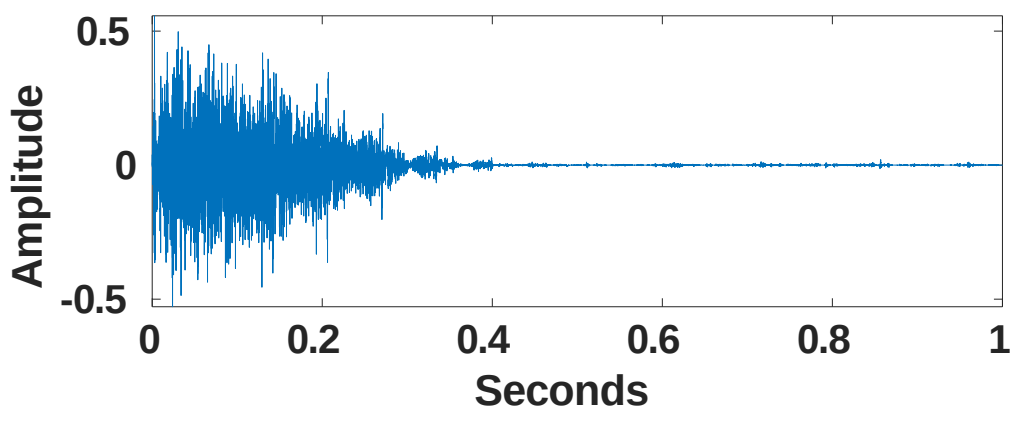

(a) RIR generated from $z_1$.

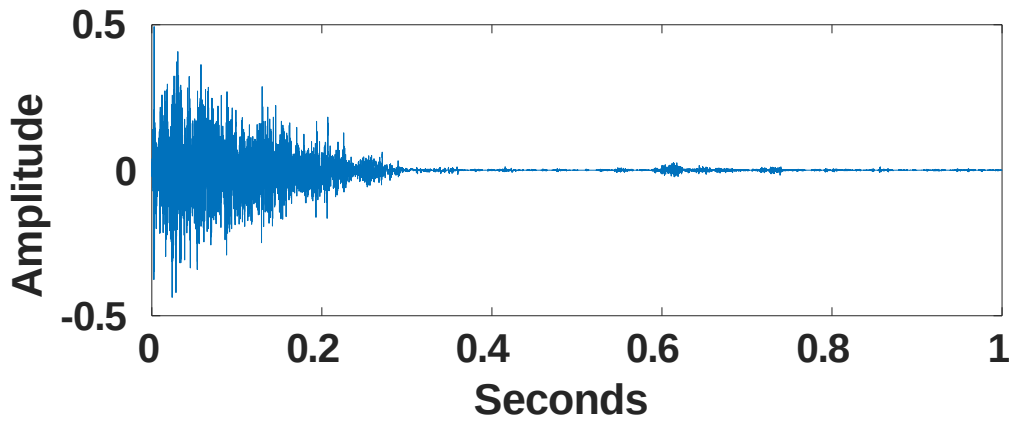

(b) RIR generated from $z'_3$.

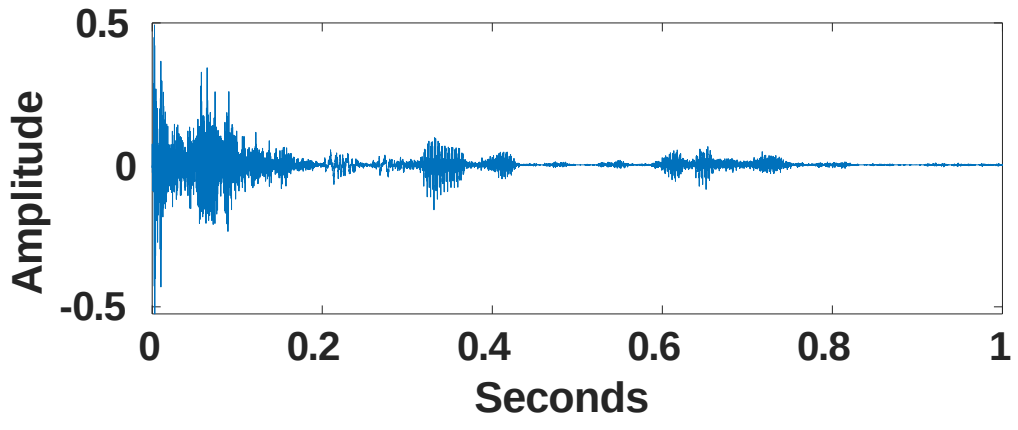

(c) RIR generated from $z'_6$.

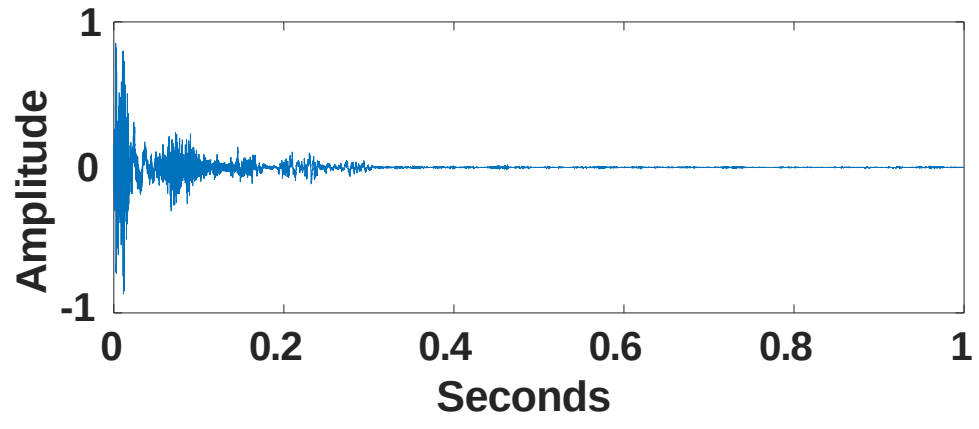

(d) RIR generated from $z_2$.

Figure 6.3: RIRs generated from randomly selected latent vectors $z_1$ and $z_2$, and two intermediate latent vectors $z'_3$ and $z'_6$ along the linear path through $z_1$ and $z_2$.

## 6.6 Discussion and Future Work

In this paper, we present an IR-GAN to generate realistic RIRs. Our proposed approach outperforms the state-of-the-art geometric acoustic simulator (GAS) by up to 8.95% in far-field ASR tests. When we combine our RIRs with RIRs generated using GAS, we can see a total reduction in word error rate by up to 14.3% in far-field ASR tests. This reduction in word error rate indicates that synthetic data generated using IR-GAN and GAS can be combined to boost the performance of far-field ASR systems. We have tested our approach only on indoor scenes, and extending it to outdoor scenes is a good topic for future work.

# Chapter 7: Improving the Quality of Synthetic Room Impulse Responses

## 7.1 Motivation

The RIR can be captured from an acoustic environment using different techniques [4, 11–13]. Recording real RIRs require a lot of human labor and special hardware. As a result, many far-field automatic speech recognition systems use synthetic RIRs for training [1, 17–19]. Synthetic RIRs can be generated using physically-based acoustic simulators for different scenes [14–16]. The current acoustic simulators have resulted in considerable improvement in far-field speech recognition [1]. However, there is still a gap between the performance of RIRs generated using acoustic simulators and the performance of real RIRs. Most commonly used acoustic simulators are unable to model all the acoustic effects in the environment, which can be captured by real RIRs. For example, ray-tracing-based acoustic simulators [1] can only simulate high-frequency acoustic effects but are not accurate in terms of low-frequency effects like diffraction or interference.

To overcome the problem, we present our TS-RIR-GAN architecture that can translate an imprecise synthetic RIR to a real RIR. We also propose a scheme to further improve the wave effects of synthetic RIRs by performing sub-band room equalization.

## 7.2 Main Contributions

We present a novel approach to improve the accuracy of synthetic RIRs. We design a TS-RIRGAN architecture to translate the synthetic RIR to a real RIR. TS-RIRGAN takes synthetic RIRs as 1x16384 audio samples to translate them into real RIRs and use multiple loss functions. We also perform real-world sub-band room equalization to the translated RIRs to further improve their quality. We also demonstrate the benefits of our post-processed RIRs in far-field ASR systems. Our main contributions include:-

- We present our TS-RIRGAN architecture, which is used to translate an imprecise synthetic RIR to a real RIR.
- We propose a scheme to further improve the wave effects of synthetic RIRs by performing sub-band room equalization.
- We show that on a modified Kaldi LibriSpeech far-field ASR benchmark [24], far-field speech augmented using our improved RIRs outperforms the far-field speech augmented using unmodified RIRs by up to 19.9%.

## 7.3 Our Approach

We translate synthetic RIRs to real RIRs and perform sub-band room equalization to improve the quality of synthetic RIRs. We compute the spectrogram for post-processed and real RIRs. We evaluate the quality of post-processed synthetic RIRs by comparing their mean value for a set of acoustic parameters (Table 7.2) and their energy distribution (Figure 7.2) with real RIRs.

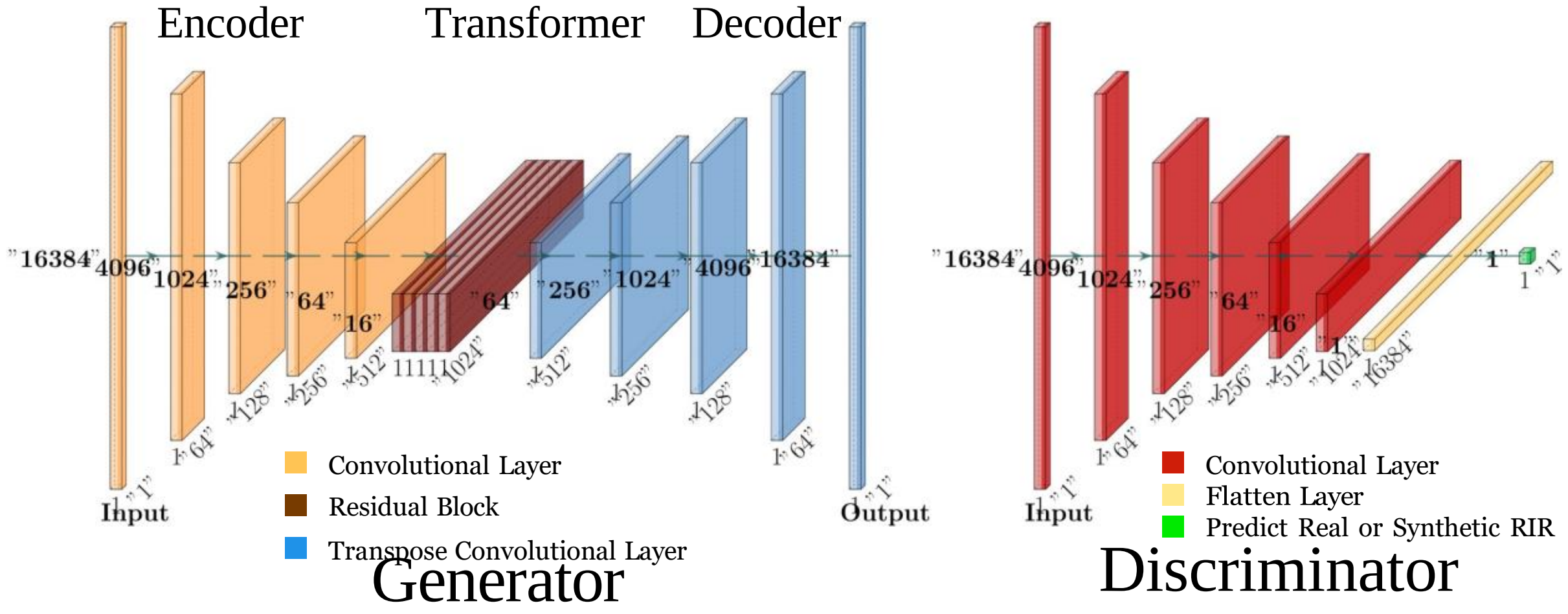

Figure 7.1: The architecture of the generator and discriminator of TS-RIR-GAN. Our generator takes a synthetic RIR as 1x16384 audio samples and translates it into a real RIR of the same dimension. The discriminator network discriminates between real RIRs and translated synthetic RIRs during training by maximizing the adversarial loss (Equation 7.1).

### 7.3.1 Translation: Synthetic RIR to Real RIR

Our TS-RIR-GAN (Figure 7.1) architecture is based upon CycleGAN [226], and WaveGAN [152]. CycleGAN learns to translate two-dimensional images from a source domain $X$ to a target domain $Y$ using unpaired image datasets. Similarly, we design a TS-RIR-GAN architecture that learns mapping functions between one-dimensional (1D) synthetic RIRs ($S$) and real RIRs ($R$) in the absence of paired training examples. Inspired by WaveGAN, which applies generative adversarial networks (GANs) to raw-waveform audio, we directly input RIRs as raw audio samples to our network to learn the mapping functions. In most cases, real and synthetic RIRs are less than one second in duration. Therefore, we re-sample the synthetic and real RIR datasets without loss of generality to 16 kHz and pass them as a one-dimensional input of length 16384.

We represent the real RIR training samples as $\{r_i\}_{i=1}^{N}$ where $r_i \in R$ and the synthetic RIR training samples as $\{s_i\}_{i=1}^{N}$ where $s_i \in S$. The data distributions of the training samples are $r \sim p_{data}(r)$ and $s \sim p_{data}(s)$. We use 2 generators to learn the mappings $G_{SR} : S \to R$ and $G_{RS} : R \rightarrow S$. Our goal is to learn the mapping $G_{SR} : S \rightarrow R$. We use the inverse mapping $G_{RS} : R \rightarrow S$ with cycle-consistency loss [227] to preserve the acoustic characteristics in Synthetic RIRs during translation. We use discriminator $D_R$ to differentiate real RIRs $\{r_i\}_{i=1}^{N}$ and translated synthetic RIRs $\{G_{SR}(s_i)\}_{i=1}^{N}$. Similarly, we use $D_S$ to discriminate $\{s_i\}_{i=1}^{N}$ and $\{G_{RS}(r_i)\}_{i=1}^{N}$ Our objective function consists of adversarial loss [148], cycle-consistency loss and identity loss [228] to learn the mapping functions.

**Adversarial Loss:** To ensure the synthetic RIRs are translated to real RIRs, we use the following objective for the mapping function $G_{SR} : S \rightarrow R$ and the discriminator $D_R$.

$$\begin{aligned} \mathcal{L}_{adv}(G_{SR}, D_R, S, R) = & \ \mathbb{E}_{r \sim p_{data}(r)}[\log D_R(r)] \\ & + \mathbb{E}_{s \sim p_{data}(s)}[\log(1 - D_R(G_{SR}(s))]. \end{aligned} \tag{7.1}$$

The discriminator $D_R$ tries to distinguish between translated RIRs using the mapping function $G_{SR} : S \rightarrow R$ from the real RIRs by maximizing ($max$) the adversarial loss. The generator $G_{SR} : S \rightarrow R$ attempts to generate real RIRs that tend to minimize ($min$) the adversarial loss, i.e., $\min_{G_{SR}} \max_{D_R} \mathcal{L}_{adv}(G_{SR}, D_R, S, R)$. Similarly, we train the mapping function $G_{RS} : R \rightarrow S$ and the discriminator $D_S$ with the objective $\mathcal{L}_{adv}(G_{RS}, D_S, R, S)$.

**Cycle Consistency Loss:** We use cycle consistency loss to preserve the acoustic characteristics in the RIRs during the translation. The cycle consistency loss ensures that $G_{RS}(G_{SR}(s)) \sim s$ and $G_{SR}(G_{RS}(r)) \sim r$.

$$\mathcal{L}_{cyc}(G_{SR}, G_{RS}) = \mathbb{E}_{s\sim p_{data}(s)}[||G_{RS}(G_{SR}(s)) - s||_1]$$
$$+\mathbb{E}_{r\sim p_{data}(r)}[||G_{SR}(G_{RS}(r)) - r||_1]. \tag{7.2}$$

**Identity Mapping Loss:** Identity mapping loss preserves the amplitude of input RIRs:

$$\mathcal{L}_{id}(G_{SR}, G_{RS}) = \mathbb{E}_{s\sim p_{data}(s)}[||G_{RS}(s) - s||_1]$$
$$+\mathbb{E}_{r\sim p_{data}(r)}[||G_{SR}(r) - r||_1]. \tag{7.3}$$

**Full Objective:** The overall objective function can be given as

$$\mathcal{L}(G_{SR}, G_{RS}, D_S, D_R) = \mathcal{L}_{adv}(G_{SR}, D_R, S, R)$$
$$+\mathcal{L}_{adv}(G_{RS}, D_S, R, S)$$
$$+\lambda_{cyc}\mathcal{L}_{cyc}(G_{SR}, G_{RS})$$
$$+\lambda_{id}\mathcal{L}_{id}(G_{SR}, G_{RS}), \tag{7.4}$$

where $\lambda_{cyc}$ and $\lambda_{id}$ control the relative importance of cycle consistency loss and identity mapping loss, respectively. We train our TS-RIR-GAN to find the optimal mapping functions $G^*_{SR}$ and $G^*_{RS}$ by solving

$$G^*_{SR}, G^*_{RS} = \arg \min_{G_{SR},G_{RS}} \max_{D_S,D_R} \mathcal{L}(G_{SR}, G_{RS}, D_S, D_R).$$

We use $G^*_{SR}$ to translate imprecise synthetic RIRs to real RIRs.

### 7.3.2 Implementation

**Network Architecture:** We adapt the discriminator architecture from Donahue et al. [152]. We did not use the phase shuffle operation proposed in WaveGAN [152] because this operation did not improve our results. Inspired by Johnson et al. [229], we designed our generator network consisting of an encoder, a transformer and a decoder. Figure 7.1 highlights our generator and discriminator architectures. Similar to WaveGAN, we use 1D filters of length 25 to perform convolution and transposed convolution operations in our TS-RIR-GAN architecture.

**Dataset:** We use an equal number of real RIRs from BUT ReverbDB [30] and synthetic RIRs generated using the geometric acoustic simulator [1] to train our TS-RIR-GAN architecture. The BUT ReverbDB consists of 1891 RIRs covering the office, hotel room, conference room, lecture room, meeting room, and stairs. We remove repeated RIRs and RIRs recorded in stairs. Among 1209 retained RIRs in BUT ReverbDB, we train our network using 967 RIRs and keep 242 RIRs for testing purposes. Room dimensions, loudspeaker location, and microphone location corresponding to each real RIRs are documented in BUT ReverbDB dataset. We use this information to generate synthetic RIRs using the geometric acoustic simulator. We use random surface absorption/reflection coefficients to generate synthetic RIRs because we do not have room-material information. Therefore one-to-one mapping between synthetic and real RIRs should not be expected.

### 7.3.3 Sub-band Room Equalization (EQ)

Sub-band room equalization bridges the gap in the frequency gain of real and synthetic RIRs over the entire frequency range. Our formulation is based on the sub-band room equalization approach described in [24]. Sub-band relative gain calculation and equalization matching are the two stages in sub-band room equalization.

**Sub-band relative gain calculation:** We calculate the re-sampled relative gains to compensate for the difference in relative gains between synthetic and real RIRs. We compute the frequency response of every RIR in a real-world dataset [30]. We compute the relative gain from the frequency response by taking the gain at 1000Hz as the reference for each real RIR. Then we extract the relative frequency gain at 7 unique sample points (62.5Hz, 125Hz, 250Hz, 500Hz, 2000Hz, 4000Hz, 8000Hz) for every real RIR. The mean and standard deviations of the relative gains for each sample point are different. Therefore we use a Gaussian mixture model to model 7 Gaussian distributions using the relative gains from the sampled points. We re-sample equal numbers of relative gains for each sample point as the input to the Gaussian mixture model. Instead of using the relative gains of the real RIRs, we use the re-sampled relative gains. We use re-sampled relative gains to avoid duplicating the real RIRs during equalization matching. We choose the reference and the number of sample points as proposed in [24].

**Equalization matching:** We match the relative gains of synthetic RIRs with the re-sampled relative gains calculated from real RIRs. We compute the relative frequency gains for the synthetic RIRs at the chosen sample points (62.5Hz, 125Hz, 250Hz, 500Hz, 2000Hz, 4000Hz, 8000Hz), taking gain at 1000Hz as the reference. We calculate the difference in the relative gains of synthetic RIRs and the re-sampled relative gains. Next, we design a finite impulse response (FIR) filter using the window method [230] to compensate for the difference in the relative gains. We filter the synthetic RIRs using our designed FIR filter to match the sub-band relative gains of synthetic RIRs with the re-sampled relative gains.

Table 7.1: Different combinations of our post-processing methods studied in this paper. The best combination is marked in **bold**.

| Combination | Description |
|---|---|
| GAS+EQ | Only perform room equalization. |
| $G^*_{SR}$ (GAS+EQ) | First, perform room equalization, then translate the equalized synthetic RIR to a real RIR. |
| $G^*_{SR}$ (GAS) | Only translate synthetic RIR to real RIR. |
| **$G^*_{SR}$ (GAS)+EQ** | **First, translate a synthetic RIR to a real RIR, then perform room equalization to the translated RIR.** |

### 7.3.4 Optimal Combination

We tried different combinations (Table 7.1) of our post-processing approach to come up with the optimal combination. We estimated 4 different acoustic parameter values from synthetic RIRs generated using the geometric acoustic simulator [1] (GAS), post-processed synthetic RIRs using a different combination of our post-processing approach and real RIRs to evaluate how much the post-processed RIRs are closer to real RIRs. Reverberation time ($T_{60}$), direct-to-reverberant ratio (DRR), early-decay-time (EDT), and early-to-late index (CTE) are four acoustic parameters used for our evaluation. $T_{60}$ is the time required to decay the sound pressure by 60 decibels (dB). The ratio of the sound pressure level of a direct sound source to the sound pressure level of reflected sound in dB is called DRR [128]. EDT is calculated by multiplying the time taken for the sound source to decay by 10 dB by a factor of 6. The proportion of the total sound energy received in the first 50ms to the energy received during the rest of the period is called CTE [221]. Synthetic RIRs and real RIRs used to train TS-RIR-GAN and to perform sub-band room equalization do not have one-to-one mapping. Therefore, we calculated the mean values

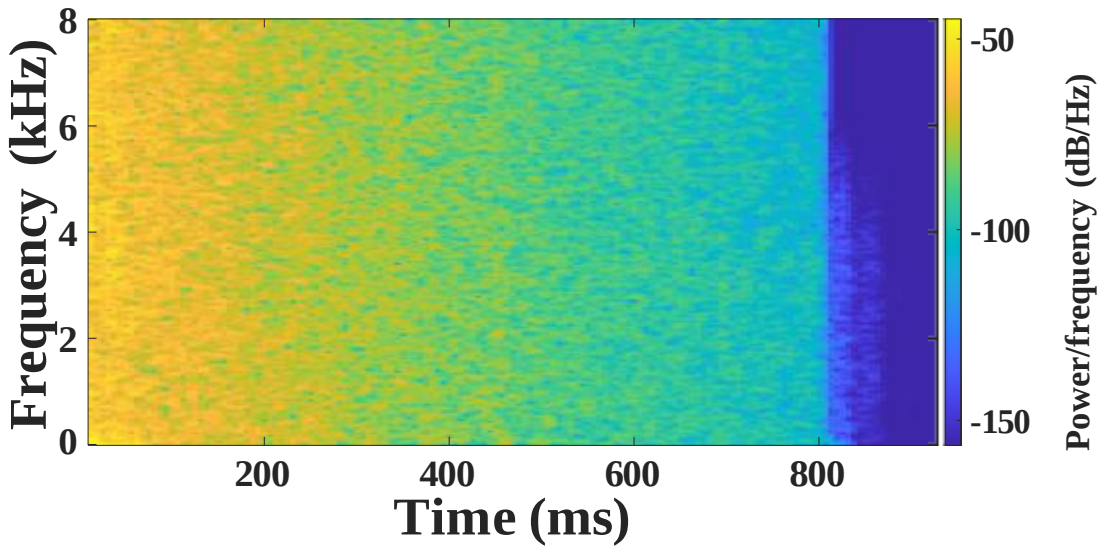

(a) GAS.

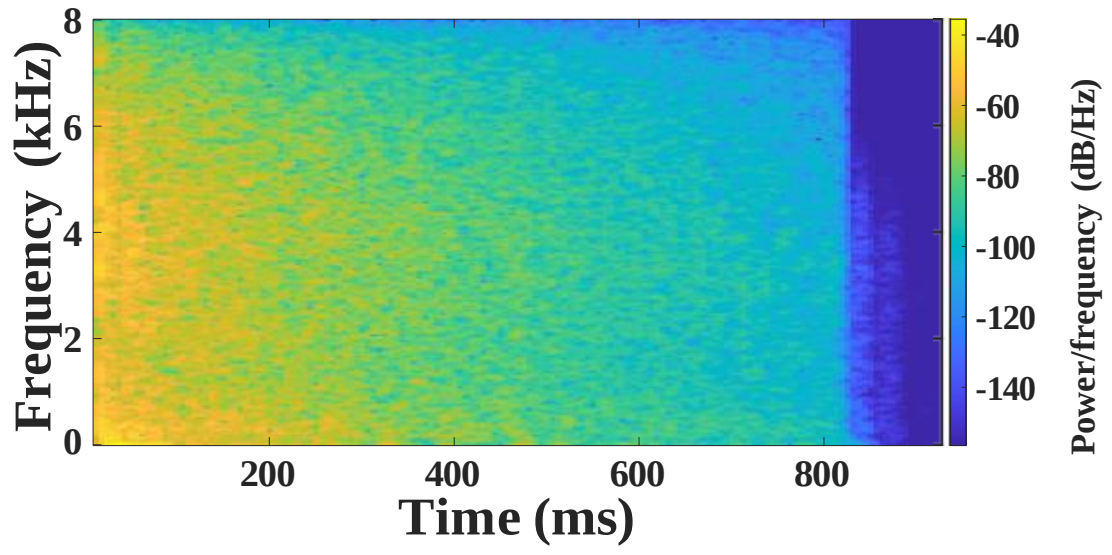

(b) GAS+EQ.

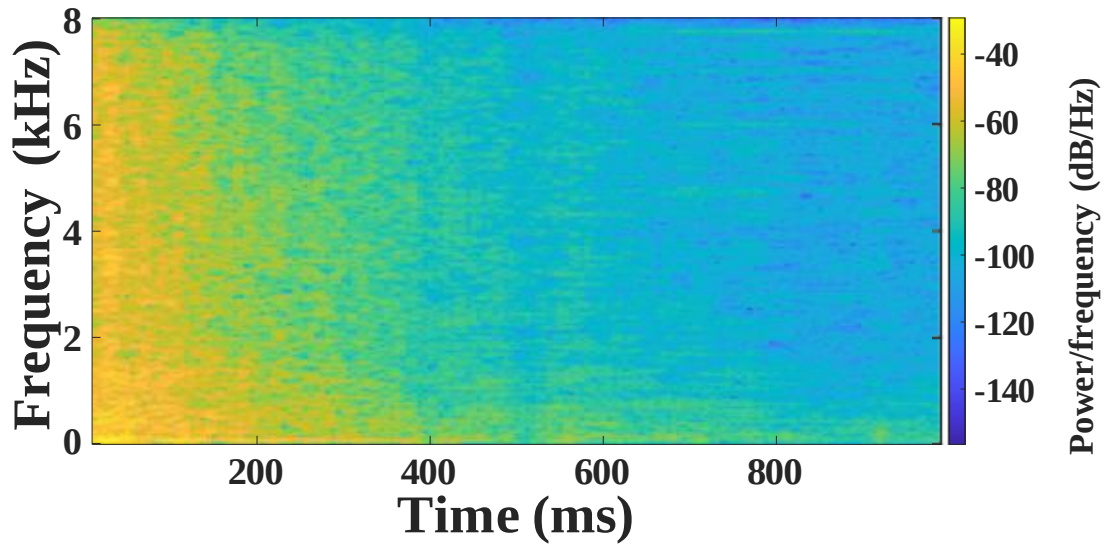

(c) $G^*_{SR}$ (GAS+EQ).

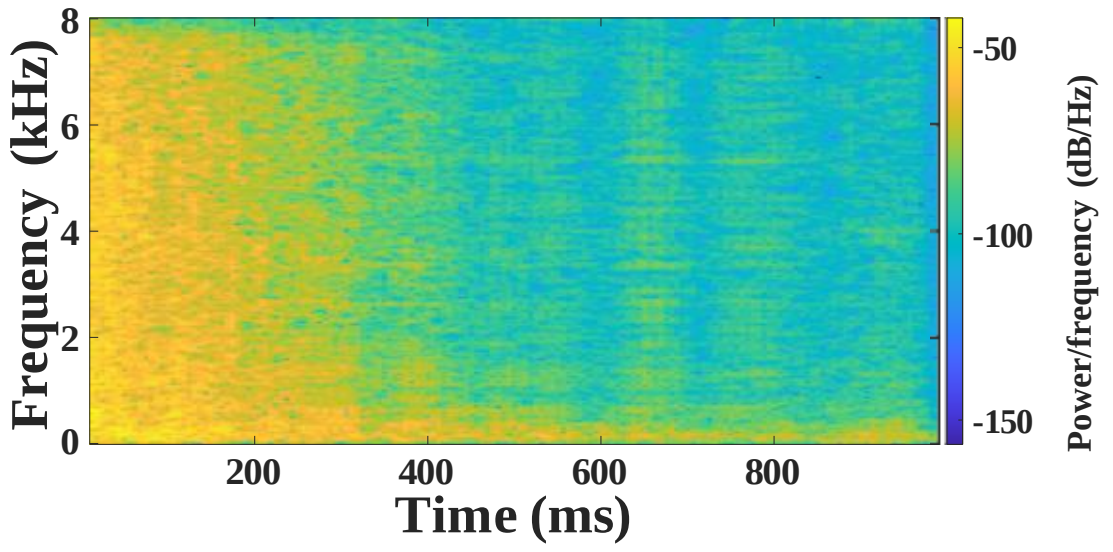

(d) $G^*_{SR}$ (GAS).

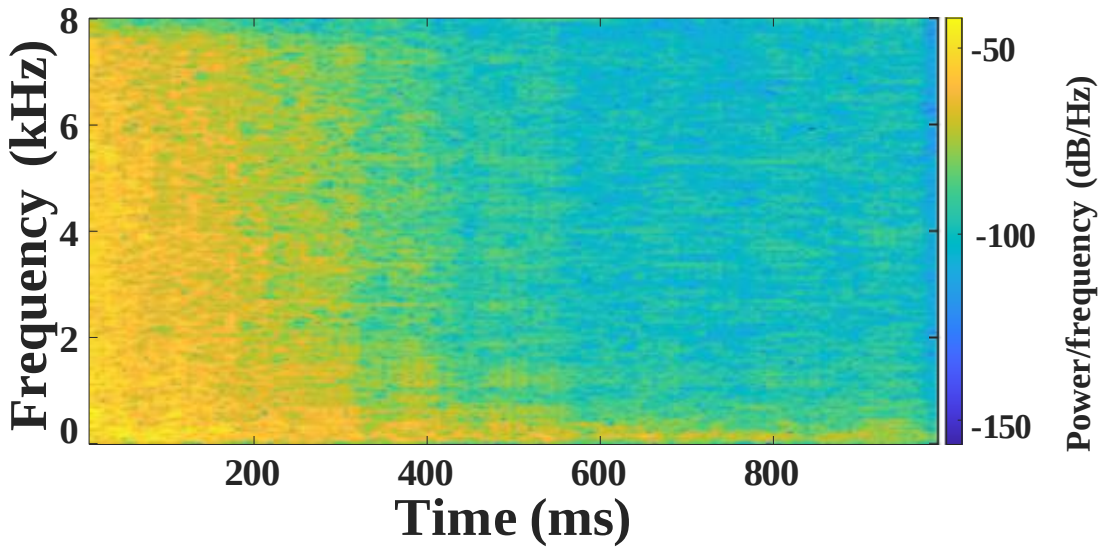

(e) $G^*_{SR}$ (GAS)+EQ.

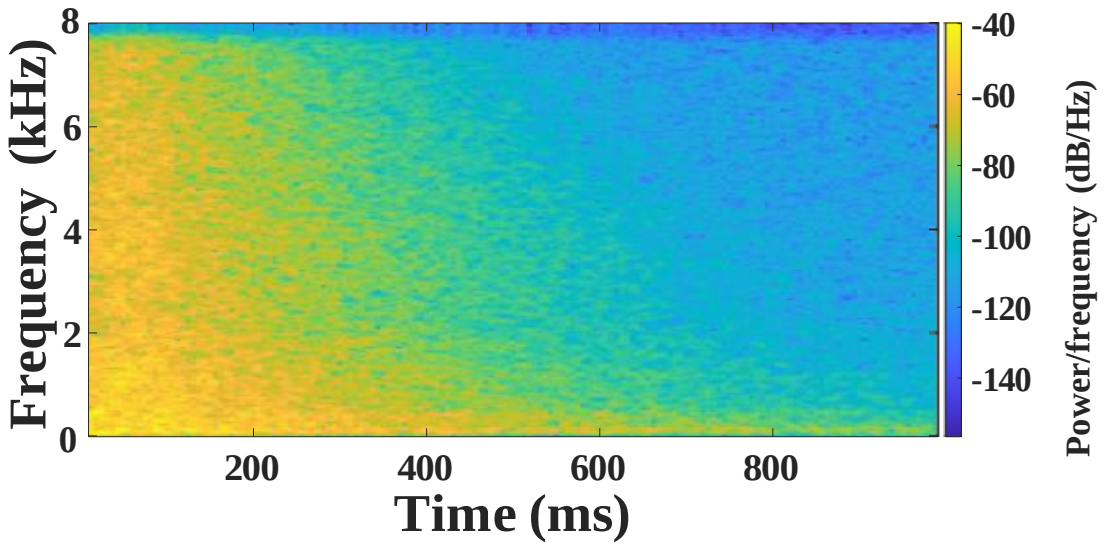

(f) Real RIR.

Figure 7.2: The spectrogram of a synthetic RIR generated using the geometric acoustic simulator [1] (Figure 7.2a), post-processed synthetic RIRs (Figure 7.2b-7.2e), and a real RIR (Figure 7.2f). Sub-band room equalization (EQ) and synthetic RIR to real RIR translation ($G^*_{SR}$()) are the two methods used to post-process the synthetic RIR in different combinations (Table 7.1). Among the energy distribution in spectrograms of post-processed synthetic RIRs, the energy distribution in $G^*_{SR}$(GAS)+EQ is closest to the spectrogram of a real RIR. We can observe that the energy distribution over the low-frequency and high-frequency region in $G^*_{SR}$(GAS)+EQ is similar to a real RIR.

for different acoustic parameters to evaluate our post-processing approach.

Table 7.2 presents the mean values of the acoustic parameters for different sets of RIRs and the absolute difference between the mean values of the acoustic parameters of synthetic and post-processed synthetic RIRs and real RIRs. We can see that mean DRR, mean EDT, and mean

Table 7.2: Mean values of the acoustic parameters. We calculated the mean reverberation time ($T_{60}$), mean direct-to-reverberant ratio (DRR), mean early-decay-time (EDT), and mean early-to-late index (CTE) for real, synthetic and post-processed synthetic RIRs. We also report the absolute mean difference of the acoustic parameters between synthetic and post-processed synthetic RIRs and real RIRs. The acoustic parameter values with the least absolute mean difference are shown in **bold**.

| **RIRs** | **$T_{60}$ (seconds)** | | **DRR (dB)** | | **EDT (seconds)** | | **CTE (dB)** | |
|---|---|---|---|---|---|---|---|---|
| | **Mean** | **Diff** | **Mean** | **Diff** | **Mean** | **Diff** | **Mean** | **Diff** |
| Real RIRs | 1.0207 | | -6.3945 | | 0.8572 | | 3.4886 | |
| GAS | **0.9553** | **0.0654** | -4.7277 | 1.6668 | 0.8846 | 0.0274 | 4.7536 | 1.265 |
| GAS+EQ | 0.9540 | 0.0667 | -7.4246 | 1.0301 | 0.8912 | 0.0340 | 5.6404 | 2.1518 |
| $G^*_{SR}$(GAS+EQ) | 1.5493 | 0.5286 | -8.3879 | 1.9934 | 1.046 | 0.1888 | 2.6562 | 0.8324 |
| $G^*_{SR}$(GAS) | 1.6433 | 0.6226 | **-6.6491** | **0.2546** | **0.8483** | **0.0089** | **3.4907** | **0.0021** |
| $G^*_{SR}$(GAS)+EQ | 1.6364 | 0.6157 | -6.7234 | 0.3289 | 0.8323 | 0.0249 | 3.5367 | 0.0481 |

CTE values of $G^*_{SR}$(GAS) and $G^*_{SR}$(GAS)+EQ are closer to the real RIRs when compared with the other combinations of our post-processing approach. Therefore our proposed TS-RIR-GAN is capable of improving the quality of synthetic RIRs by translating the wave effects present in real RIRs to synthetic RIRs. However, we can see a deviation in the mean $T_{60}$ values for the post-processed synthetic RIRs using TS-RIR-GAN.

Figure 7.2 shows the spectrogram of a synthetic RIR generated using the geometric acoustic simulator [1] (GAS), post-processed synthetic RIRs using a different combination of our post-processing approach and a real RIR. From the spectrograms, we can see that by translating a synthetic RIR to a real RIR, we improve the energy distribution in the low-frequency region (Figure 7.2d) by compensating low-frequency wave effects present in real RIRs. When we perform sub-band room equalization after translation, we observe further refinement in the spectrogram (Figure 7.2e), especially around 600ms to 800ms. After trying all the combinations, we highlight the optimal combination in Figure 7.3. We chose the optimal combination based on the set of acoustic parameter values and the energy distribution of the post-processed RIRs.

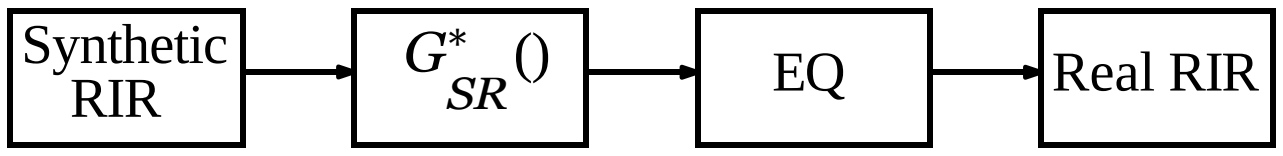

Figure 7.3: Our overall pipeline to improve the quality of synthetic RIRs. We translate synthetic RIRs to real RIRs using our learned mapping function $G^*_{SR}()$, then we augment the wave effects in translated synthetic RIRs by performing real-world sub-band room equalization (EQ).

## 7.4 Implementation and Results

### 7.4.1 Benchmark

We evaluate our approach on the Kaldi LibriSpeech far-field ASR recipe [24]. We convolve clean speech $x_c[t]$ from LibriSpeech [29] with different sets of RIRs $r[t]$ and add environmental noise $n[t]$ from BUT ReverbDB [30] to augment a far-field speech $x_f[t]$ training dataset. The environmental noise is started at a random position $l$ and repeated in a loop to fill the clean speech. In Equation 7.5, $\lambda$ is calculated for different signal-to-noise ratios, which ranges from 1dB to 2dB:

$$x_f[t] = x_c[t] \circledast r[t] + \lambda * n[t + l]. \quad (7.5)$$

We train time-delay neural networks [231] using our augmented training dataset. After training the network, we decode the identity vectors [225] (i-vectors) of a far-field speech test set using phone language models. We calculate word error rate (WER) for large four-gram (fglarge), large tri-gram (tglarge), medium tri-gram (tgmed), and small tri-gram (tgsmall) phone language models, as well as online decoding using a tgsmall phone language models. During online decoding, the i-vectors extracted from the far-field speech test set are passed in real-time. We use WER to evaluate the far-field speech augmented using different sets of RIRs.

Training and testing on the benchmark for each far-field speech training dataset take around 4 days. We ran all the experiments in the same environment to perform a fair comparison.

### 7.4.2 Data Preparation

We use real RIRs and environmental noise from BUT ReverbDB [30] and clean speech (test-clean) from LibriSpeech [29] to augment a real-world far-field speech test set using Equation 7.5. We evaluate our proposed method using the real-world far-field speech test set. We randomly split 1209 RIRs in BUT ReverbDB [30] into subsets of {773,194,242} to create training, development, and test far-field speech datasets.

We use the meta-info accompanying each real RIR to generate synthetic RIRs using the state-of-the-art geometric acoustic simulator (GAS). We post-process the synthetic RIRs by translating synthetic RIRs to real RIRs and performing real-world sub-band room equalization in different combinations (Table 7.1).

We also generated RIRs using the pre-trained IR-GAN [48] on BUT ReverbDB dataset [1]. IR-GAN is a neural network based RIR generator that can generate realistic RIRs corresponding to different acoustic environment by parametrically controlling acoustic parameters.

We created different far-field speech training set by convolving LibriSpeech training datasets (train-clean-{100,360}) with different RIRs and adding environmental noise from BUT ReverbDB [30] set using Equation 7.5. We use synthetic RIRs generated using GAS, post-processed synthetic RIRs, RIRs generated using IR-GAN and real RIRs to augment different training far-field speech datasets.

[1] https://gamma.umd.edu/pro/speech/ir-gan

Table 7.3: Word error rate (WER) reported by the Kaldi far-field ASR system. We trained the Kaldi model using the different augmented far-field speech training sets and tested it on a real-world far-field speech. The training sets are augmented using synthetic RIRs generated using GAS, post-processed synthetic RIRs (Table 7.1), synthetic RIRs generated using IR-GAN and real RIRs. We report WER for fglarge, tglarge, tgmed, and tgsmall phone language models and online decoding using tgsmall phone language model. Our best results are shown in **bold**.

| **Training data** | | **Test Word Error Rate (WER)** [%] | | | | |
|---|---|---|---|---|---|---|
| | | fglarge | tglarge | tgmed | tgsmall | online |
| | clean (Baseline) | 77.15 | 77.37 | 78.00 | 78.94 | 79.00 |
| | real (Oracle) | 12.40 | 13.19 | 15.62 | 16.92 | 16.88 |
| | GAS [1] | 16.53 | 17.26 | 20.24 | 21.91 | 21.83 |
| | GAS+EQ [24] | 14.51 | 15.37 | 18.33 | 20.01 | 19.99 |
| | $G^*_{SR}$ (GAS+EQ) | 14.27 | 14.98 | 17.79 | 19.37 | 19.36 |
| **Ours** | $G^*_{SR}$ (GAS) | 14.12 | 14.70 | 17.44 | 19.08 | 19.06 |
| | $\boldsymbol{G^*_{SR}}$ **(GAS)+EQ** | **13.24** | **14.04** | **16.65** | **18.40** | **18.39** |
| | GAS [1] | 16.53 | 17.26 | 20.24 | 21.91 | 21.83 |
| | IR-GAN [48] | 14.99 | 15.93 | 18.81 | 20.28 | 20.24 |
| | GAS+IR-GAN [48] | 14.16 | 14.99 | 17.56 | 19.21 | 19.21 |
| **Ours** | $\boldsymbol{G^*_{SR}}$ **(GAS)+EQ** | **13.24** | **14.04** | **16.65** | **18.40** | **18.39** |

### 7.4.3 Results and Analysis

Table 7.3 shows the word error rate (WER) reported by the Kaldi LibriSpeech far-field ASR benchmark [24]. We can see that the augmented far-field speech training sets perform well compared to our baseline model trained on a clean Librispeech dataset. The lowest WER is reported by our oracle model trained on real-world far-field speech. In our work, we aim to minimize the gap in the performance between real RIRs and synthetic RIRs.

The WERs for tgsmall reported by GAS+EQ and $G^*_{SR}$(GAS) are 18.33% and 17.44%, respectively. We observe that our approach outperforms the prior methods by up to 4.8%. We see an interesting observation with $G^*_{SR}$(GAS+EQ) and $G^*_{SR}$(GAS) datasets. When compared to translated synthetic RIRs ($G^*_{SR}$(GAS)), translated room equalized RIRs ($G^*_{SR}$(GAS+EQ)) perform poorly.

**Optimal Approach:** We can see that translating imprecise synthetic RIRs to real RIRs and performing real-world sub-band room equalization on the translated RIRs ($G^*_{SR}$(GAS)+EQ) gives the lowest WER. When compared to training sets created using unmodified RIRs (GAS) and room equalized RIRs (GAS+EQ), we observe a relative reduction in WER by up to 19.9% and 9.1%, respectively.

Physical-based acoustic simulators (GAS) and neural-network-based RIR generators (IR-GAN) generate RIRs using two different approaches. GAS models RIR corresponding to a particular scene by considering room dimension, speaker, listener position, etc. IR-GAN uses acoustic parameters to generate an RIR for a particular scene. In previous work, [48], far-field speech augmented using synthetic RIRs from GAS and IR-GAN are used to train a robust far-field ASR system (GAS+IR-GAN). From Table 7.3, we can observe that our post-processed RIRs using our optimal approach ($G^*_{SR}$(GAS)+EQ) outperforms the combination of RIRs generated using GAS and IR-GAN (GAS+IR-GAN).

## 7.5 Conclusion

We present a new architecture to translate synthetic RIRs to real RIRs and perform real-world sub-band room equalization on the translated RIRs to improve the quality of synthetic RIRs. We evaluate the quality of our post-processed synthetic RIRs using a set of acoustic parameter values and the energy distribution of the post-processed RIRs. The set of acoustic parameter values indicates how much the wave effects in post-processed RIRs are closer to real

RIRs. We show that the mean direct-to-reverberant ratio, mean early-decay-time, and mean early-to-late index of the post-processed synthetic RIRs are closer to the real RIRs when compared to the unmodified synthetic RIRs. We also evaluate our post-processing approach on the Kaldi LibriSpeech far-field automatic speech recognition benchmark and observe that our post-processed RIRs outperform unmodified synthetic RIRs by up to 19.9%. In the future, we would like to explore improving the quality of synthetic RIRs based on improved techniques to model acoustic wave effects and translation architectures. We would also like to evaluate their benefits for other applications, including speech separation [106, 232] and audio-visual speech recognition [233] tasks.

# Chapter 8: Room Impulse Response Estimator for Automatic Speech Recognition

## 8.1 Motivation

Accurate measurement of RIRs is prohibitively expensive and requires expert guidance [30]. On the other hand, simulating accurate RIRs requires 3D mesh representations of the underlying scenes [3] and complete knowledge of material properties [26, 27]. Therefore, it is only practical to precisely simulate RIRs for a real-world environment with their 3D representation. As an alternative, we propose an approach to directly estimate the RIR from single-channel speech with the corresponding visuals captured using a camera in household devices like Amazon Echo Show, META Portal etc.

It is well known that reverberant speech inputs affect the performance of ASR systems in home voice assistants, particularly in the presence of a domain mismatch between the training and test data in terms of reverberation effects [234]. Such a mismatch is a direct consequence of the need for augmenting training data similar to the test data by estimating the RIRs in the testing environment to ensure learnability when large neural networks are involved. Additionally, while room acoustics-agnostic domain adaptation strategies have been explored to handle the domain mismatch issue [235], we draw and study an explicit connection between improved RIR estimation and improved ASR performance.

## 8.2 Main Contributions

Overall, we propose an approach for speech-to-RIR estimation that directly benefits ASR. Our main contributions are twofold: We first propose a novel GAN-based RIR estimator (S2IR-GAN) using an energy decay relief loss, which we expect to capture the energy-based acoustic properties of input reverberant speech, and a discriminator loss to estimate the fine structure of the RIR. We demonstrate 22% improvement in capturing early reflection energy (ERE) and 72% improvement in energy decay relief (EDR). We then evaluate the benefits of the proposed model on an ASR task and demonstrate a 6.9% reduction of word error rate relative to existing RIR estimators when handling reverberant speech.

## 8.3 Dataset

We generate 98,316 one-second duration synthetic training examples of reverberant speech by convolving a subset of 360 hours of clean speech (train-clean-360) from the LibriSpeech dataset [29] with recorded RIRs from the Meta RIR (MRIR) dataset.

## 8.4 Our Approach

### 8.4.1 RIR Estimation from Reverberant Speech

We alternatingly train our RIR estimation network ($E_N$) and a discriminator network ($D_N$) using reverberant speech ($S_R$) and the corresponding RIR ($R_G$) in the data distribution $p_d$ at each iteration. Our RIR estimation network estimates the RIR of the input reverberant speech and our discriminator network is optimized to differentiate the estimated RIR from the ground truth RIR.

The objective function of $E_N$ consists of the EDR error (see below), the modified conditional GAN (CGAN) [2] error, and the mean square error (MSE). We use a modified CGAN objective function to train $D_N$.

**EDR Error:** The EDR describes the energy remaining in the RIR in a specific frequency band centered at $b_k$ Hz at time $t$ seconds. In the following EDR equation, $H(r, t, k)$ is the bin $k$ of the short-time Fourier transform of the RIR $r$ at time $t$. The total number of time frames is $T$.

$$EDR(r, t, b_k) = \sum_{t=t}^{T} |H(r, t, k)|^2. \tag{8.1}$$

We calculate the $EDR$ of the estimated RIR ($R_E$) and the ground truth RIR ($R_G$) at a set of octave frequency bands (B) with center frequencies from 16Hz to 4000 Hz. We calculate the EDR loss as follows:

$$\mathcal{L}_{EDR} = \mathbb{E}_{(S,R)\sim p_d}[\mathbb{E}[(EDR(R_E, t, b_k) - EDR(R_G, t, b_k))^2]]. \tag{8.2}$$

Our EDR loss helps the RIR estimator to capture energy-based acoustic properties of the RIR [236].

**CGAN Error (RIR Estimator):** The CGAN error is used to estimate the RIR from the reverberant speech ($S_R$) using our RIR estimator ($E_N$) that is difficult to differentiate from the ground truth RIR by the $D_n$ during training.

$$\mathcal{L}_{CGAN} = \mathbb{E}_{S_R\sim p_d}[\log(1 - D_N(E_N(S_R)))]. \tag{8.3}$$

**MSE:** For each reverberant speech example ($S_R$), we calculate the squared difference of each time sample ($t$) in the estimated RIR $R_E$ and the ground truth RIR $R_G$.

$$\mathcal{L}_{MSE} = \mathbb{E}_{(S_R)\sim p_d}[\mathbb{E}[(R_G(S_R, t) - R_E(S_R, t))^2]]. \quad (8.4)$$

We alternatingly train $E_N$ to minimize the objective function $\mathcal{L}_{E_N}$ and $D_N$ to maximize the objective function $\mathcal{L}_{D_N}$. We use the weights $\lambda_{EDR}$ and $\lambda_{MSE}$ to control the contribution of $\mathcal{L}_{EDR}$ and $\mathcal{L}_{MSE}$ respectively in $\mathcal{L}_{E_N}$:

$$\mathcal{L}_{E_N} = \mathcal{L}_{CGAN} + \lambda_{EDR}\,\mathcal{L}_{EDR} + \lambda_{MSE}\,\mathcal{L}_{MSE}, \quad (8.5)$$

$$\begin{aligned}\mathcal{L}_{D_N} = {} & \mathbb{E}_{(R_G,S_R)\sim p_d}[\log(D_N(R_G(S_R)))] \\ & + \mathbb{E}_{S_R\sim p_d}[\log(1 - D_N(E_N(S_R)))]. \end{aligned} \quad (8.6)$$

## 8.4.2 Network Architecture

**RIR Estimator:** We propose an encoder-decoder architecture to estimate the RIR from the reverberant speech (Figure 8.1). The reverberant speech ($S_R$) can be described as a convolution of clean speech ($S_C$) and an RIR as follows:

$$S_R[n] = \sum_{s=1}^{S} RIR[s] * S_C[n - s], \quad (8.7)$$

where $S$ is the total number of samples in the RIR. To capture the features of the RIR of length 4096 from the reverberant speech, we use a large convolutional layer of length 8193 in the first layer. Later, we reduce the dimension of the extracted RIR features from 64 to 16 while increasing the number of channels from 512 to 1024.

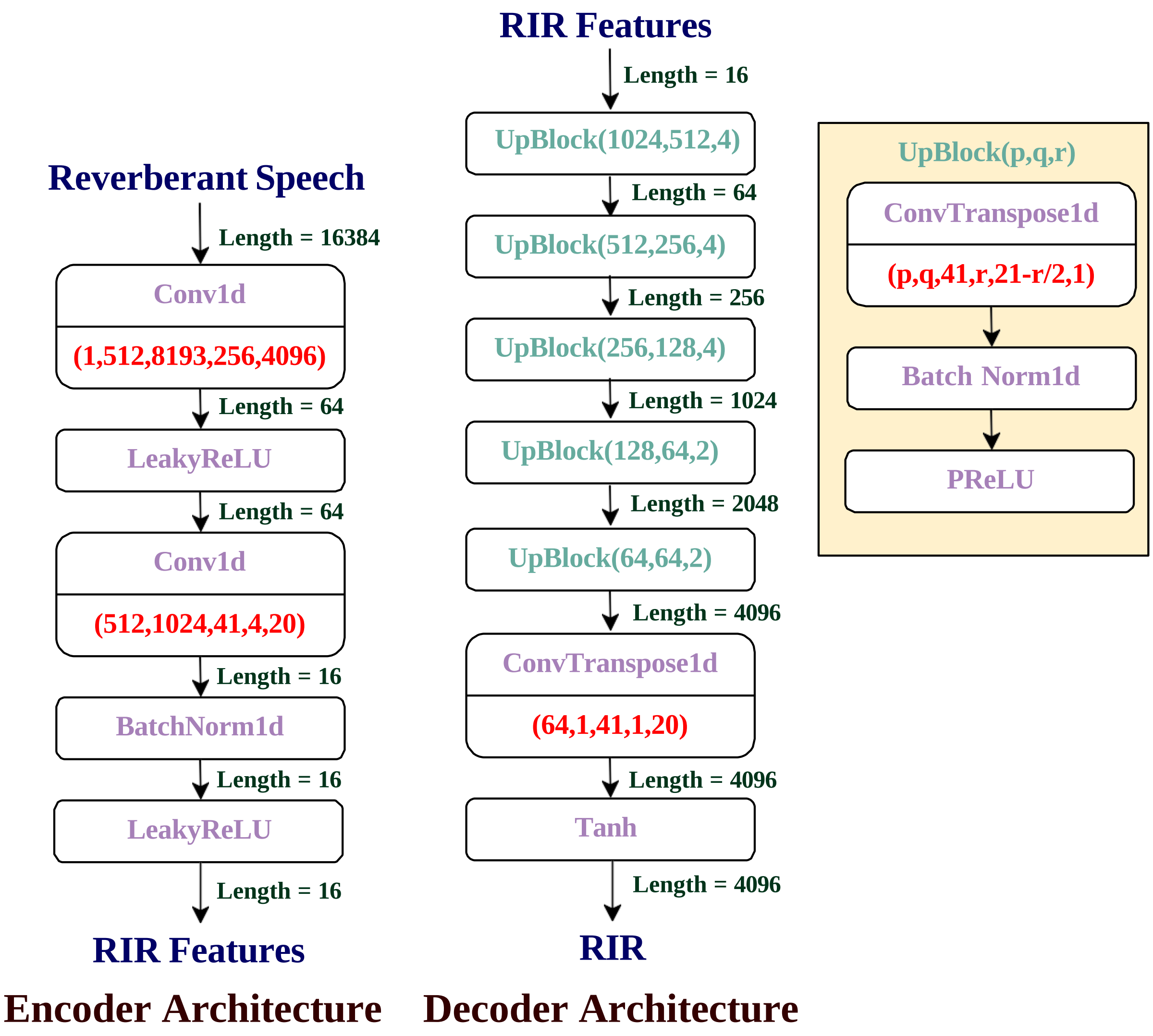

Figure 8.1: The encoder-decoder architecture of our S2IR-GAN. Our encoder network extracts the RIR features, and our decoder network constructs an RIR from the extracted features. For Conv1d layers and ConvTranspose1d layers, the parameters are input channels, output channels. kernel size, stride and padding. The last parameter of ConvTranspose1d is output padding. We use a negative slope of 0.2 for the Leaky ReLU layers.

We construct the RIR from the encoded features in the decoder network. We use 5 sets of transpose convolution layers, batch normalization and PReLU layers to gradually increase the length from 16 to 4096 and reduce the number of channels from 1024 to 64. We use a transpose convolution layer with stride 1 to collapse the number of channels from 64 to 1 and obtain the weighted averaged estimated RIR. In the last layer, we use the Tanh activation function because the RIR contains both negative and positive values. Figure 8.1 shows our encoder-decoder architecture in detail. We adapt the discriminator ($D_N$) network architecture from FAST-RIR [2].

### 8.4.3 Training

To pick the best model during training, we use the validation EDR loss. We trained our network for 200 epochs using the RMSprop optimizer. We used a batch size of 128 and an initial learning rate of $8 \times 10^{-5}$. The learning rate is decayed by 0.7 after every 40 epochs.

## 8.5 Acoustic Evaluation

We evaluate the performance of our proposed approach using EDR loss (Equation 8.2), DRR error, early reflection energy (ERE) loss and MSE (Equation 8.4) of the estimated RIR and the ground truth RIR. Finally, we also compare the ground truth RIR and the estimated RIR using our S2IR-GAN in the time domain.

The EDR of the RIR contains enough information to construct an equivalent RIR with the same gross temporal and spectral features while the fine structure can be different [176]. Acoustic parameters such as $T_{60}$ and early decay time are measured on the slope of EDR. Therefore, we use EDR loss for evaluation. DRR, the log ratio of the energy of the direct sound and the sound arriving after it, is an acoustic parameter often used in ASR applications to measure the amount of distortion introduced into speech by the RIR [237]; DRR estimation is also directly integrated into some ASR systems [90]. Here, we compute the DRR error as the mean absolute error (MAE) of the DRR between the estimated and the ground-truth RIRs. We also include the ERE loss, which is the MAE of the early reflection energy between the estimated and the ground truth RIR. ERE measures the total early sound energy between 0 and 80ms:

$$ERE = 10 \log Energy(0 - 80ms). \tag{8.8}$$

We created 5000 reverberant test data by convolving clean speech from the LibriSpeech dataset (train-clean-100) with RIRs from the MRIR dataset not used for training. We compared our performance against a baseline model and the state-of-the-art RIR estimator (FiNS). We modified the input and output dimensions of the FiNS model to match our S2IR-GAN. For a fair comparison, we have trained FiNS and our S2IR-GAN using the same training data (Section 8.4.3). We use a pre-trained open-source speech enhancement toolkit named ESPNET-SE [238] as a baseline model.

**Baseline:** ESPNET-SE can perform speech dereverberation and denoising. We pass the reverberant speech test data as the input to the network and get clean speech as the output. We compute the RIR using the reverberant speech ($S_R$) and the corresponding output clean speech ($S_C$) as follows:

$$RIR = \mathcal{F}^{-1}\left(\frac{\mathcal{F}(S_R)}{\mathcal{F}(S_C)}\right), \tag{8.9}$$

where $\mathcal{F}$ and $\mathcal{F}^{-1}$ are the Fourier transform and inverse Fourier transform, respectively.

Table 8.1 shows the EDR loss and ERE loss for the baseline model, FiNS and our S2IR-GAN. We calculate the loss over the octave-bands with center frequencies up to 4000 Hz. We can see that on average, our proposed S2IR-GAN outperforms the FiNS by 72% in EDR loss and 22% in ERE loss. In Figure 8.2, we can see the ground truth EDR and the EDR of the estimated RIRs using FiNS and S2IR-GAN at 250 Hz and 2000 Hz. The EDR of our S2IR-GAN is closest

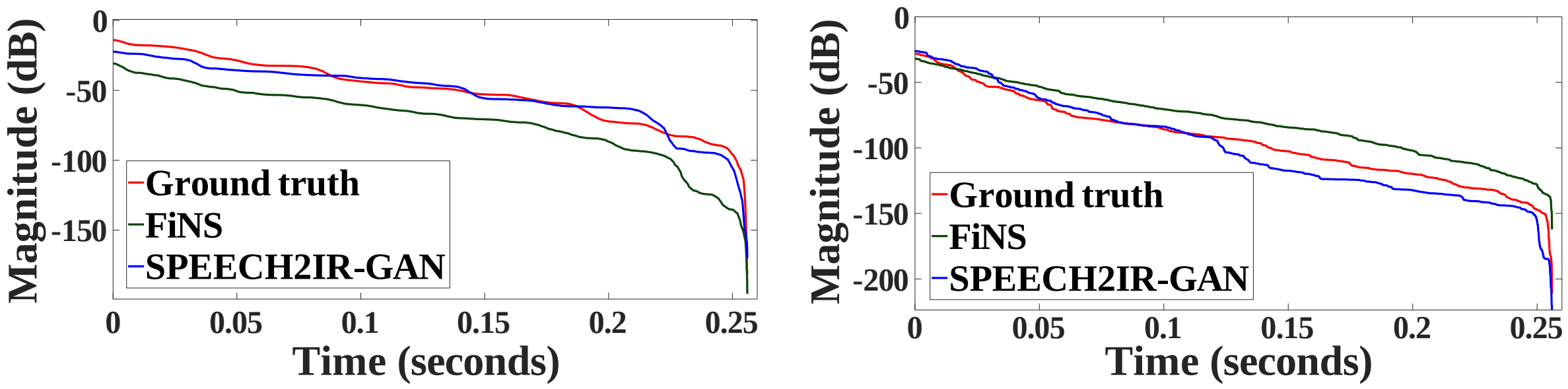

(a) 250 Hz.　　(b) 2000 Hz.

Figure 8.2: The EDR (Equation 8.1) of the ground truth RIR, and the estimated RIR using FiNS and our S2IR-GAN models at 250 HZ and 2000 Hz. We can see that the EDR of S2IR-GAN is closest to the ground truth EDR.

Table 8.1: Energy decay relief (EDR) loss (Equation 8.2) and the mean absolute error of the early reflection energy (ERE loss). In this table, we compare the baseline model, FiNS [4] and our proposed S2IR-GAN. The best results are shown in **bold**.

| Loss | Method | Frequency | | | | | | | | |
|---|---|---|---|---|---|---|---|---|---|---|
| | | **16Hz** | **32Hz** | **63Hz** | **125Hz** | **250Hz** | **500Hz** | **1000Hz** | **2000Hz** | **4000Hz** |
| **EDR Loss** | Baseline | $8.3\times10^{-1}$ | $1.7\times10^{-5}$ | $4.4\times10^{-6}$ | $7.7\times10^{-5}$ | $2.2\times10^{-3}$ | $1.4\times10^{-3}$ | $8.4\times10^{-2}$ | $1.6\times10^{-3}$ | $1.0\times10^{-1}$ |
| | FiNS | $7.9\times10^{-10}$ | $1.6\times10^{-10}$ | $4.6\times10^{-10}$ | $6.8\times10^{-5}$ | $1.0\times10^{-3}$ | $2.3\times10^{-4}$ | $6.9\times10^{-4}$ | $1.0\times10^{-3}$ | $3.3\times10^{-2}$ |
| | **Ours** | $\mathbf{1.8\times10^{-11}}$ | $\mathbf{1.9\times10^{-11}}$ | $\mathbf{3.4\times10^{-10}}$ | $\mathbf{2.0\times10^{-5}}$ | $\mathbf{3.8\times10^{-4}}$ | $\mathbf{5.9\times10^{-5}}$ | $\mathbf{1.4\times10^{-4}}$ | $\mathbf{2.2\times10^{-4}}$ | $\mathbf{8.3\times10^{-3}}$ |
| **ERE Loss** | Baseline | 8.75 | 7.66 | 4.47 | 13.31 | 11.41 | 6.44 | 5.65 | 7.67 | 5.80 |
| | FiNS | 7.41 | 6.89 | 4.69 | 3.77 | 4.21 | 3.62 | 3.81 | 3.75 | 3.67 |
| | **Ours** | **5.88** | **5.33** | **3.34** | **3.27** | **3.11** | **2.88** | **2.87** | **2.83** | **2.91** |

to the ground truth EDR.

We calculated the MSE and DRR error of the estimated RIRs for 3 different methods (Table 8.2). We can see that Baseline and S2IR-GAN give the lowest MSE error. Baseline gives higher DRR error when compared to S2IR-GAN. Figure 8.3 shows an example of the ground truth RIR and the estimated RIR from our S2IR-GAN. We can see that our network can estimate general patterns in the ground truth RIR. However, our network cannot accurately estimate the fine temporal structure of the ground truth RIR. The estimated RIRs are perceptually similar to the ground truth RIR, as can be heard through audio samples of ground truth RIRs and estimated RIRs[1].

[1] `https://anton-jeran.github.io/S2IR/`

Table 8.2: MSE (Equation 8.4) and DRR error of the estimated RIRs from the baseline model, FiNS [4] and S2IR-GAN.

| METHOD | MSE | DRR (dB) |
|---|---|---|
| Baseline | **$3.0\text{x}10^{-4}$** | 6.63 |
| FiNS | $3.7\text{x}10^{-4}$ | 3.31 |
| **S2IR-GAN (ours)** | **$3.0\text{x}10^{-4}$** | **3.28** |

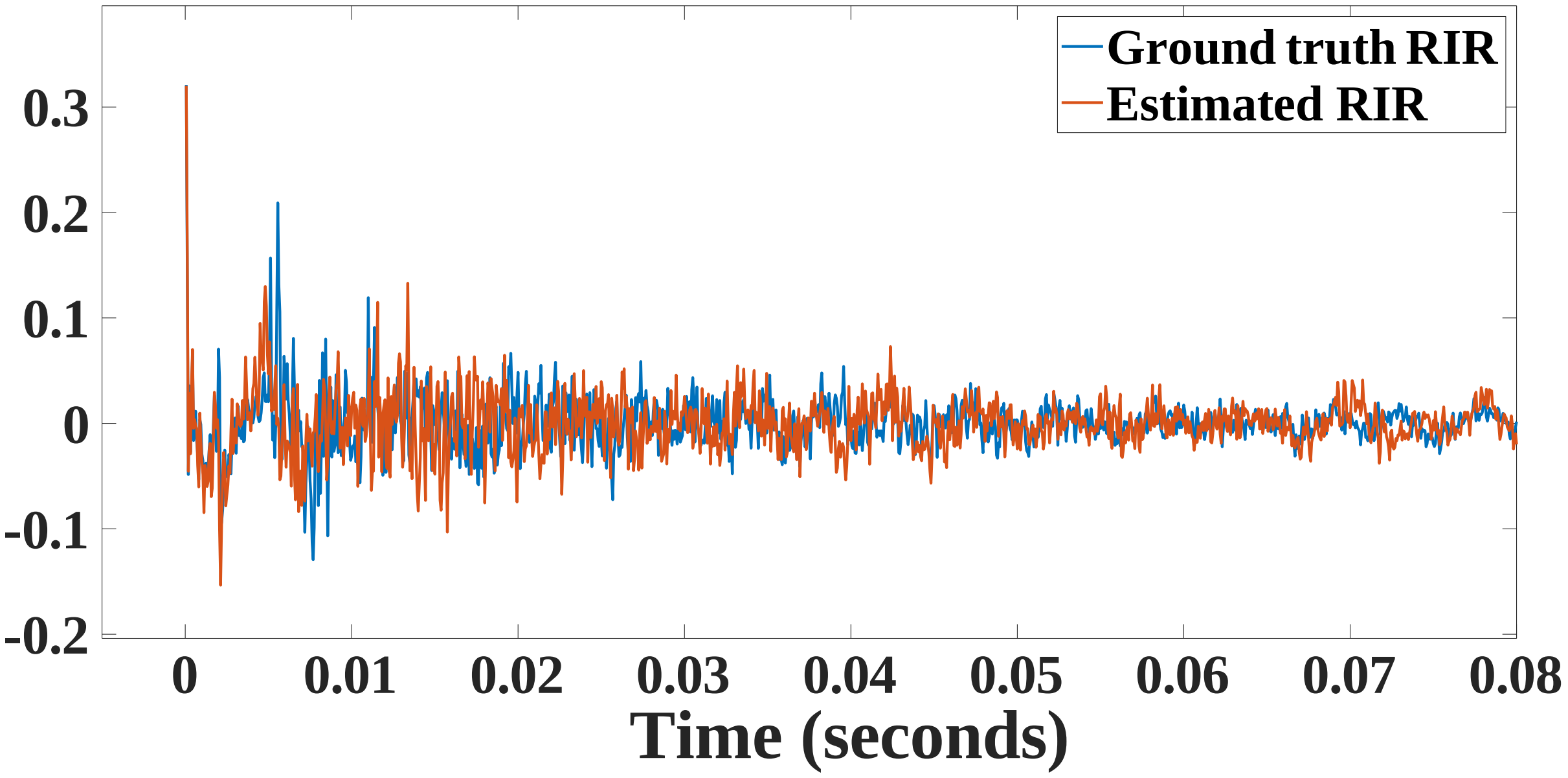

Figure 8.3: Time domain plot of the ground truth RIR and the estimated RIR from our S2IR-GAN. The estimated RIR has similar macroscopic structure as the ground truth RIR while the fine structure differs from the ground truth RIR.

## 8.6 ASR Evaluation

We evaluate the performance of our S2IR-GAN in the Kaldi automatic speech recognition (ASR) experiment[2]. We use close-talk speech data (IHM) and far-field speech data (SDM) in the AMI corpus [155] for our experiment. From the SDM corpus, we sample 2000 reverberant speech examples of approximately one second duration. We input the sampled reverberant speech to the FiNS model and our proposed S2IR-GAN and estimate the RIRs from the input speech.

[2] `https://github.com/RoyJames/kaldi-reverb/`

Table 8.3: Far-field ASR results were obtained for far-field speech data recorded by single distance microphones (SDM) in the AMI corpus. The best results are shown in **bold**.

| **Training Dataset** **Clean Speech** $\circledast$ **RIR** | **Word Error Rate** [%] |
|---|---|
| IHM $\circledast$ None | 64.2 |
| IHM $\circledast$ FiNS | 60.9 |
| **IHM** $\circledast$ **S2IR-GAN (ours)** | **54.0** |

The anechoic IHM speech is convolved with the estimated RIRs to create synthetic reverberant speech training data.

We train the modified Kaldi ASR recipe with the synthetic training data and test the ASR model on the real-world reverberant SDM data. We also train the ASR model with unmodified IHM data as our baseline model. We use word error rate (WER) to evaluate the performance of the ASR system. Lower WER indicates that the reverberation effects in the training speech data are closer to the test speech data (SDM) [26]. From Table 8.3, we can see that our S2IR-GAN outperforms FiNS by 6.9%.

## 8.7 Discussion

In this work, we present a method for the improved estimation of RIRs in the context of far-field speech recognition. Our S2IR-GAN model outperforms the state-of-the-art RIR estimator (FiNS) in acoustic metrics such as energy decay relief loss, early reflection energy loss, DRR error, and MSE. It is also shown to outperform FiNS in a downstream ASR task. The main limitation of this work is that the network cannot capture the fine temporal structure of an RIR from reverberant speech, though the perceptual implications of this shortcoming are unclear. In the future, we would like to extend this work to improve performance of the current model by expanding the set of RIRs used to augment training data; consider RIR estimation while leveraging the goals of other downstream tasks such as dereverberation or source separation; or estimate RIRs from a larger set of input modalities, such as multi-channel speech signals or audio-visual data.

# Chapter 9: Audio-Visual Room Impulse Responses Estimator for AR/VR Applications

## 9.1 Motivation

Reverberation, caused by sound reflecting off surrounding surfaces, transforms how a listener perceives the sound once it is released from a sound source. The transformation is influenced by specific properties of the surrounding area, like spatial geometry, the composition and material properties of surfaces and objects within the environment, and the positioning of various sound sources in proximity. For example, someone speaking or playing music in a large auditorium is perceptually significantly different from someone speaking in a small classroom [239, 240]. The environmental effect that any sound goes through because of the transformation can be quantitatively described by the room impulse response (RIR). RIR is a fundamental concept that characterizes how an acoustic space affects sound, essentially representing the transfer function between a sound source and a receiver, encapsulating all the direct and reflective paths that sound can travel within any indoor or outdoor environment.

RIR estimation, defined as estimating the RIR component from a given reveberant speech signal (see Equation 9.1), finds its major application in augmented reality (AR) and virtual reality (VR) [158, 159, 200]. Usually, when sound effects do not align acoustically with the visual scene, it can disrupt the audio-visual human perception. In AR and VR settings, discrepancies between the acoustics of the real environment and the virtually simulated space lead to cognitive dissonance. This phenomenon, known as the "room divergence effect", can significantly detract from the user experience [68, 241]. RIR estimation from real-world speech can help overcome these problems.

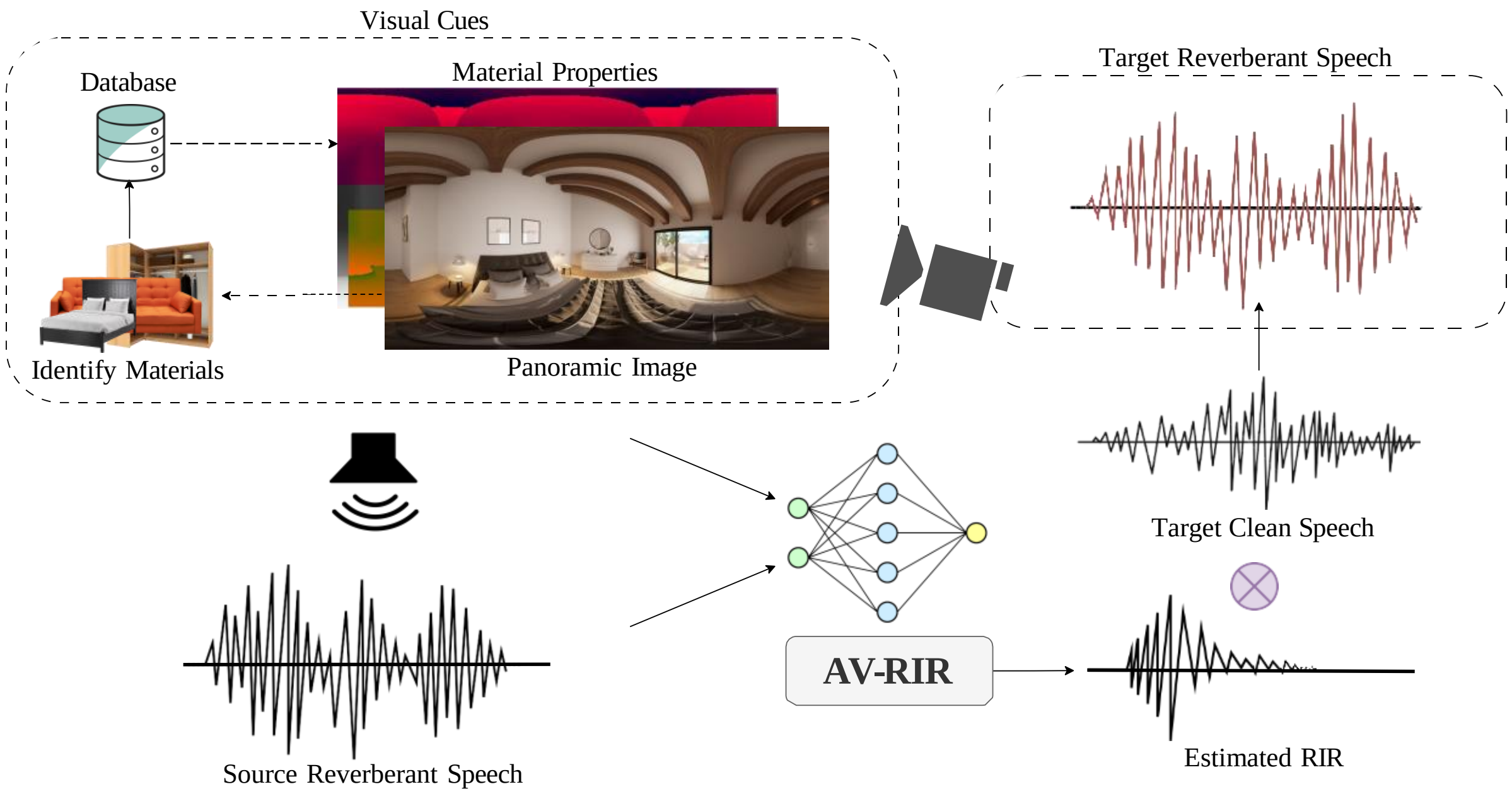

Figure 9.1: Overview of **AV-RIR**: Given a source reverberant speech in any environment, AV-RIR estimates the RIR from the reverberant speech using additional visual cues. The estimated RIR can be used to transform any target clean speech as if it is spoken in that environment.

$$S_R = S_C \circledast RIR, \tag{9.1}$$

Prior work on RIR estimation mainly deals with recorded audio signals and does not take into account visual cues. Directly estimating the RIR from source reverberant speech has been extensively studied using traditional signal processing methods [242–245]. However, these approaches may not work well in some real-world applications, mainly because they are based on the assumption that the source is a modulated Gaussian pulse, not actual speech, [244, 245] or they require pre-knowledge of the specific attributes about the speaker or the microphone used for recording [242, 243]. Recently, neural learning-based RIR estimation techniques have been proposed to estimate RIR from reverberant speech [4, 6]. These techniques are capable of estimating early components (i.e., the direct response and early reflections of RIR) and are not very effective in estimating late components (i.e., the late reverberation of RIR) because the early components of the RIR have impulsive sparse components, while the late components have a noise-like structure with significantly lower magnitude compared to early components. Therefore audio-only approach approximates the late components using a sum of decaying filtered noise signal [4, 246].

Pioneering learning-based work on generating RIR from a single RGB image of physical environments includes Image2Reverb [5], which is based on a conditional GAN-based architecture. Generating RIR from a single RGB image might not be the most effective as it does not have enough information, like 3D geometry, information about the material properties of objects in the environment, speaker position, etc. To overcome the limitations of audio-only and visual-only RIR estimation approaches, we propose a novel audio-visual RIR estimation method.

## 9.2 Main Contributions

We propose AV-RIR, a novel multi-modal multi-task learning approach for RIR estimation. AV-RIR employs a novel neural codec-based multi-modal architecture that takes as input a reverberant speech uttered in a source environment, the panoramic image of the environment, and a novel

Geo-Mat feature that incorporates information about room geometry and the materials of surfaces and objects. The multi-modal architecture consists of carefully designed encoders, decoders, and a Residual Vector Quantizer that learns rich task-specific (RIR estimation and speech deverberation) features while discarding the noise in training data [137]. Additionally, AV-RIR incorporates a dual-branch structure for multi-task learning, where we solve an auxiliary speech dereverberation task alongside the primary RIR estimation task. This approach effectively redefines the ultimate learning objective as decomposing reverberated speech into its constituent anechoic speech and RIR components. Furthermore, we propose Contrastive RIR-Image Pre-training (CRIP) to improve late reverberation of the estimated RIR during inference time using image-to-RIR retrieval. To summarize, our main contributions are as follows:

1. We propose AV-RIR, a novel multi-modal multitask learning approach for RIR estimation.

2. AV-RIR employs a neural codec-based multi-modal architecture that takes as input audio, visual cues and a novel Geo-Mat feature. We also propose CRIP to improve late reverberation effects using retrieval.

3. During training, AV-RIR solves an auxiliary speech dereverberation task for learning RIR estimation. Through this, AV-RIR essentially learns to separate anechoic speech and RIR.

4. We perform extensive experiments to prove the effectiveness of AV-RIR. AV-RIR outperforms prior works by significant margins both quantitatively and qualitatively. We achieve 36% - 63% on RIR estimation on the SoundSpaces dataset [83], and 56% - 79% people find that AV-RIR is closer to the ground-truth in the visual acoustic matching task over our baselines. Additionally, the dereverbed speech predicted by AV-RIR improves performance across various spoken language processing (SLP) tasks. We also perform extensive ablation experiments to demonstrate the critical role of each modules within the AV-RIR framework.

## 9.3 Dataset

For training and evaluation, we use the widely adopted SoundSpaces dataset [27, 83]. The SoundSpaces dataset provides paired reverberant speech and its RIR. The data is sourced by convolving simulated RIRs with clean speech from the LibriSpeech [29] dataset. The RIRs are simulated using the geometric acoustic simulator in the SoundSpaces platform [27, 83] for 82 Matterport [247] 3D environments. SoundSpaces can simulate highly realistic RIR for any arbitrary camera views and microphone positions by considering direct sounds, early reflections, late reverberations, material and air absorption properties, etc. The panoramic images in the SoundSpaces dataset contain 3D humanoids of the same gender as the speaker in each data. In some data, the speaker is out of view and will not be visible in the panoramic image. The sound spaces dataset has 49,430/2700/2,600 train/validation/test samples respectively.

To additionally evaluate how AV-RIR fairs in speech dereverberation in real-world scenarios, we use web videos in the filtered AVSpeech dataset [248] proposed in VAM [9]. The filtered dataset contains 3-10 seconds YouTube clips with reverberant audio recordings. Also, the filtered dataset microphone and the camera are co-located and placed at a different position than the sound sources. The cameras in the filtered videos are static. Since AVSpeech does not have ground truth (GT) RIR, we only used it to evaluate our speech dereverberation pipeline. Our datastore DS comprises synthetic RIRs generated from SoundSpaces, excluding test set RIRs.

## 9.4 Methodology

### 9.4.1 Overview: AV-RIR

Figure 9.2 gives an overview of our approach. Given a reverberant speech $S_R$, the task of AV-RIR is to learn to estimate the RIR from $S_R$. To achieve this, we propose a novel multi-modal neural architecture and solve two parallel tasks for learning accurate RIR estimation. As input, together with the $S_R$, AV-RIR also receives the RGB panoramic image $I_P$ of the source environment and our proposed Geo-Mat feature map $I_G$. The construction of $I_G$ is illustrated in Figure 9.3. The $S_R$ is first passed through a reverberant speech encoder, after which AV-RIR breaks into two branches that solve two different tasks. The bottom branch, which also receives ResNet-18 encoded features of $I_G$, solves our primary RIR estimation task. The other branch, which receives the encoded features of $I_P$ , solves an auxiliary speech dereverberation task to predict enhanced speech $S_C$. After the multi-modal feature fusion step, both branches employ a Residual Vector Quantizer module. During inference, to synthesize any target speech as if spoken in the source environment, we convolve the estimated RIR with the target clean speech. Additionally, we propose CRIP (Figure 9.4) to retrieve an RIR from a datastore DS, conditioned on the $I_P$ of the source environment, to improve late components. Next, we will describe each module in detail.

### 9.4.2 AV-RIR Architecture

**Reverberant Speech Encoder ($E_R$).** Our $E_R$ consists of a simple CNN-based architecture with a single 1-D CNN layer and a single input and output channel. As speech dereverberation and RIR estimation are similar learning problems and based on convolution operations (the latter learns deconvolution, and the former learns the inverse), the latent output $S_E$ from the encoder serves as efficient representations for both tasks AV-RIR solves.

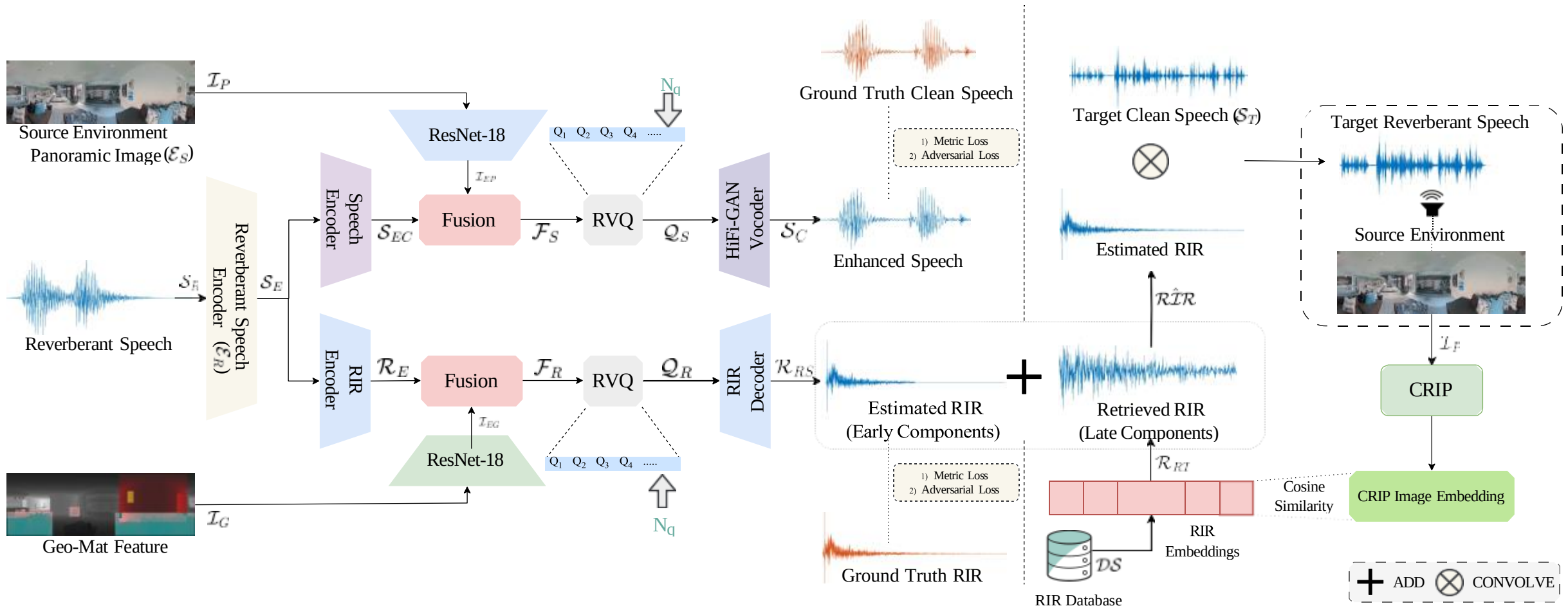

Figure 9.2: Overview of our AV-RIR learning method: Given the input reverberant speech $S_R$ from any source environment $E_S$, the primary task of AV-RIR is to estimate the room impulse response RIR by separating it from the clean speech $S_C$ (see Equation 9.1). The input $S_R$ is first encoded using a Reverberant Speech Encoder $E_R$. The latent output $S_E$ is then passed to two different encoders in two different branches. While one of these branches solves the RIR estimation task, the other solves the speech dereverberation task by estimating $S_C$. Outputs from both the Speech Dereverberation Encoder $S_{EC}$ and RIR Encoder $R_E$ are fused with ResNet-18 encodings from the panoramic image $I_P$ and Geo-Mat feature $I_G$ respectively. The output latent multi-modal encodings $I_{EP}$ and $I_{EG}$ are then passed to a trainable Residual Vector Quantization module (RVQ), which quantizes $F_S$ to latent codes $Q_S$, and $F_R$ to latent codes $Q_R$. Finally, the HiFi-GAN vocoder decodes the enhanced speech $S_C$ from $Q_S$ and the RIR decoder decodes estimated early components of RIR $R_{RS}$ from $Q_R$ which are used to calculate losses for training. At inference time, our CRIP retrieves an RIR from a database DS and is used to improve late reverberation in the estimated RIR. Finally, post addition, the final estimated RIR is convolved with any $S_C$ to make it sound like it was uttered in $E_S$.

**Room Impulse Response Encoder ($E_{IR}$).** $E_{IR}$ is adapted and modified from the S2IR-GAN encoder [6]. Our three-layer $E_{IR}$ has 256, 512, and 1024 output channels, 14401, 41, and 41 kernel lengths, and 225, 2, and 2 strides, respectively. The large kernel length in the first layer encodes the RIR features efficiently. We significantly reduce the input dimension by a factor of 900. We process reverberant speech $S_R$ segments of $\mathbb{R}^{1\times 14400}$ samples. Therefore, every reverberant speech sample is encoded into $\mathbb{R}^{1024\times 16}$ RIR temporal features.

**Vision Encoders ($E_P$, $E_{GM}$).** Prior work on audio-visual speech dereverberation and localization has shown that ResNet-18 [249] is capable of extracting strong cues from image $I_P$ and depth maps [97, 140, 236]. Therefore, we employ two separate ResNet-18-based feature encoders $E_P$ and $E_{GM}$ to encode the $I_P$ and the Geo-Mat feature $I_G$, respectively, and reshape the features to $\mathbb{R}^{1024\times 4}$.

**Multi-modal Fusion Modules (M).** Similar to previous neural audio codec architectures [137], we fuse the visual features with the audio stream along the temporal axis. The combined audio-visual encoded representation is projected into the designed multi-dimensional space [250] and passed to the next stage to quantize into codes.

**Residual Vector Quantizer (RVQ).** RVQ are used in neural audio codecs to compress the encoder output into a discrete set of code vectors to transmit the data at a fixed target bitrate R (bits/second). For AV-RIR, we modify the RVQ proposed in SoundStream [251]. SoundStream proposes a VQ with trainable codebooks that are trained together with the model end-to-end. SoundStream [251] adapts the VQ proposed in [252, 253] and improves the codebook with the multi-stage VQ [254]. Our primary modification involves relaxing the constraints (i.e., the target bitrate) to improve the speech dereverberation performance. SoundStream is designed for real-time transmissions and streaming compressed audio at 3-18 Kbps. For our task, we relaxed the compression to ≈59 Kbps. We observed that relaxing beyond 59 Kbps did not significantly improve the performance. Audio codecs have shown increased performance by increasing the birates in subjective tests [255]. Our RVQ cascades $N_q = 64$ layers of VQ and uses a large codebook size $N = 8192$. Having a larger $N/N_q$ ratio has been shown to achieve higher coding efficiency [251].

**Decoders ($D_{IR}$, $D_S$).** As the two branches of AV-RIR are responsible for decoding separate outputs from the compressed codes (i.e., enhanced speech and RIR), we use two different decoders for this task. We use a HiFi-GAN vocoder for the speech dereverberation branch to decode enhanced speech from the compressed code. HiFi-GAN [256] has shown impressive performance in generating high-fidelity speech, especially with audio codecs [250]. For the RIR estimation branch, we use a modified SoundStream decoder [251]. We modify the decoder to have 6 transposed convolutional (Conv) blocks with output channels of (256, 128, 64, 32, 32, 16) and strides of (5, 5, 2, 2, 1, 1). The output from the last transposed Conv block is passed to a final 1D Conv layer with kernel size 1 and stride 1 to project the code to the waveform domain.

### 9.4.3 Geo-Mat Features

The Geo-Mat feature ($I_G$) represents the geometry and sound absorption properties of materials in the environment. Physics-based RIR simulators take the 3D geometry of the environment and the material absorption coefficients (AC) of each material present in the environment as input to accurately generate RIR for that environment [15, 27, 257]. On the other hand, Changhan *et al.* [140] show that leveraging depth maps improves the performance of audio-visual dereverberation. Inspired by these techniques, we propose Geo-Mat $I_G$ to improve AV-RIR's understanding of geometry and material information of the environment. Our proposed $I_G$ is a 3-channel feature map constructed before training our AV-RIR using AC as the first two channels and the depth map as the last channel. We represent $I_G$ with 3 channels to encode $I_G$ using commonly used image encoder [249].

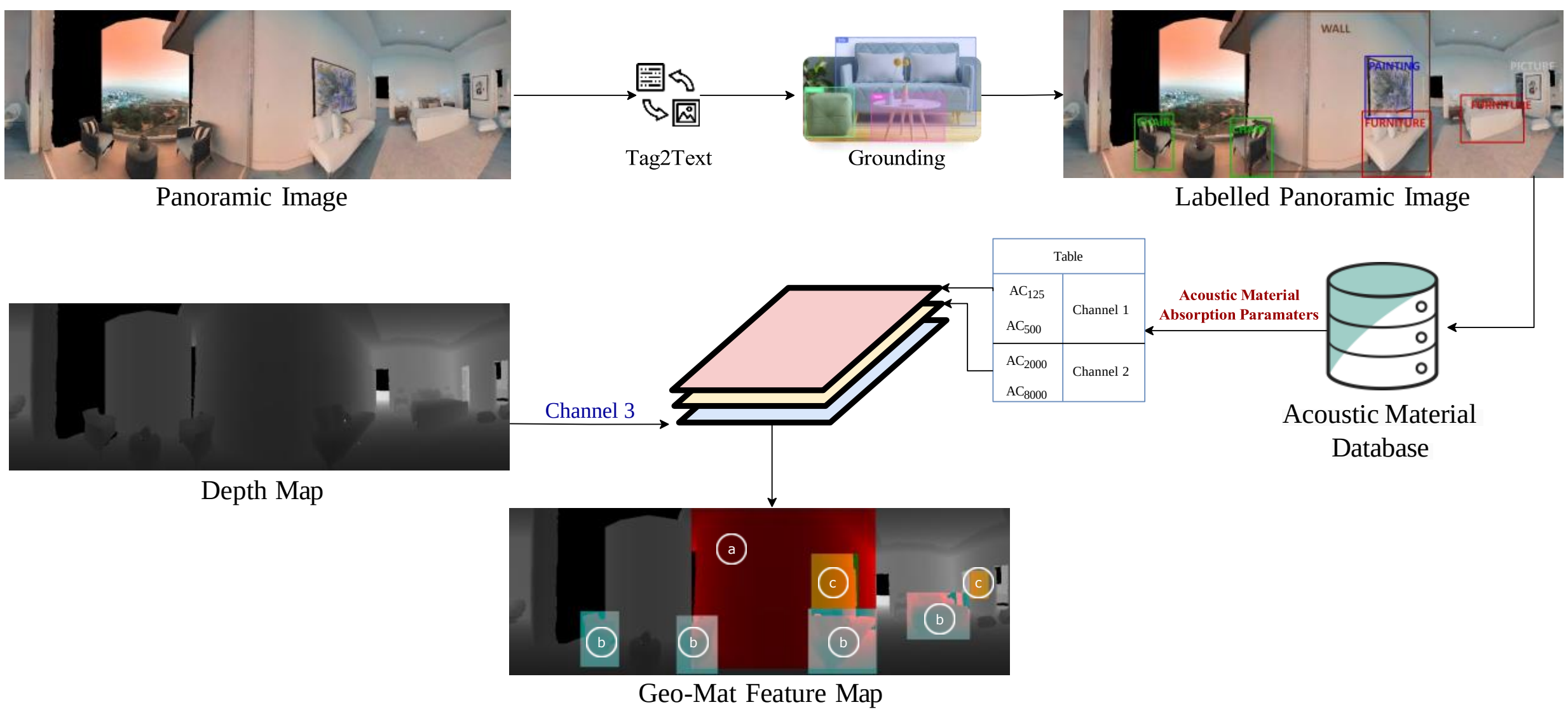

Figure 9.3: The computation pipeline of **Geo-Mat feature map**. The first two channels of the Geo-Mat feature ($I_G$) comprise the absorption coefficients (AC) of each acoustic material. The third channel comprises the depth map. We illustrate objects in the environment having similar AC with similar colors: chairs and furniture with similar materials are represented in light blue, painting, and wall pictures with similar materials are represented in yellow, and the rest in grey. More details on the method to obtain AC is described in Section 9.4.3.

**Obtaining material absorption coefficients (AC).** To obtain absorption coefficients AC of all materials in the environment from image $I_P$ , we employ a language-guided pipeline with SOTA pre-trained models. We first use Tag-2-Text [258], a SOTA object tagging model that identifies all objects in the image. This is followed by Grounding DINO [259, 260], which provides bonding box locations of each object identified by Tag-2-Text. Next, we use a large-scale room acoustic database with measured frequency-dependent AC to match the AC for each detected object [209]. For the matching operation, we adhere to a simple semantic matching technique to match the material names in the database to the ones detected using Tag-2-Text using embeddings from sentence transformer [261]. Precisely, we calculate an embedding $e_{I_P} \in \mathbb{R}^{768}$ for every object detected by Tag-2-Text and an embedding $e_M \in \mathbb{R}^{768}$ for every object in the database. Then we take the coefficient of the material in the database with the highest cosine similarity to $e_{I_P}$. The acoustic AC are frequency-dependent [26] and, therefore, we use the sub-band AC at 125 Hz, 500 Hz, 2000 Hz, and 8000 Hz to create the Geo-Mat feature $I_G$.

**Feature Map Construction.** After obtaining the absorption coefficients (AC) for each material in the panoramic image ($I_P$), we finally construct our 3-channel Geo-Mat feature $I_G$ (Equation 9.2). **Channel 1:** The material AC at low frequencies (i.e., 125 Hz and 500 Hz) **Channel 2:** The AC of the materials at high frequencies (i.e., 2000 Hz and 8000 Hz). **Channel 3:** The last channel of the $I_G$ represents the monocular depth map ($I_D$) of the $I_P$. Most datasets used in our experiments provide depth maps; however, for datasets that do not, we use the system provided by Godard *et al.* [262] to compute the depth map from a single image. We notice that the order of the channels does not matter.

$$
\begin{aligned}
I_G[:, :, 0] &= AC_{125} + AC_{500} * 16 \\
I_G[:, :, 1] &= AC_{2000} + AC_{8000} * 16 \\
I_G[:, :, 2] &= I_D
\end{aligned}
\tag{9.2}
$$

### 9.4.4 Training AV-RIR

$D_{QIR}$ and $D_{QS}$ are RIR and clean speech quantizers followed by a decoder, respectively, and CRIP represents CRIP. We estimate clean speech ($\hat{S}_C$), RIR ($\hat{RIR}$), and reverberant speech ($\hat{S}_R$) as shown below.

$$\hat{S}_C = D_{QS}(E_P(I_P), E_S(S_R)).$$

$$R\hat{I}R = D_{QIR}(E_{GM}(I_{GM}), E_{IR}(S_R)) + CRIP(I_P).$$

$$\hat{S}_R = \hat{S}_C \circledast R\hat{I}R. \tag{9.3}$$

**RIR Estimation Loss.** We calculate the time-domain mean squared error (MSE) for the estimated RIR as follows.

$$L_{MSE} = \mathbb{E}[||RIR - R\hat{I}R||_2]. \tag{9.4}$$

To train our RIR estimation RVQ codebook, we use the exponential moving average loss proposed in [252] as our vector quantizer (VQ) loss $L_{VQ}$.

**Speech Dereverberation Loss.** For solving the speech dereverberation task, we optimize three losses:

**(1) Mel-Spectrogram (Mel) loss ($L_{Mel}$).** The mel-spectrogram loss helps improve the perceptual quality of the predicted enhanced speech [263]. The 1D waveform ($\hat{S}_R$) output from the HiFi-GAN vocoder is first converted to the Mel-spectrogram, transforming it from the time domain to the frequency domain representation. A drawback of the Mel loss is its fixed resolution. The window length determines whether it has a good frequency or time resolution (i.e., a wide window has good frequency resolution and poor time resolution). Therefore, we calculate $L_{Mel}$, over a range of window lengths $W_L$ = {64, 128, 256, 512, 1024, 2048, 4096}. Formally, $L_{Mel}$ can be defined as:

$$L_{MEL} = \mathbb{E}[||MEL(S_R)-MEL(\hat{S}_R)||_1 + ||MEL(S_C)-MEL(\hat{S}_C)||_1], \tag{9.5}$$

where MEL() is the operation that converts time-domain speech into its mel-spectrogram representation.

**(2) Short-Time Fourier Transform (STFT) loss ($L_{STFT}$).** The STFT loss helps in the high-fidelity reconstruction of the predicted enhanced speech [140]. The 1D waveform is first converted to the frequency domain by applying the STFT() operation. The STFT of a waveform can be represented as a complex spectrogram where $M_S$ represents the magnitude of the STFT and $P_S$ is the phase of the STFT. We map the phase angle of the STFT to the rectangular coordinate on the unit circle to avoid phase wraparound issues [140]. Our $L_{STFT}$ is the sum of magnitude loss $L_{MAG}$ and phase loss $L_{PH}$. Similar to $L_{Mel}$, we calculate $L_{STFT}$ over a range of window lengths in a similar setting.

$$L_{MAG} = \mathbb{E}[||M_S(S_R)-M_S(\hat{S}_R)||_2 + ||M_S(S_C)-M_S(\hat{S}_C)||_2].$$

$$L_P(x, \hat{x}) = \mathbb{E}[|| \sin(P_S(x)) - \sin(P_S(\hat{x})))||_2 + || \cos(P_S(x)) - \cos(P_S(\hat{x}))||_2].$$

$$L_{PH} = L_P((S_R, (\hat{S}_R) + L_P(S_C, \hat{S}_C).$$

$$L_{STFT} = L_{MAG} + L_{PH} \tag{9.6}$$

Our total metric loss $L_{METRIC}$ (including RIR estimation and speech derverberation) is described in Equation 9.7.

$$L_{METRIC} = L_{MEL} + \lambda_1 L_{STFT} + \lambda_2 L_{MSE}, \tag{9.7}$$

where $\lambda_1$and $\lambda_2$ are the weights.

**(3) Adversarial loss ($L_{ADV}$)** In addition to metric loss, we train our network using adversarial loss. We train separate discriminator networks $D_R$ and $D_S$ for reverberant and clean speech respectively. Our $L_{ADV}$ is described in Equation 9.8.

$$L_{ADV} = \mathbb{E}[\max(0, 1\text{-}D_R(\hat{S}_R)) + \max(0, 1\text{-}D_S(\hat{S}_C))], \tag{9.8}$$

We train speech dereverbaration RVQ codebook using VQ loss $L_{VQ2}$ [252]. Equation 9.9 presents our total generator loss $L_{Gen}$. In Equation 9.9, $\lambda_1$ and $\lambda_2$ are the weights.

$$L_{GEN}(x) = L_{METRIC} + \lambda_1 L_{ADV} + \lambda_2(L_{VQ} + L_{VQ2}), \tag{9.9}$$

**Discriminator ($D_R$, $D_S$ ).** We use the multi-period discriminator network (MPD) proposed by HiFi-GAN [256] and the multi-scale discriminator network (MSD) proposed in MelGAN. MPD effectively captures the periodic details by having several sub-discriminators, each handling different parts of the input audio. MSD captures consecutive patterns and long-term dependencies.

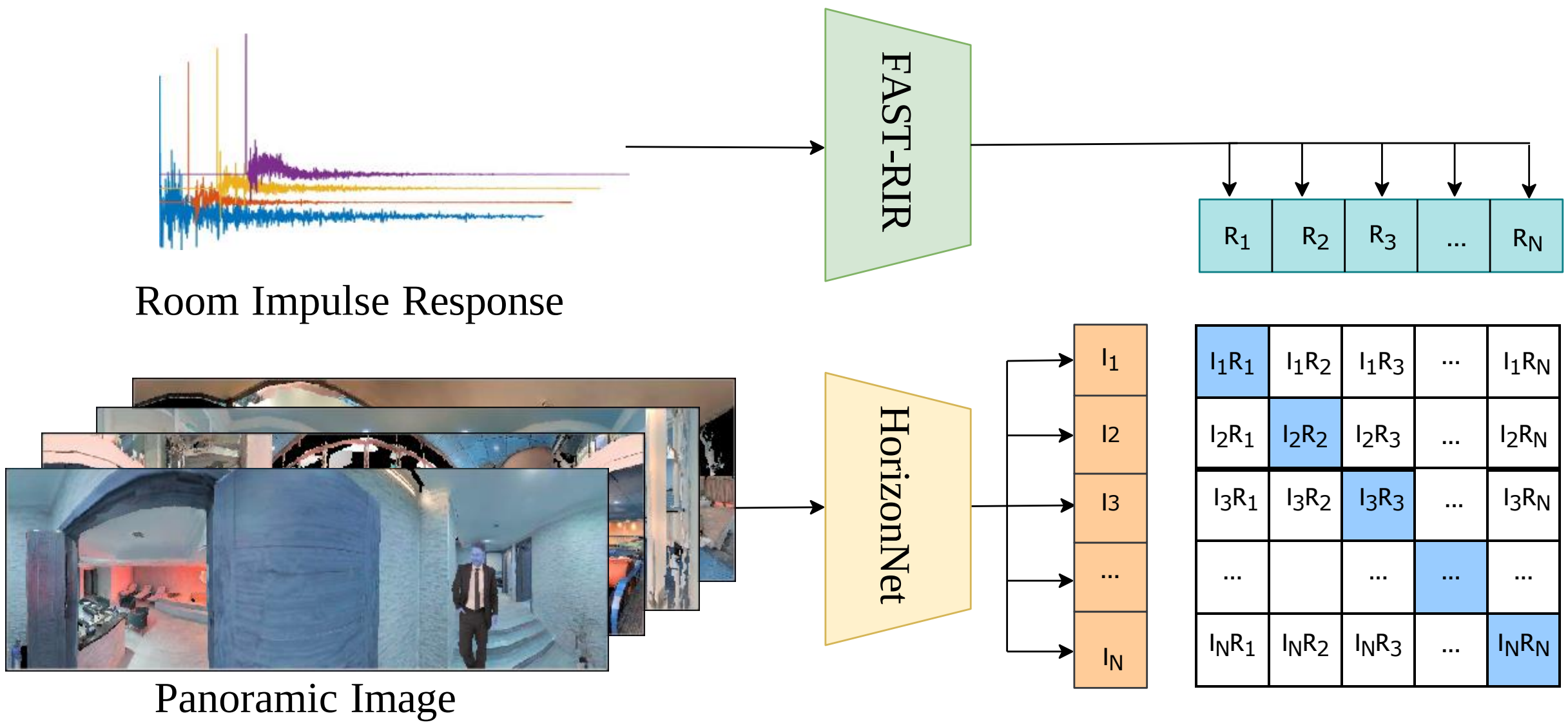

Figure 9.4: Illustration of **CRIP** training. Like CLIP [7], we propose two networks, one to encode a panoramic image and the other to encode the RIR to learn a joint embedding space between both. We use our CRIP-based image-to-RIR retrieval during inference to improve late reverberation in the estimated RIR from AV-RIR.

### 9.4.5 Contrastive RIR-Image Pretraining (CRIP)

We propose CRIP, a model built on the fundamentals of CLIP [7], that learns a joint embedding space between the panoramic image ($I_P$) and their corresponding RIRs. Similar to CLIP, CRIP employs two encoders, a pre-trained HorizonNet encoder [264] $E_H$ which serves as our $I_P$ encoder, and the discriminator network proposed in FAST-RIR [2], which serves as our RIR encoder $E_F$. Formally, $E_H$ takes as input $I_P$ and outputs an embedding $I_{EM} \in \mathbb{R}^{N \times 1024}$, and $E_F$ takes as input RIR $\in \mathbb{R}^{N \times 1 \times 4096}$ and output an embedding $R_{EM} \in \mathbb{R}^{N \times 1024}$. Finally, we measure similarity by calculating the dot product between the two as follows:

$$
\begin{aligned}
C_{r2i} &= \tau * \left(\text{RIR} \cdot \text{I}_{EM}^{\top}\right) \\
C_{i2r} &= \tau * \left(\text{I}_{EM} \cdot \text{RIR}^{\top}\right)
\end{aligned}
\tag{9.10}
$$

where $\tau$ is the temperature. This is followed by calculating the RIR-to-Image loss $\ell_{r2i}$ and the Image-to-RIR loss $\ell_{r2i}$ as follows:

$$\ell_{r2i} = \frac{1}{N}\sum_{i=0}^{N} \log \text{diag}(\text{softmax}(C_{r2i}))$$
$$\ell_{r2i} = \frac{1}{N}\sum_{i=0}^{N} \log \text{diag}(\text{softmax}(C_{i2r})) \tag{9.11}$$

Finally, we optimize the average of both losses:

$$L = 0.5 * (\ell_{r2i} + \ell_{i2r}) \tag{9.12}$$

**Why CRIP?** Neural-network-based RIR estimators are known to inaccurately approximate the late components of the RIR as a sum of decaying filtered noise [4]. Similarly, while our codec-based AV-RIR can accurately estimate the early components with structured impulsive patterns, it cannot precisely estimate late reverberation, which generally contains noise-like components. Thus, we propose CRIP to fill this gap. CRIP uses HorizonNet that captures room geometry/layout information [264]. The late components of RIR depend on the geometry of the room [265].

**CRIP for AV-RIR Inference.** During inference, for $I_P$ of the target scene, we retrieve an RIR from a datastore DS. The retrieval is performed by calculating cosine similarity between the CRIP embeddings of $I_P$, denoted as $I^t{}_{EM}$, and the RIR embeddings for all RIRs in datastore DS. DS is generally a large collection of RIRs in the wild, which in our case is composed of synthetic RIRs, more details on which can be found in Section 9.5. The final estimated RIR is obtained by replacing the late components of the original estimated RIR $R\hat{I}R$ with the late components of the retrieved RIR $RIR_{CRIP}$. We perform hyper-parameter tuning to find the optimal number of samples $S$ from $R\hat{I}R$ to replace with $R_{CRIP}$, and we found $S = 2000$ to give us the best improvements. This whole process can be formalized as:

$$R\hat{I}R[2000:4000] = RIR_{CRIP}[2000:4000]. \tag{9.13}$$

## 9.5 Experiments and Results

**Datasets.** For training and evaluation, we use the widely adopted SoundSpaces dataset [27, 83]. The SoundSpaces dataset provides paired reverberant speech and its RIR. The data is sourced by convolving simulated RIRs with clean speech from the LibriSpeech [29] dataset. The RIRs are simulated using a geometrical acoustic simulation techniques [35, 44] with environments taken from the Matterport3D dataset [247]. To additionally evaluate how AV-RIR fairs in speech dereverberation in real-world scenarios, we use web videos in the filtered AVSpeech dataset [248] proposed in VAM [9]. Since AVSpeech does not have ground truth (GT) RIR, we only used it to evaluate our speech dereverberation pipeline. Our datastore $DS$ comprises synthetic RIRs generated from SoundSpaces, excluding test set RIRs.

**Hyperparameters.** We train AV-RIR with a batch size of 16 for 400 epochs with only metric loss (Equation 9.7) and VQ loss ($L_{VQ}$, $L_{VQ_2}$). Later, we train with total loss (Equation 9.9) for 1K epochs. We use Adam Optimizer [266] with $\beta_1 = 0.5$, $\beta_2 = 0.9$ and learning rate $5 \times 10^{-5}$. For every 200K steps, we decay the learning rate by 0.5.

### 9.5.1 RIR Estimation

**Evaluation Metrics.** We quantitatively measure the accuracy of estimated RIR using standard room acoustic metrics. Reverberation time ($T_{60}$), direct-to-reverberant ratio (DRR), and early decay time (EDT) are the commonly used room acoustic statistics. We calculate the mean absolute difference between the acoustic statistics of estimated and ground truth (GT) RIRs as the error. $T_{60}$ measures the time taken for the sound pressure to decay by 60 decibels (dB), and EDT is 6 times the time taken for the sound pressure to decay by 10 dB. $T_{60}$ depends on the room size and room materials, and EDT depends on the type and location of the sound source [48]. DRR is the ratio between the sound pressure level of the direct sound source and the reflected sound [128]. We also report the mean square difference (MSE) between the GT and estimated early component (EMSE) and late component (LMSE) of the RIR in time domain. We show the benefit of RIR estimated from our approach in SLP tasks in our supplementary.

**Baselines.** We compare the performance of AV-RIR with six other baselines.

(1) **Image2Reverb [5]:** Predicts RIR from the camera-view image.

(2) **Visual Acoustic Matching (VAM) [9]:** Takes as input the source audio and the target environment image and outputs resynthesized audio matching the target environment. VAM does not explicitly estimate the RIR of the target environment. Therefore, we compare VAM only on our perceptual evaluation.

(3) **FAST-RIR++ [2]**: Takes as input the room geometry, source and listener positions, and $T_{60}$ to generate the RIR. We modified the architecture to the input panoramic image ($I_P$) of the environment and estimate the RIR (FAST-RIR++).

(4) **CRIP-only** *(ours)*: CRIP retrieves the closest RIR from the large synthetic dataset $DS$ using a panoramic image $I_P$ as input. We use it as a baseline to evaluate how much improvement our audio-visual network contributes.

(5) **Filtered Noise Shaping Network (FiNS) [4]:** An audio-only time domain RIR estimator from reverberant speech.

(6) **S2IR-GAN [6]:** An audio-only GAN-based reverberant speech-to-IR estimator.

**Results.** Table 9.1 compares our approach AV-RIR with audio-only and visual-only baselines. We can see that the audio-only baselines outperform the visual-only baselines. Audio and visual cues provide complementary information for RIR estimation, and we can see a significant boost in performance in our AV-RIR, which inputs audio and visual cues. AV-RIR outperforms the SOTA audio-only approach S2IR-GAN by 36%, 42%, 63%, 89% and 98% on $T_{60}$, DRR, EDT, EMSE, and LMSE, respectively.

### 9.5.2 Speech Dereverberation

**Evaluation Metrics.** We evaluate speech dereverberation using our AV-RIR on two downstream tasks: Automatic Speech Recognition (ASR) and Speaker Verification (SV). We use standard metrics such as Word Error Rate (WER) to measure ASR performance and Equal Error Rate (EER) to measure the SV. Following prior works [140, 141], we used pre-trained models from SpeechBrain [273] to evaluate the ASR and SV performance. AVSpeech dataset does not have parallel clean speech to perform ASR and SV tasks and we use the Reverberation Time Error

Table 9.1: We compare the RIR estimated using our AV-RIR with prior visual-only method (Image2Reverb [5]) and audio-only methods (FiNS [4] and S2IR-GAN [6]). We perform an ablation study to show the benefit of each component of our network.

| | Method | $T_{60}$ Error (ms) | DRR Error (dB) | EDT Error (ms) | EMSE ($x10^{-5}$) | LMSE ($x10^{-5}$) |
|---|---|---|---|---|---|---|
| | Image2Reverb [5] | 131.7 | 4.94 | 382.1 | 4907 | 1126 |
| | FAST-RIR++ [2] | 126.4 | 3.62 | 334.2 | 2630 | 990 |
| | FiNS [4] | 87.7 | 3.30 | 235.7 | 924 | 561 |
| | S2IR-GAN [6] | 63.1 | 3.04 | 168.3 | 730 | 310 |
| **ABLATION** | AV-RIR (Audio-Only) | 88.8 | 2.96 | 122.4 | 176 | 51 |
| | AV-RIR w Random | 77.6 | 2.67 | 109.2 | 124 | 6 |
| | AV-RIR w/o CRIP | 61.7 | 2.07 | 79.8 | 79 | 42 |
| | AV-RIR w/o Geo-Mat | 55.7 | 1.98 | 74.1 | 104 | 6 |
| | AV-RIR w/o Multi-Task | 77.8 | 2.56 | 105.4 | 144 | 6 |
| | AV-RIR w/o STFT Loss | 59.4 | 1.94 | 77.2 | 123 | 6 |
| | CRIP-only *(ours)* | 118.9 | 3.14 | 298.4 | 212 | 6 |
| | **AV-RIR *(ours)*** | **40.2** | **1.76** | **62.1** | **82** | **6** |

(RTE) metric [141] to evaluate the estimated clean speech from our AV-RIR.

**Baselines.** We compared the speech dereverberation using our approach with the following prior audio-only and audio-visual speech enhancement networks. (1) **Audio-Only:** WPE [270] is a statistical-model-based speech dereverberation network that cancels late reverberation without the knowledge of RIR. MetricGan+ [267], DEMUCS [268], HiFi-GAN [269], VoiceFixer [271], SkipConvGAN [134] and Kotha *et al.* [272] are learning-based speech dereverberation networks. (2) **Audio-Visual :** VIDA [140] is the first audio-visual speech dereverberation network that takes $I_P$ as visual cues. Recently, geometry-aware AdVerb [141] has been shown to achieve SOTA results in downstream speech tasks.

**MetricGAN++ [267]:** MetricGAN++ is an improvised version of the MetricGAN framework

Table 9.2: Performance comparison of AV-RIR with audio-only baselines (marked with ‡) and audio-visual baselines (marked with ∗) on the SLP tasks on the SoundSpaces dataset. "Reverberant" refers to clean speech convolved with ground-truth RIR. We also report RTE for real-world audio from the AVSpeech dataset.

| | **Method** | **WER (%) ↓** | **EER (%) ↓** | **(in sec)** |
|---|---|---|---|---|
| | Clean (Upper bound) | 2.89 | 1.53 | - |
| | Reverberant | 8.20 | 4.51 | 0.382 |
| | MetricGAN+ [267]‡ | 7.48 (+9%) | 4.67 (-4%) | 0.187 (+51%) |
| | DEMUCS [268]‡ | 7.97 (+3%) | 3.82 (+15%) | 0.129 (+66%) |
| | HiFi-GAN [269]‡ | 9.31 (-14%) | 4.32 (+4%) | 0.196 (+49%) |
| | WPE [270]‡ | 8.43 (-3%) | 5.90 (-31%) | 0.173 (+55%) |
| | VoiceFixer [271]‡ | 5.66 (+31%) | 3.76 (+16%) | 0.121 (+68%) |
| | SkipConvGAN [134]‡ | 7.22 (+12%) | 4.86 (-8%) | 0.119 (+69%) |
| | Kotha *et al.* [272]‡ | 5.32 (+35%) | 3.71 (+17%) | 0.124 (+68%) |
| | VIDA [140]∗ | 4.44 (+46%) | 3.97 (+12%) | 0.155 (+59%) |
| | AdVerb [141]∗ | **3.54 (+57%)** | 3.11 (+31%) | 0.101 (+74%) |
| **ABLATION** | AV-RIR (Audio-Only) | 5.24 (+36%) | 2.67 (+41%) | 0.055 (+86%) |
| | AV-RIR w Random Image | 4.85 (+41%) | 2.56 (+43%) | 0.049 (+87%) |
| | AV-RIR w/o Multi-Task | 4.57 (+44%) | 2.55 (+43%) | 0.048 (+87%) |
| | AV-RIR w/o CRIP | 4.67 (+43%) | 2.66 (+41%) | 0.049 (+87%) |
| | AV-RIR w/o Geo-Mat | 4.54 (+45%) | 2.21(+51%) | 0.044 (+88%) |
| | AV-RIR w/o MEL Loss | 4.44 (+46%) | 2.44 (+46%) | 0.048 (+87%) |
| | **AV-RIR *(ours)*** | 4.17 (+49%) | **2.02 (+55%)** | **0.042 (+89%)** |

where the discriminator network is trained with noisy speech. We use the implementation of MetricGAN in Speechbrain for our comparison [273].

**DEMUCS [268]:** DEMUCS is the music source separation architecture in the time-domain modified into a time-domain speech enhancer. DEMUCS can work in real-time on consumer-level CPUs.

**HiFi-GAN [269]:** HiFi-GAN is GAN-based architecture trained on multi-scale adversarial loss in both the time domain and time-frequency domain to enhance real-world speech recording to studio quality.

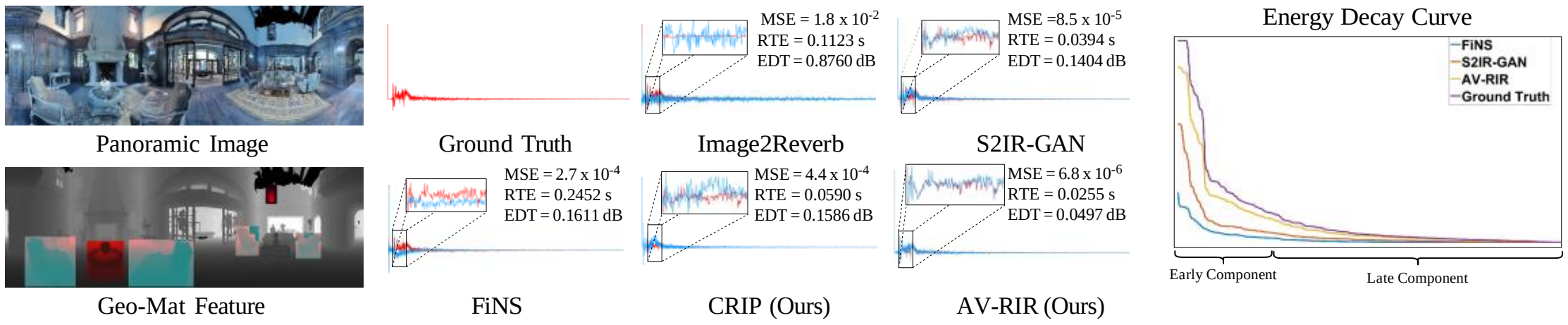

Figure 9.5: **Qualitative Results**. (Left) We show the Geo-Mat feature generated using our approach. The cushion chairs with a similar material absorption property are represented in green. The table and window with similar material are represented in red. (Right) We plot the time-domain representation of the RIRs estimated using prior methods and our approach with the ground truth (GT) RIR (GT: Red, Estimated: Blue). We also report the MSE (Equation 9.4), $T_{60}$ error (RTE), and EDT error (EDT). It can be seen that the RIR estimated using our AV-RIR matches closely with GT RIR when compared with the baseline. Also, we can see that the RIR retrieved from our CRIP has similar late components as the GT RIR. However, the early component of the retrieved RIR (shown in zoom) significantly differs from the GT. Our full AV-RIR pipeline estimates the early components of the RIR using audio-visual features and adds the late component of the RIR from our CRIP to accurately predict the full RIR. The energy decay curve (EDC) depicts the energy remaining in the RIR over time [8]. We can see that the EDC of the late component of RIR estimated from AV-RIR (yellow) matches closely with the GT RIR (purple).

**WPE [270]:** WPE is a statistical model-based speech dereverberation approach. WPE can perform dereverberation by removing late reverberation in a reverberant speech signal.

**VoiceFixer [271]:** Voice-fixer is a two-stage speech dereverberation approach. The analysis stage of the VoiceFixer is modelled using ResUNet and the synthesis stage is modelled using TF-GAN.

**SkipConvGAN [134]:** SkipConGAN is the GAN-based speech enhancement architecture where the Generator network estimates the complex time-frequency mask and the discriminator network helps to restore the formant structure in the synthesized enhanced speech.

**Kotha *et al.* [272]:** This speech enhancement network integrates the complex-valued TFA module with the deep complex convolutional recurrent network to improve the overall speech quality of the enhanced speech.

**VIDA [140]:** VIDA is the audio-visual speech dereverberation network that enhances reverberant speech. Visual input gives valuable information about the room geometry, materials and speaker positions.

**AdVerb [141]:** Adverb is a geometry-aware cross-modal transformer architecture, that predicts the complex ideal ratio mask. Clean speech is estimated by applying the complex ideal ratio mask to reverberant speech.

**Results.** Table 9.2 shows the benefits of speech dereverberation from our AV-RIR in the ASR and SV task. We can see that AV-RIR outperforms SOTA network AdVerb in SV tasks by 35% and outperforms all the baselines except for AdVerb in the ASR. To evaluate the robustness of AV-RIR, we test our network on recorded speech not used for training in the AVSpeech using the RTE metric. We can see that our work outperforms all the baselines by 60%.

**Ablation Study.** We perform a comprehensive ablation study to show the benefit of different components of our AV-RIR.

(1) **Multi-task learning.** To prove the effectiveness of the multi-task learning approach, we train the branches separately (AV-RIR w/o Multi-Task). Table 9.1 and Table 9.2 show that our multi-task learning approach benefits both tasks mutually. While the RIR estimation performance improves by 31% - 48%, speech dereverberation performance improves by 13% - 21%.

(2) **CRIP.** Table 9.1 shows that adding CRIP during inference for late reverberation improves

the late component MSE (LMSE) by 86%.

(3) **Geo-Mat feature.** Table 9.1 shows that the Geo-Mat feature improves RIR estimation accuracy by 11% - 28%.

(4) **Visual cues.** To prove the effectiveness of visual cues, we discard them while training and directly pass RIR and speech encoder outputs to RVQ. Additionally, we also discard CRIP and directly estimate the full RIR. Table 9.1 and Table 9.2 show that AV-RIR outperforms our audio-only AV-RIR variation in RIR estimation tasks by 41% - 55% and speech dereverbation task by around 24%.

### 9.5.3 Perceptual Evaluation

In Table 9.3 we report scores for perceptual evaluation of our estimated RIRs. For each environment, we provide the ground truth (GT) speech, generated speech using Image2Reverb [5], VAM [9], our AV-RIR and the environment image. The participants were asked to select the generated speech that sounds closer to the GT speech.

We performed our user study on 50 participants. We only allow participants to perform the survey on a laptop or desktop with headphones to get accurate results. Among the 50 responses, we filtered out noisy responses from our first questions. In the first question, we ask the participants which of the three synthetic reverberant speech matches closely to the ground-truth speech. We place ground truth speech among the synthetic speech and expect the participants with good hearing to choose the ground truth speech. We only counted the responses of 43 participants who chose the ground truth speech.

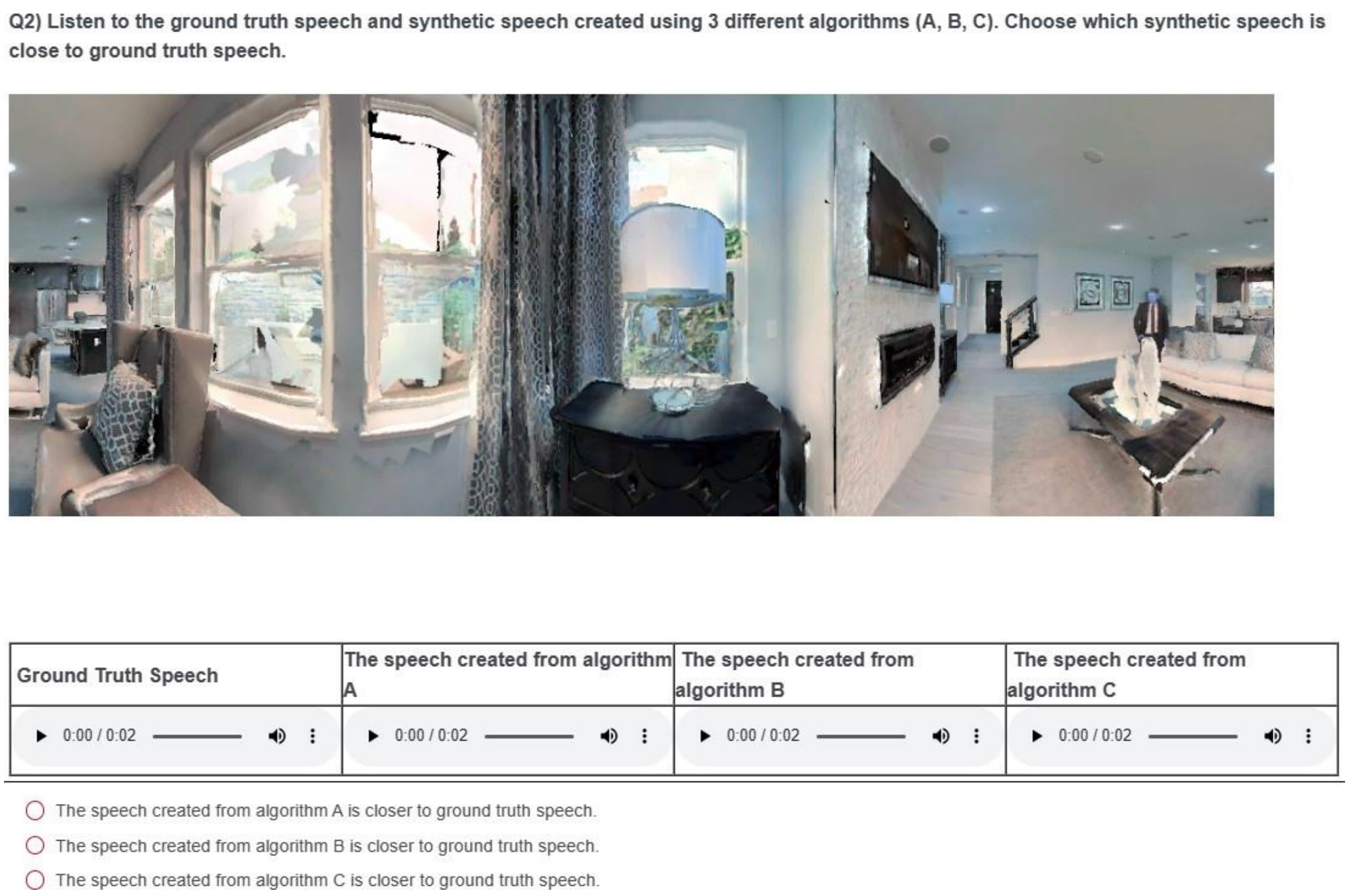

Figure 9.6: User study interface. We created 3 synthetic speech using our AV-RIR, Image2Reverb [5] and Visual Acoustic Matching [9] and asked the participants, which synthetic reverberant speech matches closely with ground-truth reverberant speech. For each question, we randomly shuffle the order of the synthetic reverberant speech from different approaches.

Out of 50 participants, 33 are male and 17 are female. The six participants are aged between 18-24 years, 28 participants are between 25-34 years and 16 participants are older than 34 years. Figure 9.6 shows the second question from our user study interface.

We select 6 scenes with varying complexity with $T_{60}$ ranging from 0.2 seconds to 0.7

Table 9.3: **Perceptual Evaluation**. Participants find that the reverberant speech generated using our AV-RIR is closer to GT reverberant speech when compared to visual-only baselines.

| Scene | $T_{60}$ | Image2Reverb [5] | VAM [9] | AV-RIR (Ours) |
|---|---|---|---|---|
| Scene 1 | 0.22 | 2% | 19% | **79%** |
| Scene 2 | 0.31 | 16% | 26% | **58%** |
| Scene 3 | 0.35 | 14% | 16% | **70%** |
| Scene 4 | 0.38 | 5% | 40% | **56%** |
| Scene 5 | 0.47 | 16% | 16% | **67%** |
| Scene 6 | 0.65 | 14% | 12% | **74%** |

seconds. We can see that, irrespective of the $T_{60}$ and the environment complexity, 56% to 79% of the participants said that the generated speech from AV-RIR closely matched the GT speech.

### 9.5.4 Supplementary Video

We provide a supplementary video[1] showing the qualitative results of RIR estimation with AV-RIR when applied to three different tasks. In addition, we compare the enhanced speech from our AV-RIR with ground truth clean speech. We also demonstrate our approach's failure cases in the supplementary video.

The RIRs estimated from our approach are evaluated in three practical tasks, that are:

**Novel View Acoustic Synthesis :** In the novel view acoustic synthesis task, given the audio-visual input from the source viewpoint, we modify the reverberant speech from the source viewpoint to sound as if it is recorded from the target viewpoint. We use reverberant speech as audio input, and the panoramic image and our proposed Geo-Mat feature as our visual input. We use SoundSpaces [83] dataset to perform this task.

[1] https://anton-jeran.github.io/AVRIR/

To perform this task, we estimate the enhanced speech using audio-visual input from the source viewpoint. We estimate the RIR corresponding to the target viewpoint from audio-visual input. We convolve the enhanced speech from the source viewpoint with RIR from the target viewpoint to make the speech from the source viewpoint sound as if it is recorded from the target viewpoint.

**Visual-Acoustic Matching :** In the visual-acoustic matching task, we resynthesize the speech from the source environment to match the target environment. We combine the enhanced source environment speech from AV-RIR and the estimated RIR from the target environment to perform this task. Convolving the estimated RIR with clean speech leads to synthesizing speech from the source environment to match the target environment. All our experiments on this task are performed on the SoundSpaces [83] dataset.

**Voice Dubbing :** Voice dubbing is replacing dialogue in one language with another in a video. Voice dubbing is commonly used to dub movies from one language to another. To test the robustness of RIR estimation from our AV-RIR, we estimated RIR using our AV-RIR on recorded video clips on YouTube. We chose two English video clips in the AVSpeech dataset [248]. We dubbed the video clips with French clean speech from Audiocite [274]. We convolved the French clean speech with the estimated RIR from the YouTube clip to match the room acoustics of French dialogue with the original English dialogue. We replaced the English dialogue with modified

French dialogue using our approach.

### 9.5.5 Far-field Automatic Speech Recognition

In order to evaluate the effectiveness of RIR estimated from our AV-RIR, we performed a Kaldi Far-field Automatic Speech Recognition (ASR) experiment using a modified KALDI ASR recipe[2]. For our experiment, we use the AMI corpus [155]. The AMI corpus has 100 hours of meeting recording. The meeting is recorded using both an individual headset microphone (IHM) and a single distant microphone (SDM). The IHM data has a high signal-to-distortion ratio when compared to the SDM data. Therefore, IHM data can be considered as clean speech.

To evaluate the benefit of RIRs estimated from our AV-RIR, we take a subset of SDM data with 300 speech samples and estimate the RIRs of the subset of SDM data. We create synthetic reverberant speech data by convolving clean speech from IHM data with the estimated RIR. We train the KALDI ASR recipe with and without modifying the IHM using our audio-only AV-RIR. We evaluated the audio-only version of our AV-RIR because there are no corresponding visual inputs in the AMI corpus. We test the trained model on far-field SDM data. We use word error rate as our metric to evaluate the performance of the speech recognition system. A lower word error rate indicates improved performance.

Modifying IHM data using our audio-only AV-RIR will bridge the domain gap between the training and test data. From Table 9.4 we can see that modifying the IHM data with our audio-only AV-RIR improves the word error rate by 12%.

[2] https://github.com/RoyJames/kaldi-reverb/tree/ami/

Table 9.4: Far-field ASR results. We train the Kaldi ASR recipe with and without modified IHM data and test on SDM data. We modify the IHM data by convolving RIR estimated using our audio-only AV-RIR.

| **Training Dataset** | **Word Error Rate** ↓ [%] |
|---|---|
| IHM without Modification | 64.2 |
| **IHM ⊛ AV-RIR (ours)** | **52.1** |

### 9.5.6 Societal Impact

Our model to estimate the RIR and enhance speech can have positive impacts on real-world applications. For example, the model can give an immersive experience in AR/VR applications and improve voice dubbing in movies. Also, our can be useful for different speech processing applications such as automatic speech recognition systems, telecommunication systems etc. We trained and evaluated our network on open-sourced publicly available datasets. We got the certification and license to perform user studies from the Institutional Review Board and we followed their protocols. We did not collect any personal information from the participants.

## 9.6 Conclusion, Limitations and Future Work

We propose AV-RIR, a novel multi-modal multi-task learning approach for RIR estimation. AV-RIR leverages both audio and visual cues using a novel neural codec-based multi-modal architecture and solves speech dereverberation as the auxiliary task. We also propose Contrastive RIR-Image Pre-training (CRIP), which improves late reverberation components in estimated

RIR using retrieval. Both quantitative metrics and perceptual studies show that our AV-RIR significantly outperforms all the baselines. We evaluate the speech dereverberation performance on the recorded AVSpeech dataset not used for training and observe that our approach outperforms the baselines by 60%.

AV-RIR assumes stationary single-talker input speech or single-source audio without noise. Future work aims to tackle multi-channel RIR estimation and RIR estimation from noisy, multi-source environments with moving sources.

# Chapter 10: Summary and Future Work

## 10.1 Summary

In this dissertation, we utilize learning-based techniques to render realistic audio and augment far-field speech at interactive rates in both real and virtual environments, with the overall goal of enhancing the performance of gaming, augmented reality, virtual reality, and speech applications. We proposed innovative neural network models for fast room impulse response generation, aiming to produce real-time and plausible acoustic effects. Our fast room impulse response generator demonstrates improved performance over the previous GPU-based interactive RIR generator [28] on the Kaldi ASR benchmark [1] for AMI Corpus [155]. We conducted perceptual evaluations and analyzed the acoustic characteristics of room impulse responses generated by our proposed learning-based technique, finding that our approach's performance is comparable to previous interactive geometric sound propagation methods while being two orders of magnitude faster.

We also developed a GAN-based model called IR-GAN to augment room impulse responses (RIRs) by learning the data distribution of real-world RIRs and generating new RIRs with desired acoustic characteristics. Our results demonstrate that RIRs produced by IR-GAN outperform those generated by the state-of-the-art geometric acoustic simulator in a modified Kaldi LibriSpeech ASR recipe [24].

[1] `https://github.com/RoyJames/kaldi-reverb/`

The exact positions of the speaker and listener in real RIRs recorded from real-world environments are not always known. In contrast, we know the precise speaker and listener positions of synthetic RIRs generated using room impulse response simulators and learning-based RIR generators. While thousands of synthetic RIRs can be easily produced for any given 3D environment, a performance gap exists between synthetic and real RIRs. To bridge this gap, we introduce the TS-RIR-GAN architecture for RIR-to-RIR translation between the synthetic and real domains without the need for paired examples, ensuring reversible and consistent translation between synthetic and real RIRs using cycle consistency loss. Additionally, we perform sub-band equalization on the translated RIRs to bridge the frequency gain gap between real and synthetic RIRs across the entire frequency range. Our approach maintains the precise speaker and listener positions while enhancing the quality of synthetic RIRs. We demonstrate the improved performance of the ASR system trained on the enhanced synthetic RIRs using our approach on the modified Kaldi LibriSpeech ASR Benchmark [24].

Finally, we developed an RIR estimator to estimate RIRs from a reverberant speech signal, with or without visual cues of the corresponding environment. Unlike traditional RIR generators, RIR estimators do not require a 3D representation of the environment. We demonstrate the benefits of RIR estimation in the context of an ASR system using the Kaldi AMI ASR Benchmark [146], as well as in augmented and virtual reality applications through perceptual evaluation.

Our results demonstrate that our techniques for room impulse response generation, augmentation, improvement, and estimation can render thousands of high-quality audio outputs at interactive rates, significantly enhancing the performance of real-world audio processing tasks in real-time. Each chapter provides a detailed discussion of the limitations associated with our methodology.

## 10.2 Future Work

In the future, we aim to overcome the limitations outlined in each chapter and investigate the following prospective directions.

### 10.2.1 RIR Estimation in a Multiple Moving Source Environment

Estimating RIRs in environments with multiple moving sound sources is particularly challenging because each sound source has a unique RIR that depends on its specific location, and these locations can change over time. Due to the inherent complexity, previous studies have typically simplified the problem by estimating a single representative RIR from a source positioned 1.5 meters away from the listener [275], without considering the varying locations of multiple sound sources. In our dissertation, we proposed and evaluated a novel audio-visual RIR estimator (AV-RIR) specifically designed for a single-speaker scenario. Looking forward, we aim to leverage visual cues to identify the number of sound sources, their relative positions, and their movement. This approach will enable the estimation of unique RIRs for each moving sound source, offering a more accurate and dynamic understanding of sound propagation in complex environments.

### 10.2.2 Learning-Based RIR Generator for Dynamic Environment

Our proposed learning-based RIR generator can efficiently produce 10,000 impulse responses per second for various speaker and listener positions within a given indoor 3D scene. However, the current approach requires approximately 0.04 seconds for mesh-to-graph conversion, which limits its efficiency in dynamic environments where object positions (e.g., furniture) within the indoor 3D scene change over time. To address this, instead of regenerating the entire mesh for each new scene, we plan to explore representing the relative changes in the 3D scene using a simplified model. This could significantly accelerate our method, making it more suitable for dynamic scenarios.

### 10.2.3 Differentiable Physics-Based Room Impulse Response Generators

A differentiable room impulse response (RIR) generator is a mathematical model that simulates the room impulse response of an acoustic environment, enabling gradient-based optimization. Differentiable physics-based RIR generators have shown promising results in addressing inverse problems, such as optimizing source and receiver placement through gradient-based methods. Moreover, parametric differentiable RIR generators require significantly fewer RIRs for training [276]. However, previous physics-based differentiable RIR generators were designed for empty, shoebox-shaped rooms and only modeled specular reflections. These assumptions do not hold in practical scenarios. Extending this approach to complex, real 3D environments and incorporating acoustic wave properties is challenging due to the lack of a closed-form solution.

### 10.2.4 Novel-View Acoustic Synthesis

In novel-view acoustic synthesis tasks, we aim to estimate the sound at any point in a scene from audio recordings captured by ”M” microphones containing unknown ”S” sound sources within the same real 3D environment [277]. Previous approaches have addressed this task under the assumption that the RIRs are available, which is impractical in real-world scenarios. A promising direction for future research is to estimate the ”S” number of RIRs and their locations directly from the audio recordings captured by the ”M” microphones containing ”S” sources.